\documentclass[11pt,a4paper]{article}
\usepackage[margin=1in]{geometry} 
\usepackage{graphicx}
\usepackage{amssymb,amsmath,amsfonts}
\usepackage{amsthm}
\usepackage{xurl}
\usepackage{siunitx}
\usepackage{listings} 
\usepackage{enumitem}
\usepackage{bm}

\usepackage{lmodern} 
\usepackage{anyfontsize} 

\newcommand{\yp}[1]{{\color{black}{#1}}}
\newcommand{\yb}[1]{{\color{black}{#1}}}

\usepackage{hyperref}
\urldef{\myurla}\url{https://depts.washington.edu/bdecon/workshop2012/auto-tutorial/documentation/overview\%20auto07p\%20essentials.pdf}
\urldef{\myurlb}\url{https://sites.pitt.edu/\~phase/bard/bardware/xpp/help/xppauto.html}

\newcommand{\ve}{\varepsilon}
\newcommand{\pa}{\partial}

\newcommand{\vc}{\text{vec}}

\newcommand{\tr}{\top}
\newcommand{\dd}{\,\mathrm{d}}

{
	\theoremstyle{plain}
	\newtheorem{assumption}{Assumption}
}

\newtheorem{remark}{Remark}

\usepackage{multirow}%
\usepackage{mathrsfs}%
\usepackage[title]{appendix}%
\usepackage{xcolor}%
\usepackage{textcomp}%
\usepackage{manyfoot}%
\usepackage{booktabs}%
\usepackage{authblk}

\usepackage{caption}
\usepackage{easy-todo}

\title{n:m Phase-Locking of Coupled Oscillators with Nonlinearities in Coupling Strength and Heterogeneity}
\author[1]{Youngmin Park\footnote{Corresponding author park.y@ufl.edu}}

\affil[1]{Department of Mathematics, University of Florida, Gainesville, FL 32611}

\date{}

\begin{document}

\maketitle





\abstract{We introduce a scalar reduction method for forced or coupled systems with nonlinearities in both heterogeneity and coupling strength. Heterogeneity is formulated as a relatively weak but nonlinear alteration of the vector field(s). The method can be used to determine the existence and stability of $n{:}m$ phase-locked states in a variety of forced or coupled biological oscillator models, including the nonradial isochron clock, a thalamic neural oscillator, and the Van der Pol oscillator. The proposed scalar reduction successfully captures the emergence and disappearance of phase-locked states as a function of nonlinear coupling strength and nonlinear heterogeneity. We find that even small amounts of heterogeneity can significantly alter phase-locked states in ways that cannot be captured by assuming identical oscillators. The proposed method enables a reduction and analysis of high-dimensional systems of coupled oscillators in more biologically realistic settings.}

\maketitle

\section{Introduction}

Coupled oscillators are ubiquitous in natural and physical systems. Pacemaker neurons collectively generate rhythms in the nervous systems of vertebrates and invertebrates \cite{butera1999models,winfree2001geometry,izhikevich2007,marder2022new,zang2023neuronal}. Walking patterns of crowds on bridges may elicit a resonance response, where the bridge acts as a mechanical coupling between individuals' gait cycles \cite{strogatz2005crowd,ott2008low}. Chemical preparations such as the famed Belousov–Zhabotinsky reaction exhibit spatio-temporal patterns through diffusive coupling \cite{kuramoto84,epstein1998introduction}. In particular, $n{:}m$ phase-locking, where one coupled oscillator traverses $n$ periods in the same time that the other traverses $m$ periods, is a natural feature of physical systems such as microelectromechanical system (MEMS) oscillators \cite{bhaskar2021integer}, pulse-coupled cardiac pacemakers \cite{mirollo1990synchronization}, and Belousov–Zhabotinsky chemical oscillators with asymmetric coupling \cite{horvath2019phase}. 

Homogeneity and symmetry are common assumptions in the study of coupled oscillators that turn relatively complex models and behaviors into amenable forms for mathematical analysis. Examples include the celebrated Ott-Antonsen ansatz \cite{ott2008low} and recent fashionable works on $N$-body interactions on simplicial complexes \cite{skardal2023multistability,adhikari2023synchronization,zhang2024deeper}. Homogeneity and symmetry enable detailed analytical studies of the existence and stability of $n{:}m$ phase-locked states, even for strong coupling strengths and high-dimensional oscillators, e.g., through the application of group theory \cite{golubitsky1986hopf,golubitsky2003symmetry,golubitsky2006symmetry} and firing time maps \cite{canavier2009phase}. It is also common to use particular models that are analytically tractable, such as the Van der Pol oscillator \cite{bhaskar2021integer}, circle map-like systems \cite{glass1982fine}, and the R\"{o}ssler oscillator \cite{chen2001phase}.

However, in biological systems, heterogeneity is an inescapable feature that alters the behavior of coupled oscillators in the physical world. Experimental preparations of chemical oscillators with only 3\% variation in intrinsic frequencies disrupt the existence and stability of phase-locked states predicted using group theory \cite{hunter2022pattern}. Introducing heterogeneity in ecosystem models has been shown to have significant effects on phase dynamics \cite{goldwyn2009small,milne2021coupled}. Biological nervous systems exhibit high degrees of heterogeneity \cite{stein2005neuronal,urban2012circuits}, where heterogeneity may even be a necessary feature of robust brain function \cite{lengler2013reliable}. The ubiquity of heterogeneity in nature warrants exploration in this direction. However, general phase-reduction methods that incorporate heterogeneity are relatively lacking. \yp{See Section \ref{sec:nonlinear_het} for additional biological background motivating the present study along with a mathematical formulation of heterogeneity. In brief, heterogeneity is formulated as a relatively weak but nonlinear alteration of the vector field(s).}

We introduce a method to reduce a pair of coupled oscillators to a scalar equation. The reduction accounts for $n{:}m$ phase-locking and heterogeneity. The proposed method uses a combination of phase and isostable coordinates employed in recent works \cite{budi12,castejon2013phase,wils16isos,shir17,wilson2019phase,wilson2020phase,park2021high,park2024body,nicks2024insights} \yp{(see Section \ref{sec:phase-iso} for additional background on phase-isostable coordinates)}. The proposed method does not require oscillators to be strongly attracting, such as in phase-based approaches  \cite{canavier2009phase}. It also does not require perturbed solutions to be arbitrarily close to the unperturbed limit cycle.

This paper is organized as follows. In Section \ref{sec:phase-isostable-intro}, we briefly discuss the phase-isostable reduction and establish notation used throughout this paper. In Section \ref{sec:derivation}, we introduce assumptions on the types of vector fields amenable to the reduction, then derive the phase-isostable reduction for $n{:}m$ coupling. We then apply this reduction to a forced nonradial isochron clock in Section \ref{sec:nric} and a forced thalamic model in Section \ref{sec:thal1}, followed by two coupling examples: one of a pair of coupled thalamic neurons in Section \ref{sec:thal_coupling}, and another of a pair of mixed model types in Section \ref{sec:vdp_thal_coupling}. We conclude with a discussion of the method's limitations, utility, and future directions in Section \ref{sec:discussion}.

\section{Background}\label{sec:phase-isostable-intro}

\subsection{Phase Reduction}\label{sec:background:phase_reduction}

Consider a general dynamical system,
\begin{equation}\label{eq:f0}
    \frac{\dd X}{\dd t} = F(X) + \ve U(X,t),
\end{equation}
where $X\in\mathbb{R}^n$ is the vector of state variables, $F:\mathbb{R}^n \rightarrow \mathbb{R}^n$ is a smooth vector field, and $U(X,t)\in\mathbb{R}^n$ is some additive input function. The function $U$ could represent an endogenous or exogenous forcing function, or a reciprocally coupled oscillator.

Let $\Gamma(t)$ be a stable $T$-periodic limit cycle satisfying \eqref{eq:f0} for $\ve=0$, which persists for some range of $\ve\neq 0$. The classic theory of weakly coupled (or weakly forced) oscillators seeks to understand how weak perturbations, $0 < \ve \ll 1$, affect the timing of oscillators. In the case of two oscillators, this timing is intuitively defined as the time difference when the oscillators cross a Poincar\'e section. Will they synchronize, i.e., will the time difference converge to zero? Or will they exhibit some other phase-locked behavior, where the time difference converges to some nonzero constant?

This time difference can be quantified by computing the phase of each oscillator then taking the phase difference. To compute an oscillator's phase, we transform \eqref{eq:f0} to a scalar variable, where the limit cycle $\Gamma$ is associated with values on the unit circle, $\theta \in [0,2\pi)$ (modulo $2\pi$). In the absence of coupling or forcing, $\ve=0$, it is straightforward to extend the idea of phase to the basin of attraction such that manifolds of initial conditions exhibit the same phase (isochrons) with constant phase advance,
\begin{equation*}
    \frac{\dd \theta}{\dd t} = \yp{\omega = 2\pi/T}.
\end{equation*}
Then, following the classic work of Kuramoto \cite{kuramoto84}, we invoke the chain rule to derive the phase equation,
\begin{equation}\label{eq:phase0}
    \begin{split}
    \frac{\dd\theta}{\dd t}&= \nabla \theta \cdot \frac{\dd X}{\dd t}\\
    &= \nabla \theta \cdot F(X) + \ve \nabla \theta \cdot U(X,t) \\
    &= \yp{\omega} + \ve Z(\theta) \cdot U(X,t).
    \end{split}
\end{equation}
The first term of the last line follows from our choice of $d\theta/dt=\yp{\omega}$ (literally, the directional derivative of the isochron in the direction of the vector field is always constant), and the second term is a simple relabeling of $\nabla\theta$ evaluated on the limit cycle, $Z(\theta) \equiv \nabla\theta|_\Gamma(\theta)$. \yb{Note that in the derivation in Section \ref{sec:derivation}, we will factor out $\omega$ so that we work with equations of the form
\begin{equation*}
    \frac{1}{\omega}\frac{\dd \theta}{\dd t} = 1 + \ve Z(\theta)\cdot U(X,t),
\end{equation*}
where a factor of $1/\omega$ is implicitly absorbed into $U$.}

Equation \eqref{eq:phase0} is an elegant  transformation of the limit cycle $\Gamma$ to a scalar variable $\theta$, where the order $\ve$ term quantifies how incoming perturbations alter the constant phase advance ($\omega$) of the unperturbed oscillator. Equation \eqref{eq:phase0} is the first of two components of the phase-isostable reduction to be used in this paper. 

\subsection{Isostable Coordinate}\label{sec:iso}

While the phase reduction \eqref{eq:phase0} has long been an exceptionally useful means to understand the existence and stability of phase-locked states \cite{ermentrout2010,park2017utility,norton2019dynamics}, it requires that the limit cycle $\Gamma$ is strongly attracting: perturbed solutions must return to the limit cycle within a single period for the phase reduction to be accurate. Thus, a natural limitation arises when a perturbed solution $X$ takes multiple periods to decay back to the unperturbed limit cycle solution $\Gamma$. Such instances occur naturally with either strong perturbations or weakly attracting limit cycles \cite{wils16isos}. Of the many methods at our disposal to address this limitation \cite{wils16isos,rosenblum2019numerical,wedg13,wils18operat,shir17}, we choose to augment our phase variable with \textit{isostable coordinates} -- coordinates that are defined as the level sets of the slowest decaying modes of the Koopman operator \cite{maur13,mezi20}. We describe an intuitive definition that closely follows \cite{wils16isos,wilson2020phase,park2024body}.

Let $U(X,t) = 0$ and define $\Delta X = X - \Gamma(\theta)$. Then, to a linear approximation, our original system \eqref{eq:f0} becomes,
\begin{equation} \label{deltaxeq}
	\Delta \dot{X} = J(t) \Delta X,
\end{equation}
where $J(t)$ is the Jacobian matrix of the vector field $F$ evaluated at $\Gamma(\theta(t))$.  Because \eqref{deltaxeq} is a linear system with time-varying coefficients that are $T$-periodic, we may utilize Floquet theory to summarize the dynamics of \eqref{deltaxeq}. Let $\Phi$ be the fundamental matrix, i.e., $\Delta X(T) = \Phi \Delta X(0)$ with initial condition $X(0)$ such that $\theta(X(0)) = 0$, and let $w_j, v_j$, and $\lambda_j$ be left eigenvectors, right eigenvectors, and associated eigenvalues, of $\Phi$, respectively, for $j=1,\ldots,n-1$. Floquet exponents may be obtained using $\kappa_j = \log(\lambda_j)/T$, and we may sort them without loss of generality, such that $\kappa_1$ is the slowest decaying nonzero Floquet exponent. If $\kappa_1$ is unique\footnote{It is possible for more than one isostable coordinates to decay relatively slowly, in which case it is necessary to compute additional isostable terms in addition to $\psi_1$ \eqref{isostabledef}. More information about how to handle such cases can be found in \cite{wils16isos,wilson2018greater}.}, an associated isostable coordinate can be defined in the basin of attraction of the limit cycle \cite{wilson2018greater}:
\begin{equation} \label{isostabledef}
	\psi_1(X) = \lim_{k \rightarrow \infty}(w_1^\tr(\eta(t^k_\Gamma,X) - \Gamma_0   )   \exp(-\kappa_1 t_\Gamma^k)),
\end{equation}
where $t_\Gamma^k$ denotes the time of the $k$th stroboscopic transversal of the $\theta = 0$ isochron, and $^\tr$ denotes the transpose. The function $\eta(t,X)$ denotes the unperturbed flow of the vector field that takes an initial condition $X(0)$ to a state $X(t)$ at time $t$, and the vector $\Gamma_0$ denotes the intersection between the periodic orbit and the $\theta = 0$ isochron. The isostable coordinate \eqref{isostabledef} gives a sense of the distance from the periodic orbit, with greater values of $|\psi_1(X)|$ corresponding to states that will take longer to approach the periodic orbit. Using Equation \eqref{isostabledef}, it is possible to show directly that when $U(X,t) = 0$,
\begin{equation}\label{eq:psi0}
    \frac{\dd \psi_1}{\dd t} = \kappa_1 \psi_1
\end{equation}
in the basin of attraction of the limit cycle \cite{wilson2018greater}. Intuitively, the amplitude $\psi_1$ will decay exponentially towards the limit cycle at a rate $\kappa_1$. If $\kappa_1$ is small, then $\psi_1$ will decay relatively slowly.

The corresponding isostable equation is straightforward to derive using the chain rule \cite{wils16isos,wilson2020phase},
\begin{equation}\label{eq:iso0}
\begin{split}
   	\frac{\dd \psi_1}{\dd t} &=\nabla \psi_1 \cdot \frac{\dd X}{\dd t} \\
	&= \nabla \psi_1 \cdot (F(X) + \ve U(X,t)) \\
	&= \kappa_1 \psi_1 + \ve I(\theta) \cdot U(X,t), 
\end{split}
\end{equation}
where the first term follows from \eqref{eq:psi0} and the second term is a simple relabeling of $\nabla\psi$ evaluated on the limit cycle, $I(\theta) \equiv \nabla\psi|_\Gamma(\theta)$.

Equation \eqref{eq:iso0} is a useful coordinate to track solutions that decay slowly back to the underlying limit cycle $\Gamma$. The order $\ve$ term quantifies how incoming perturbations alter the exponential decay of the amplitude. Equation \eqref{eq:iso0} is the second component of the phase-isostable reduction to be used in this paper.

\subsection{Phase-Isostable Reduction}\label{sec:phase-iso}

We continue following the description in \cite{wils16isos,wilson2020phase,park2024body} to outline the phase-isostable reduction that will be utilized extensively throughout this paper. We assume that all non-zero Floquet exponents except $\kappa_1$ have a large real component (in this case, all other isostable coordinates decay rapidly and are well approximated by zero). For notational convenience, we will simply use $\psi$ and $\kappa$ to denote the sole non-trivial isostable coordinate and its Floquet exponent.

With this isostable coordinate in hand, we now consider \eqref{eq:phase0} and \eqref{eq:iso0},
\begin{equation} \label{phaseisoeq}
\begin{split}
    \dot{\theta} &= \yp{\omega} + \ve\mathcal{Z}(\theta,\psi) \cdot U(X,t),\\
    \dot{\psi} &= \kappa \psi + \ve \mathcal{I}(\theta,\psi) \cdot U(X,t)
\end{split}
\end{equation}
where the gradient of the phase $\mathcal{Z}$ and the gradient of the isostable $\mathcal{I}$ are each evaluated on the state $X(\theta)$, which is not necessarily on the periodic orbit. Note that if $\psi=0$ and $\ve=0$, then $X(\theta)=\Gamma(\theta)$, and if $\mathcal{Z}$ and $\mathcal{I}$ are evaluated on $X(\theta)=\Gamma(\theta)$, then $\mathcal{Z}\equiv Z$ and $\mathcal{I} \equiv I$.

The only unknown terms remaining are $\mathcal{Z}$ and $\mathcal{I}$. These functions are computed using a Taylor expansion centered at $\psi = 0$:
\begin{align} \label{phaseampred}
	\dot{\theta} &= \yp{\omega} + \ve (Z^{(0)}(\theta) + \psi Z^{(1)}(\theta) + \psi^2 Z^{(2)}(\theta) + \dots) \cdot U(X,t), \\
	\dot{\psi} &= \kappa \psi + \ve(I^{(0)}(\theta) + \psi I^{(1)}(\theta) + \psi^2 I^{(2)}(\theta) + \dots) \cdot U(X,t),
\end{align}
where $Z^{(j)}$ and $I^{(j)}$ correspond to the $j^{th}$ order terms in the expansions. These \yp{first- and second}-order expansion terms are straightforward to compute by solving initial value problems \cite{wilson2020phase}. These functions are useful when the nonlinear effects of coupling or forcing strength $\ve$ become nontrivial \cite{wilson2019phase,park2021high,park2024body,nicks2024insights}. From this point forward, any time we mention $\mathcal{Z}$, $\mathcal{I}$, or their Taylor expansion terms $Z^{(j)}$ and $I^{(j)}$, we assume that they have already been computed numerically.

\yp{
\subsection{Nonlinear Heterogeneity}\label{sec:nonlinear_het}

Biological systems are heterogeneous by default. Even two identical neuron types of similar morphology, identical gene expression, with the same function in the same network will exhibit similar but non-equal firing frequencies. This difference is due to cell-to-cell variability in membrane conductances \cite{gunay2008channel}, channel density \cite{golowasch1999activity,migliore2018physiological}, and activity-dependent changes \cite{desai1999plasticity,galante2001homeostatic}. The question of how nervous systems are able to generate reliable rhythms despite such individual differences is a fundamentally important biological question of broad interest \cite{golowasch2002failure,marder2011variability,marder2011multiple,goaillard2021ion}. In contrast, computational neural models often use average data in model parameters, despite strong evidence showing that individual variability is a feature of neural computation \cite{marder2011variability}. 

While there exist studies that consider various types of heterogeneity in biological oscillators (such as most works with the Kuramoto oscillator and studies using mean-field methods \cite{laing2014derivation,chen2022exact}), these are often in the limit of large neuron number. In contrast, the present study is motivated by experiments performed in the crab nervous system, where neural networks may contain fewer than a dozen neurons \cite{nusbaum2002small}. In this context, mathematical results that rely on arbitrarily large numbers of neurons are invalid, and it is a natural question to ask how to predict phase-locked states of small numbers of neurons that have similar periods but potentially very different vector fields.}

\yp{There are two common biological reasons why two neural oscillators may have similar periods: in the first case, the neurons differ by small amounts in their vector fields or parameters, and in the second case, the neurons have very different underlying parameters, but operate in a robust parameter regime that gives rise to similar behavior. We mathematically formulate these two cases as follows.}

\yp{\paragraph{Case 1: Two almost identical neurons}
We first motivate our formulation of heterogeneity through a concrete neurobiologically realistic example. Consider the Morris-Lecar model \cite{morris1981voltage},
\begin{equation}\label{eq:ml}
\begin{split}
    C\frac{\dd V}{\dd t} &= I -\bar{g}_\text{L} (V-E_\text{L})-\bar{g}_\text{Ca}m_{\infty}(V)(V-V_\text{Ca}) - \bar{g}_\text{K}n(V-V_\text{K}),\\
    \frac{\dd n}{\dd t} &= \frac{n_\infty(V)-n}{\tau_n},
\end{split}
\end{equation}
where
\begin{align*}
    m_\infty(V) &= [1+\tanh((V-U_1)/U_2)]/2,\\
    n_\infty(V) &= [1+\tanh((V-U_3)/U_4)]/2,\\
    \tau_N &= 1/[\phi \cosh((V-U_3)/(2U_4)].
\end{align*}
By assumption, $\bar{g}_\text{L}$, $\bar{g}_\text{Ca}$, and $\bar{g}_\text{K}$ are averaged after taking measurements across multiple cells. However, in a biological preparation, there exists a distribution of conductances for a collection of $N$ neurons, where the variability is not due to measurement error. For concreteness, suppose that we know the distribution for the parameter $\bar{g}_\text{L}$, and let $\delta_i = \bar{g}_\text{L} - g_{\text{L},i}$, where $g_{\text{L},i}$ is the particular conductance for cell $i$ for $i=1,\ldots,N$. Thus, for a particular neuron, its voltage equation becomes,
\begin{align*}
    C\frac{\dd V_i}{\dd t} &= I -\bar{g}_\text{L} (V_i-E_\text{L})-\bar{g}_\text{Ca} m_{\infty}(V_i)(V_i-V_\text{Ca}) - \bar{g}_\text{K}(V_i-V_\text{K})\\
    &=I -(\bar{g}_{\text{L},i}+\delta_i) (V_i-E_\text{L})-\bar{g}_{\text{Ca}}m_{\infty}(V_i)(V_i-V_\text{Ca}) - \bar{g}_{\text{K}}(V_i-V_\text{K})\\
    &=I -\bar{g}_{\text{L},i}(V_i-E_\text{L})-\delta_i(V_i-E_\text{L})-\bar{g}_{\text{Ca}}m_{\infty}(V_i)(V_i-V_\text{Ca}) - \bar{g}_{\text{K}}(V_i-V_\text{K}).
\end{align*}
If we define $J(V):= -(V-E_L)$, then
\begin{equation*}
    C\frac{\dd V_i}{\dd t} = I -\bar{g}_{\text{L},i}(V_i-E_\text{L})-\bar{g}_{\text{Ca}}m_{\infty}(V_i)(V_i-V_\text{Ca}) - \bar{g}_{\text{K}}(V_i-V_\text{K})+\delta_iJ(V_i).
\end{equation*}
In other words, the vector field for this particular neuron is identical to the vector field using only the average parameters (which we will refer to as the average neuron), except for a possibly nonlinear term that depends on the parameter difference $\delta_i$. Thus, variability in membrane conductances and channel density may be captured as an additive term in the Morris-Lecar model. This parameter difference generally yields a different firing rate between the particular neuron $i$ and the average neuron. 

This example calculation generalizes naturally to any conductance-based neural model and we are also not restricted to models of neural oscillators. To simplify the discussion, we restrict our attention to at most one heterogeneous parameter per neuron, although a similar observation holds in the case of multiple heterogeneous parameters. There is also the caveat that $\delta_i$ must not be too large, but this is also a natural assumption for the following reasons.

In general conductance-based neural models (including the Morris-Lecar model \eqref{eq:ml}), nonlinear heterogeneity may naturally appear in any of the $U_i$ parameters, which correspond to thresholds for various ion channels and are generally non-equal from cell-to-cell. Standard deviations for parameters directly related to $U_i$ are on the order of $\pm$\SI{5}{mV} \cite{boulet2017predictions}. Given that membrane potentials oscillate on the order of \SI{100}{mV}, this type of parameter variability is a natural example of relatively small and nonlinear heterogeneity.

Temperature changes are another natural example where conductances may deviate from given values in real neurons and conductance-based models \cite{alonso2020temperature}. Alonso and Marder (2020) explore the effect of alterations to conductances due to temperature by modifying conductance values $g_i$ by $g_i R_i(T)$ and all timescales $\tau_i$ by $\tau_i R_i(T)^{-1}$ where
\begin{equation}\label{eq:shock}
    R_i(T) = Q_{10_i}^{(T-T_\text{ref})/\SI{10}{\degree C}},
\end{equation}
where $T$ is the temperature in \SI{}{\degree C}, $T_\text{ref} = \SI{10}{\degree C}$ is a reference temperature, and $ Q_{10_i}$ is defined as the fold change per \SI{10}{\degree C} from the reference temperature. Different neurons in the same pyloric network exhibit small relative variations in intrinsic frequencies due to changes in temperature  \cite{alonso2020temperature}; this example is also relevant in Case 2, which we will discuss shortly. When temperature changes are not too large, which is a common state for animals (particularly compared to the timescale of neural oscillations), then these alterations yield small, nonlinear changes to model conductances.}

\yp{
To formalize these small deviations mathematically, consider a sufficiently smooth vector field $F$. If $\delta$ is a parameter representing some deviation from some average parameter value, then
\begin{equation}\label{eq:background_expansion}
    \frac{\dd X}{\dd t} = F(X,\delta) = F(X,0) + \delta F_\delta(X,0) + \delta^2/2 F_{\delta\delta}(X,0) + O(\delta^3), 
\end{equation}
for some time-dependent state vector $X\in\mathbb{R}^n$, where $F_\delta$ and $F_{\delta\delta}$ are the first and second partial derivatives of the vector field with respect to $\delta$, respectively. In this paper, ``nonlinear heterogeneity'' is in reference to Equation \eqref{eq:background_expansion} (for the case of linearly heterogeneous coupled oscillators, where terms of $O(\delta^2)$ are neglected, see \cite{park2017utility,park2018multiple}).  For additional concrete examples, we refer the reader to Section \ref{sec:thal_coupling}, where we study $n{:}m$ phase-locked states of a pair of identical coupled oscillators with small, nonlinear deviations in their parameters.}


\yp{
\paragraph{Case 2: Two different neurons with similar periods}
Two functionally identical neurons with similar periods may have parameters that differ nontrivially, even by orders of magnitude \cite{lemasson1993activity,foster1993significance,goldman2001global,golowasch2002failure,alonso2019visualization}.  A common example is the temperature shock \eqref{eq:shock}, where large changes to the temperature yield large corresponding changes to conductances, yet result in strikingly little change in neural oscillations \cite{alonso2020temperature}. It is a general feature of coupled biological neural oscillators to maintain similar frequencies relative to each other despite significant differences in their underlying currents, i.e., despite neurons obeying different vector fields \cite{alonso2019visualization}. 

This feature, sometimes called ``robustness'' by neuroscientists, may be quantified as follows. Consider two different neural oscillator models,
\begin{align*}
    \frac{\dd X}{\dd t} &= F_X(X),\\
    \frac{\dd Y}{\dd t} &= F_Y(Y),
\end{align*}
where there exists a $T_X$-periodic and a $T_Y$-periodic limit cycle solution satisfying each vector field, respectively. Since we are operating in the regime of robustness, we are given that $T_X \approx T_Y$ (or more generally, $nT_X \approx m T_Y$). We then ask under what conditions neural oscillator $X$ and $Y$ will phase-lock under small, possibly nonlinear, deviations to parameters in either vector field $F_X$ or $F_Y$, along with nonlinear synaptic interactions. These small, nonlinear alterations are quantified using the framework:
\begin{align*}
    \frac{\dd X}{\dd t} &= F_X(X) + \delta J_X(X) + \ve G_X(X,Y),\\
    \frac{\dd Y}{\dd t} &= F_Y(Y)+ \delta J_Y(Y) + \ve G_Y(X,Y),
\end{align*}
where $\delta J_P$ quantifies deviations of the vector field(s) due to parameter perturbations, and $\ve G_P$ quantifies deviations of the vector field(s) due to synaptic interactions, for each oscillator $P\in\{X,Y\}$. We refer the reader to Section \ref{sec:vdp_thal_coupling} for an example where we study $n{:}m$ phase-locked states of a pair of coupled heterogeneous oscillators satisfying different vector fields.
}



\section{Derivation}\label{sec:derivation}

We seek phase-locked solutions of the coupled system \yp{exhibiting relatively weak and potentially nonlinear heterogeneity in  $\delta=O(\ve)$:}
\begin{equation}\label{eq:1}
    \begin{split}
        \frac{1}{\omega_X}\frac{\dd X}{\dd t} &= F_X(X) +\delta J_X(X) + \ve G_{X}(X,Y),\\
        \frac{1}{\omega_Y}\frac{\dd Y}{\dd t} &= F_Y(Y) +\delta J_Y(Y) + \ve G_{Y}(X,Y),
    \end{split}
\end{equation}
where $\yp{F_P}:\mathbb{R}^{n_P} \rightarrow \mathbb{R}^{n_P}$ is a smooth vector field, $\yp{J_P}:\mathbb{R}^{n_P}\rightarrow \mathbb{R}^{n_P}$ is a smooth function describing some additive heterogeneity, \yp{$G_P$ is a smooth coupling function, and $n_P \in \mathbb{N}$ for each oscillator $P\in\{X,Y\}$}. The scalar $\ve$ (not necessarily small) modulates the overall coupling strength of the network, and the scalar $\delta=O(\ve)$ controls the magnitude of heterogeneity.  We also assume the following:
\begin{itemize}
    \item \yp{For the systems $\frac{\dd P}{\dd t} = F_P(P)$ where $P\in \{X, Y\}$ in isolation with $\epsilon=\delta=0$ there is a $T_P$ periodic solution $\Gamma_P$ which may be parameterized by a phase variable $\theta_P \in [0,2\pi)$ such that when $\epsilon=0$ and $\delta=0$, $\Gamma_P= \Gamma_P(\omega_P t)$ where $\omega_P=2\pi/T_P$ and $\omega_X/\omega_Y= n/m$.}
    \item The isolated limit cycle(s) at $\ve=\delta=0$ \yp{may be used to approximate limit cycle dynamics for $\ve,\delta\neq0$ under sufficiently small perturbations}.
    \item All Floquet eigenmodes decay rapidly except in one direction. We denote this nontrivial Floquet exponent by $\kappa_X$ ($\kappa_Y$) for oscillator $X$ ($Y$). 
\end{itemize}

\begin{remark} We focus on one source of heterogeneity to simplify calculations, but the proposed method may be extended to any number $M_i\in\mathbb{N}$ of heterogeneous parameters for each oscillator $i$, e.g., $\delta_{ij}$ for $j=1,\ldots, M_i$, with possible nonlinear dependencies on $\delta_{ij}$. Nonlinearities may be handled by truncating Taylor expansions in $\delta_{ij}$. Such a formulation for $N\geq 2$ oscillators for $1{:}1$ phase-locking is discussed in \cite{park2024body}. $\blacksquare$
\end{remark}

\begin{remark} Our goal is to verify that the proposed scalar reduction is valid for non-linear coupling strengths $\ve$ and non-linear heterogeneity $\delta$ in the case of $n{:}m$ phase-locking. As such, we will typically take expansions up to order $O(\ve^2)$. \yp{Scalar reductions in the coupling strength for order $O(\ve^3)$ and above} for $1{:}1$ phase-locking may be found in \cite{park2021high,park2024body}. $\blacksquare$
\end{remark}

\subsection{Phase-Isostable Reduction}

To reduce the system to a set of lower-dimensional equations, we transform \eqref{eq:1} into phase coordinates using the chain rule \cite{wilson2020phase}:
\begin{equation}\label{eq:phase_0X}
    \begin{split}
        \frac{\dd \theta_X}{\dd t} &= \omega_X \nabla \theta_X \cdot[F_X(X) +\delta J_X(X)  + \ve G_X(X,Y)]\\
        &= \omega_X [\nabla \theta_X \cdot F_X(X) +\delta\nabla \theta_X \cdot J_X(X) + \ve \nabla \theta_X \cdot G_X(X,Y)]\\
        &=\omega_X [1 + \delta\mathcal{Z}_X(X) \cdot J_X(X) + \ve \mathcal{Z}_X(X) \cdot G_X(X,Y)],
    \end{split}
\end{equation}
where $\mathcal{Z}_i$ is the general phase response function valid for stronger perturbations of the limit cycle beyond the linear regime in \yp{perturbation strength $\ve$} \cite{wilson2020phase}. Note that the solution $X$ ($Y$) may be expressed purely in terms of phases $\theta_X$ ($\theta_Y$) and isostables $\psi_X$ ($\psi_Y$), but we keep \eqref{eq:phase_0X} in terms of $X$ and $Y$ for notational compactness.

The remaining phase-isostable equations for \eqref{eq:1} are obtained using the chain rule \cite{wilson2020phase,park2021high,park2024body}:
\begin{equation}\label{eq:phase-isostable}
    \begin{split}
        \frac{1}{\omega_X}\frac{\dd\theta_X}{\dd t} &=1 + \mathcal{Z}_X(X) \cdot [\delta J_X(X)+ \ve G_X(X,Y)],\\
        \frac{1}{\omega_X}\frac{\dd \psi_X}{\dd t} &= \kappa_X \psi_X + \mathcal{I}_X(X) \cdot[\delta J_X(X) + \ve G_X(X,Y)],\\
        \frac{1}{\omega_Y}\frac{\dd \theta_Y}{\dd t} &= 1+ \mathcal{Z}_Y(Y)\cdot [\delta J_Y(Y) + \ve G_Y(X,Y)], \\
        \frac{1}{\omega_Y}\frac{\dd \psi_Y}{\dd t} &= \kappa_Y \psi_Y + \mathcal{I}_Y(Y)\cdot [\delta J_Y(Y)+ \ve G_Y(X,Y)].
    \end{split}
\end{equation}

To simplify calculations (and following in the same spirit as \cite{ermentrout1981}), we let $s=\omega_Yt$, transforming \eqref{eq:phase-isostable} to,
\begin{equation}\label{eq:phase_isostable0}
    \begin{split}
        \frac{1}{\omega}\theta_X' &= 1 + \ve\mathcal{Z}_X(X) \cdot \hat G_X(X,Y), \\
        \frac{1}{\omega}\psi_X'&= \kappa_X \psi_X + \ve\mathcal{I}_X(X) \cdot \hat G_X(X,Y),\\
        \theta_Y' &= 1+ \ve \mathcal{Z}_Y(Y)\cdot \hat G_Y(X,Y), \\
        \psi_Y' &= \kappa_Y \psi_Y + \ve\mathcal{I}_Y(Y)\cdot \hat G_Y(X,Y),
    \end{split}
\end{equation}
where $'=d/ds$,  $\omega := \omega_X/\omega_Y$, $\hat G_X(X,Y) = bJ_X(X) + G_X(X,Y)$, $\hat G_Y(X,Y) = bJ_Y(Y) + G_Y(X,Y)$, and $b = \delta/\ve$ (because $\delta$ is at most order $O(\ve)$, it follows that $b$ is at most order $O(1)$). Note that we treat $\hat G_i$, the sum of $G_i$ and heterogeneity, as a coupling function to simplify the notation when performing the calculations to follow. However, we will be careful to treat $J_i$ and $G_i$ as distinct functions in the results, because even small values of the heterogeneity parameter $\delta$ can significantly alter phase-locking dynamics. 

We begin by expanding all terms in $\psi$ and plugging them into \eqref{eq:phase_isostable0}, similar to \cite{park2021high,park2024body,nicks2024insights}. The expansions are given by
\begin{align}
    \mathcal{Z}_i(\theta,\psi) &\approx Z_i^{(0)}(\theta) + \psi Z_i^{(1)}(\theta) + \psi^2 Z_i^{(2)}(\theta) +\cdots,\label{eq:z_exp}\\
    \mathcal{I}_i(\theta,\psi) &\approx I_i^{(0)}(\theta) + \psi I_i^{(1)}(\theta) + \psi^2 I_i^{(2)}(\theta) +\cdots,\label{eq:i_exp}\\
    X(t) &\approx X_0(\theta_X) + \psi_X g_X^{(1)}(\theta_X)+ \psi_X^2g_X^{(2)}(\theta_X)+\cdots,\label{eq:x_exp}\\
    Y(t) &\approx Y_0(\theta_Y) + \psi_Y g_Y^{(1)}(\theta_Y)+ \psi_Y^2g_Y^{(2)}(\theta_Y)+\cdots,\label{eq:y_exp}
\end{align}
where $i=X,Y$, and the functions $Z_i^{(k)}$, $I_i^{(k)}$, and $g_i^{(k)}$ are the \yp{second}-order \yp{(or greater)} correction terms to the infinitesimal phase response curve ($Z_i^{(0)}$), infinitesimal isostable response curve ($I_i^{(0)}$), and Floquet eigenfunction ($g_i^{(1)}$), respectively. These correction terms may be computed by numerically integrating initial value problems \cite{wilson2020phase,perez2020global}.  We will assume that these calculations have been completed.

We now subtract the moving frame using $\hat\theta_X = \theta_X - \omega s$ and $\hat\theta_Y = \theta_Y - s$, and we plug in the expansions \eqref{eq:z_exp}-\eqref{eq:y_exp} into \eqref{eq:phase_isostable0}, arriving at a set of non-autonomous phase-isostable equations:
\begin{equation}\label{eq:phase_isostable_na}
    \begin{split}
        \frac{1}{\omega}\hat\theta_X' &= \ve\mathcal{Z}_X(\hat\theta_X+\omega s,\psi_X) \cdot \hat G_X(\hat\theta_X+\omega s,\hat\theta_Y + s,\psi_X,\psi_Y), \\
        \frac{1}{\omega}\psi_X'&= \kappa_X \psi_X + \ve\mathcal{I}_X(\hat\theta_X+\omega s,\psi_X) \cdot \hat G_X(\hat\theta_X+\omega s,\hat\theta_Y + s,\psi_X,\psi_Y),\\
        \hat\theta_Y' &= \ve \mathcal{Z}_Y(\hat\theta_Y+s,\psi_Y) \cdot \hat G_Y(\hat\theta_X+\omega s,\hat\theta_Y + s,\psi_X,\psi_Y), \\
        \psi_Y' &= \kappa_Y \psi_Y + \ve\mathcal{I}_Y(\hat\theta_Y+s,\psi_Y)\cdot \hat G_Y(\hat\theta_X+\omega s,\hat\theta_Y + s,\psi_X,\psi_Y),
    \end{split}
\end{equation}

\begin{assumption}\label{as:separation}
    We assume that the moving-frame subtracted phase variables, $\hat\theta_i$, evolve on a relatively slow timescale compared to the time variable $s$.
\end{assumption}
As our results will show, \yp{second}-order averaging is not necessary for the reduced equations to capture nonlinear effects in the coupling strength $\ve$. Heuristically, this property holds because phase differences tend to evolve on significantly slower timescales than the underlying spiking dynamics. The mathematical question of how this separation of timescales is retained as a function of coupling strength will be a topic of future work. In cases where such a timescale separation does not hold, one may utilize \yp{second}-order averaging methods from, e.g., \cite{llibre2014higher,maggia2020higher}\yp{, although it will generally not be possible to obtain a phase reduction purely in terms of phase differences (the isostable coordinates may still be eliminated, however)}.

Assumption \ref{as:separation} will help us apply averaging in a later step, after we have altered the form of the $\psi_i$ equations in \eqref{eq:phase_isostable_na} so that they are directly amenable to averaging.

\subsection{Elimination of Isostable Coordinates}

We solve the isostable coordinates to further reduce \eqref{eq:phase_isostable_na} from 4 dimensions to 2 dimensions. To this end, we will perform the following steps:
\begin{enumerate}
    \item Convert all expansions in $\psi_i$ to expansions in $\ve$ using the ansatz
    \begin{align}
        \psi_i(s) &\approx \ve p_i^{(1)}(s) + \ve^2 p_i^{(2)}(s) + \ve^3 p_i^{(3)}(s) + \cdots\label{eq:psi_exp}.
    \end{align}
    \item Obtain inhomogeneous linear equations for $p_i^{(k)}$ in a hierarchy of equations in powers of $\ve$.
    \item Solve for each $p_i^{(k)}$ to the desired order in $\ve$ (this step is handled by a symbolic \yp{algebra} package).
    \item Plug in each equation for $p_i^{(k)}$ into the phase equation, thus effectively eliminating the isostable variable. 
\end{enumerate}

\yp{All details behind these steps can be found in Section \ref{sec:elimination} and in the GitHub repository for this paper\footnote{\url{https://doi.org/10.5281/zenodo.13824274}}. For brevity, we assume that this process has been completed and all isostable variables have been eliminated. We then obtain a system of autonomous phase equations} 
\begin{align}
    \frac{1}{\omega}\theta_X' &= b \sum_{\ell=1}^M \ve^{\ell}\mathcal{J}_X^{(\ell)} + \sum_{\ell=1}^M \ve^{\ell} \left[\mathcal{H}_X^{(\ell)}(\theta_X - \omega \theta_Y)\right] , \label{eq:th_avg_X}\\
    \theta_Y' &= b \sum_{\ell=1}^M \ve^{\ell}\mathcal{J}_Y^{(\ell)} + \sum_{\ell=1}^M \ve^{\ell} \left[ \mathcal{H}_Y^{(\ell)}(\theta_X - \omega \theta_Y)\right], \label{eq:th_avg_Y}
\end{align}
where
\begin{align*}
    \mathcal{J}_i^{(1)} &= \frac{1}{2\pi}\int_0^{2\pi} Z_i^{(0)}(s)\cdot J_i^{(0)}(s) \dd s,\\
    \mathcal{J}_i^{(2)} &= \frac{1}{2\pi}\int_0^{2\pi} Z_i^{(1)}(s)\cdot J_i^{(0)}(s) + Z_i^{(0)}(s)\cdot J_i^{(1)}(s) \dd s,\\
    \mathcal{J}_i^{(3)} &= \frac{1}{2\pi}\int_0^{2\pi} Z_i^{(2)}(s)\cdot J_i^{(0)}(s) +  Z_i^{(1)}(s)\cdot J_i^{(1)}(s) +Z_i^{(0)}(s)\cdot J_i^{(2)}(s) \dd s,\\
    &\quad \vdots
\end{align*}
and
\begin{align*}
    \mathcal{H}_X^{(1)}(\xi) &= \frac{1}{2\pi m}\int_0^{2\pi m}Z_X^{(0)}(\xi+\omega s)\cdot K_X^{(0)}(\xi+\omega s, s) \dd s,\\
    \mathcal{H}_X^{(2)}(\xi) &= \frac{1}{2\pi m}\int_0^{2\pi m} Z_X^{(0)}(\xi + \omega s)\cdot K_X^{(1)}(\xi + \omega s, s)\\
    &\quad\quad\quad\quad\quad\quad +p_X^{(1)}(\xi + \omega s, s) Z_X^{(1)}(\xi+\omega s)\cdot K_X^{(0)}(\xi+\omega s, s) \dd s,\\
    \mathcal{H}_X^{(3)}(\xi) &= \frac{1}{2\pi m}\int_0^{2\pi m} Z_X^{(0)}(\xi+\omega s)\cdot K_X^{(2)}(\xi+\omega s, s)\\
    &\quad\quad\quad\quad\quad\quad+ p_X^{(1)}(\xi+\omega s, s) Z_X^{(1)}(\xi+\omega s)\cdot K_X^{(1)}(\xi+\omega s, s)\\
    &\quad\quad\quad\quad\quad\quad+  p_X^{(2)}(\xi+\omega s, s)Z_X^{(1)}(\xi+\omega s) \cdot K_X^{(0)}(\xi+\omega s, s)\\
    &\quad\quad\quad\quad\quad\quad+ p_X^{(1)}(\xi+\omega s,s)^2 Z_X^{(2)}(\xi+\omega s) \cdot K_X^{(0)}(\xi+\omega s, s) \dd s,\\
    &\vdots.
\end{align*}
The heterogeneity parameter $b$ is implicit. 

\yb{\begin{remark}
    The functions $K_X^{(\ell)}$ are obtained by taking Taylor expansions in $\ve$ of the coupling function $G_X$:
    \begin{equation*}
        \begin{split}
            G_{X}(\theta_X,\theta_Y)= K_{X}^{(0)}(\theta_X,\theta_Y)+ \ve K_{X}^{(1)}\left(\theta_X,\theta_Y\right)+ \cdots.
        \end{split}
    \end{equation*}
    More details on how we calculate $K_X^{(\ell)}$ is provided in Section \ref{a:g}. In brief, $K_X^{(0)}$ is the coupling function, $G_X$, evaluated on the limit cycles $\Gamma_P$ for $P\in\{X,Y\}$, i.e., $K_X^{(0)}(\theta_X,\theta_Y) = G_X(\Gamma_X(\theta_X),\Gamma_Y(\theta_Y))$, and $K_X^{(1)}$ is the first-order Taylor expansion of $G_X$ in terms of $p_P^{(1)}$, $g_P^{(1)}$, and the partials of the coupling function $G_X$. Let $G_{X,m}$ be the $m$th output coordinate of the coupling function $G_X$ for $m=1,\ldots,n_X$, i.e., $G_X = [G_{X,1},\ldots,G_{X,n_{X}}]^\tr$. Noting that each $G_{X,m}$ has $n_X + n_Y$ total input coordinates, the $m$th output coordinate of $K_{X}^{(1)}(\theta_X,\theta_Y)$ is given by
    \begin{align*}
        K_{X,m}^{(1)}(\theta_X,\theta_Y) =& \,\,\pa_{X_1} G_{X,m}(\Gamma_X(\theta_X),\Gamma_Y(\theta_Y))p_X^{(1)}(\theta_X,\theta_Y)g_{X,1}^{(1)}(\theta_X) \\
        &+ \pa_{X_2} G_{X,m}(\Gamma_X(\theta_X),\Gamma_Y(\theta_Y))p_X^{(1)}(\theta_X,\theta_Y)g_{X,2}^{(1)}(\theta_X)\\
        &+\cdots\\
        &+ \pa_{X_{n_X}} G_{X,m}(\Gamma_X(\theta_X),\Gamma_Y(\theta_Y))p_X^{(1)}(\theta_X,\theta_Y)g_{X,n_X}^{(1)}(\theta_X)\\
        &+\pa_{Y_1} G_{X,m}(\Gamma_X(\theta_X),\Gamma_Y(\theta_Y))p_Y^{(1)}(\theta_Y,\theta_X) g_{Y,1}^{(1)}(\theta_Y)\\
        &+\pa_{Y_2} G_{X,m}(\Gamma_X(\theta_X),\Gamma_Y(\theta_Y))p_Y^{(1)}(\theta_Y,\theta_X) g_{Y,2}^{(1)}(\theta_Y)\\
        &+\cdots\\
        &+\pa_{Y_{n_Y}} G_{X,m}(\Gamma_X(\theta_X),\Gamma_Y(\theta_Y))p_Y^{(1)}(\theta_Y,\theta_X) g_{Y,n_Y}^{(1)}(\theta_Y).
    \end{align*}
    $\blacksquare$
\end{remark}}

Similarly,
\begin{align*}
    \mathcal{H}_Y^{(1)}(\xi) &= \frac{1}{2\pi m}\int_0^{2\pi m}Z_Y^{(0)}(\xi+\omega s)\cdot K_Y^{(0)}(\xi+\omega s, s) \dd s,\\
    \mathcal{H}_Y^{(2)}(\xi) &= \frac{1}{2\pi m}\int_0^{2\pi m} Z_Y^{(0)}(\xi+\omega s)\cdot K_Y^{(1)}(\xi+\omega s, s)\\
    &\quad\quad\quad\quad\quad\quad +p_Y^{(1)}(\xi+\omega s, s) Z_Y^{(1)}(\xi+\omega s)\cdot K_Y^{(0)}(\xi+\omega s, s) \dd s,\\
    \mathcal{H}_Y^{(3)}(\xi) &= \frac{1}{2\pi m}\int_0^{2\pi m} Z_Y^{(0)}(\xi+\omega s)\cdot K_Y^{(2)}(\xi+\omega s, s)\\
    &\quad\quad\quad\quad\quad\quad+ p_Y^{(1)}(\xi+\omega s, s) Z_Y^{(1)}(\xi+\omega s)\cdot K_Y^{(1)}(\xi+\omega s, s)\\
    &\quad\quad\quad\quad\quad\quad+  p_Y^{(2)}(\xi+\omega s, s)Z_Y^{(1)}(\xi+\omega s) \cdot K_Y^{(0)}(\xi+\omega s, s)\\
    &\quad\quad\quad\quad\quad\quad+  p_Y^{(1)}(\xi+\omega s,s)^2 Z_Y^{(2)}(\xi+\omega s) \cdot K_Y^{(0)}(\xi+\omega s, s) \dd s,\\
    &\vdots
\end{align*}
\yb{where the functions $K_Y^{(\ell)}$ are obtained by taking Taylor expansions in $\ve$ of the coupling function $G_Y$:
\begin{equation*}
    \begin{split}
        G_{Y}(\theta_X,\theta_Y)= K_{Y}^{(0)}(\theta_X,\theta_Y)+ \ve K_{Y}^{(1)}\left(\theta_X,\theta_Y\right)+ \cdots.
    \end{split}
\end{equation*}
The function $K_Y^{(0)}$ is the coupling function $G_Y$, and $K_Y^{(1)}$ is calculated similarly to $K_X^{(1)}$:
\begin{align*}
        K_{Y,m}^{(1)}(\theta_X,\theta_Y) =& \,\,\pa_{X_1} G_{Y,m}(\Gamma_X(\theta_X),\Gamma_Y(\theta_Y))p_X^{(1)}(\theta_X,\theta_Y)g_{X,1}^{(1)}(\theta_X) \\
        &+ \pa_{X_2} G_{Y,m}(\Gamma_X(\theta_X),\Gamma_Y(\theta_Y))p_X^{(1)}(\theta_X,\theta_Y)g_{X,2}^{(1)}(\theta_X)\\
        &+\cdots\\
        &+ \pa_{X_{n_X}} G_{Y,m}(\Gamma_X(\theta_X),\Gamma_Y(\theta_X))p_X^{(1)}(\theta_X,\theta_Y)g_{X,n_X}^{(1)}(\theta_X)\\
        &+\pa_{Y_1} G_{Y,m}(\Gamma_X(\theta_X),\Gamma_Y(\theta_Y))p_Y^{(1)}(\theta_X,\theta_Y) g_{Y,1}^{(1)}(\theta_Y)\\
        &+\pa_{Y_2} G_{Y,m}(\Gamma_X(\theta_X),\Gamma_Y(\theta_Y))p_Y^{(1)}(\theta_X,\theta_Y) g_{Y,2}^{(1)}(\theta_Y)\\
        &+\cdots\\
        &+\pa_{Y_{n_Y}} G_{Y,m}(\Gamma_X(\theta_X),\Gamma_Y(\theta_Y))p_Y^{(1)}(\theta_X,\theta_Y) g_{Y,n_Y}^{(1)}(\theta_Y).
    \end{align*}
}
Defining $\phi = \theta_X - \omega \theta_Y$, the corresponding ordinary differential equation is given by
\begin{align*}
    \phi' &= \theta_X' - \omega \theta_Y'\\
    &= \omega \sum_{\ell=1}^M \ve^{\ell}\left(\mathcal{J}_X^{(\ell)} - \mathcal{J}_Y^{(\ell)}\right)b + \omega \sum_{\ell=1}^M \ve^{\ell} \left(\mathcal{H}_X^{(\ell)}(\phi) - \mathcal{H}_Y^{(\ell)}(\phi)\right).
\end{align*}
Finally, defining $\mathcal{H}_{n,m}^{(\ell)}(\phi) =  \mathcal{H}_X^{(\ell)}(\phi) - \mathcal{H}_Y^{(\ell)}(\phi)$, and $b^{(\ell)} = (\mathcal{J}_X^{(\ell)} - \mathcal{J}_Y^{(\ell)})b$, the phase difference dynamics is given by the scalar equation,
\begin{equation}\label{eq:dphi}
    \frac{1}{\omega}\phi' = \sum_{\ell=1}^M\ve^{\ell} [b^{(\ell)}  + \mathcal{H}_{n,m}^{(\ell)}(\phi)],
\end{equation}
\yp{which captures all the dynamics of \eqref{eq:th_avg_X} and \eqref{eq:th_avg_Y} on the invariant torus.}

\begin{remark}
\begin{itemize}[noitemsep,nolistsep]
    \item \yp{In the case that $G_P(X,Y)$ represent coupling}, there is an implicit dependence on the heterogeneity term $b$ inside the functions $b^{(\ell)}$ and $\mathcal{H}_{n,m}^{(\ell)}(\phi)$ in Equation \eqref{eq:dphi}, because for nonzero $b$, each forcing function of $p_i^{(\ell)}$ will also contain $b$ (up to the power $b^{\ell}$). We show an example of the phase reduction with explicit heterogeneity $b$ up to order $O(\ve^2)$ in \ref{a:fourier}, which is used to calculate two-parameter bifurcation diagrams in Section \ref{sec:coupling}.
    \item \yp{In the case that $G_P(X,Y)$ represent forcing}, changing the forcing frequency results in a vertical shift of the $\mathcal{H}_{n,m}^{(\ell)}$ functions because the dependence on the heterogeneity parameter $\delta$ is completely described by $b^{(\ell)}$ in \eqref{eq:dphi} (see Section \ref{sec:forcing}). $\blacksquare$
\end{itemize}
\end{remark}



\section{Results}\label{sec:results}

We first examine the performance of the scalar reduction for forced systems (Section \ref{sec:forcing}), including a forced nonradial isochron clock (Section \ref{sec:nric}) and a forced thalamic neuron (Section \ref{sec:thal1}). We then test the scalar reduction on coupled systems (Section \ref{sec:coupling}), including a pair of coupled thalamic neurons with weak and nonlinear heterogeneity (Section \ref{sec:thal_coupling}), followed by a coupled system of mixed model types, where a thalamic model is synaptically coupled to the Van der Pol oscillator (Section \ref{sec:vdp_thal_coupling}).

\subsection{Forcing}\label{sec:forcing}

We use the same formulation \eqref{eq:dphi} to reduce a forced oscillator to a scalar variable that represents the phase difference between the exogenous forcing function and the phase of the forced oscillator. In particular, let $\omega_Y$ be the forcing frequency and $G_X(X,\omega_Y t)$ be the ``coupling'' function. Then our original coupled system \eqref{eq:1} becomes
\begin{equation}\label{eq:1f}
\begin{split}
    \frac{1}{\omega_X}\frac{\dd X}{\dd t} &= F_X(X) + \ve G_{X}(X, \theta_Y),\\
    \frac{1}{\omega_Y}\frac{\dd \theta_Y}{\dd t} &= 1 + \delta/\omega_Y,
\end{split}
\end{equation}
where the heterogeneity parameter $\delta$ directly modulates the forcing frequency.
\yb{
\begin{remark}
    The choice to include the heterogeneity term $\delta/\omega_Y$ directly to the right-hand side of $\dd \theta_Y/\dd t$ is made without loss of generality. Recall the original system \eqref{eq:1}:
    \begin{equation}\tag{\ref{eq:1}}
        \begin{split}
            \frac{1}{\omega_X}\frac{\dd X}{\dd t} &= F_X(X) +\delta J_X(X) + \ve G_{X}(X,Y),\\
            \frac{1}{\omega_Y}\frac{\dd Y}{\dd t} &= F_Y(Y) +\delta J_Y(Y) + \ve G_{Y}(X,Y),
        \end{split}
    \end{equation}
    To convert \eqref{eq:1} into a forced system where oscillator $Y$ forces oscillator $X$, we take $J_X \equiv 0$ and $G_Y\equiv 0$. Then \eqref{eq:1} becomes
    \begin{equation*}
        \begin{split}
            \frac{1}{\omega_X}\frac{\dd X}{\dd t} &= F_X(X) + \ve G_{X}(X,Y),\\
            \frac{1}{\omega_Y}\frac{\dd Y}{\dd t} &= F_Y(Y) +\delta J_Y(Y).
        \end{split}
    \end{equation*}
    The phase equation for oscillator $Y$ is trivial to compute because it is unperturbed and thus the amplitude equation is not necessary. Using the chain rule,
    \begin{equation}\label{eq:forcing:chain_rule}
        \begin{split}
            \frac{\dd \theta_Y}{\dd t} &= \nabla \theta_Y\cdot \frac{\dd Y}{\dd t}\\
            &= \nabla \theta_Y\cdot \left[ \omega_Y F_Y(Y) +\delta \omega_Y J_Y(Y)\right]\\
            &= \omega_Y +\delta \omega_Y  \nabla \theta_Y \cdot J_Y(Y).
        \end{split}
    \end{equation}
    The second term in the last line of \eqref{eq:forcing:chain_rule} can be further simplified by noting that oscillator $Y$ is unperturbed and therefore the solution $Y$ is the same as the unperturbed limit cycle, i.e., $Y = \Gamma$, and the gradient of the phase is just the infinitesimal phase response curve, i.e., $\nabla\theta_Y \equiv Z_Y^{(0)}$. We consider very simple forms of $J_Y$ in this paper, where only one coordinate is nonzero, e.g., $J_Y(Y) = [1,0,\ldots,0]^\tr$. Then, the phase equation for oscillator $Y$ becomes
    \begin{equation}\label{eq:forcing:dtheta_y}
        \frac{\dd \theta_Y}{\dd t} = \omega_Y +\delta \omega_Y Z_{Y,1}^{(0)}(\theta_Y),
    \end{equation}
    where $Z_{Y,1}^{(0)}$ is the first coordinate of the infinitesimal phase response curve $Z_Y^{(0)}$. While \eqref{eq:forcing:dtheta_y} captures all possible nonlinearities that contribute to changes in the natural frequency of $\theta_Y$, we only wish to understand how any arbitrary change to the natural frequency $\omega_Y$ of oscillator $Y$ affects entrainment with oscillator $X$. Thus, we consider the simplified form present in \eqref{eq:1f},
    \begin{equation*}
        \frac{\dd \theta_Y}{\dd t} = \omega_Y +\delta.
    \end{equation*} 
    $\blacksquare$
\end{remark}
}

Applying the scalar reduction in \ref{sec:elimination} to the forced system \eqref{eq:1f} yields
\begin{equation}\label{eq:phase_f}
    \begin{split}
        \frac{1}{\omega}\phi'= -\frac{\delta}{\omega_Y} + \sum_{\ell=1}^M \ve^\ell \mathcal{H}_{n,m}^{(\ell)}(\phi),
    \end{split}
\end{equation}
where we have implicitly used the same assumptions from the coupling case that there is a reasonable separation of timescales between $\hat\theta_X$ ($\hat\theta_Y$) and $s$. Note that higher powers of $b$ do not appear in the forcing case where $\delta = \ve b$, and $\delta$ remains an order $O(\ve)$ parameter. Thus, linear changes to the forcing frequency only yield linear shifts to the $\mathcal{H}$ functions in a forced system.

\begin{figure}[t]
\centering
    \includegraphics[width=.6\textwidth]{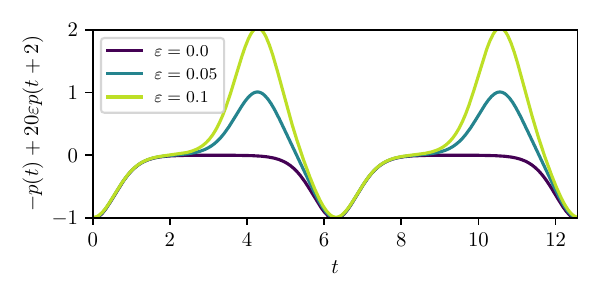}
    \caption{\yp{The forcing function \eqref{eq:force} as a function of $\ve$. The second-order term results in a significant deformation of the forcing function as $\ve$ increases from $\ve=0$ to $\ve=0.1$. $\omega_Y = 1$, $\delta=0$.}}\label{fig:forcing_function}
\end{figure}

There are no particular restrictions on the choice of forcing function besides sufficient smoothness and periodicity. While choosing a sinusoidal function such as $\cos(\omega_Y t)$ would be a good starting point, the lack of nontrivial Fourier modes simplifies the existence of phase-locking to $1{:}1$, as in the case of the nonradial isochron clock, or to $n{:}1$, as in the case of the forced thalamic model. Thus, we use \yp{an exponential function composed with a sinusoid} to ensure that there are enough nontrivial Fourier modes to explore $n{:}m$ phase-locking for $n,m\geq 1$.

\yp{We define the forcing function to depend only on time,
\begin{equation}\label{eq:force0}
    p(t) = [\exp(\cos t)/2.7]^3.
\end{equation}
The factor of 2.7 keeps the function approximately order $O(1)$ and the cubic power gives the function a slight pulsatile form to more closely resemble action potentials. We include a nonlinear effect in perturbation strength $\ve$ to reflect the assumption that perturbations in nature do not necessarily scale linearly:
\begin{equation}\label{eq:force}
    \ve G(X,\omega_Y t) \equiv \ve G(\omega_Y t) = -\ve p(\omega_Y t) + 20 \ve^2 p(\omega_Y (t+2)).
\end{equation}}
The relatively large second-order term is chosen so that the forcing function exhibits a large deformation in its second-order term. This choice will help us verify the accuracy of the non-linear terms in the scalar reduction in biologically relevant cases. A plot of the forcing function \eqref{eq:force} is shown in Figure \ref{fig:forcing_function} for various values of $\ve$.

\subsubsection{Nonradial Isochron Clock}\label{sec:nric}

\begin{figure}[ht!]
	\centering
	\includegraphics[width=\textwidth]{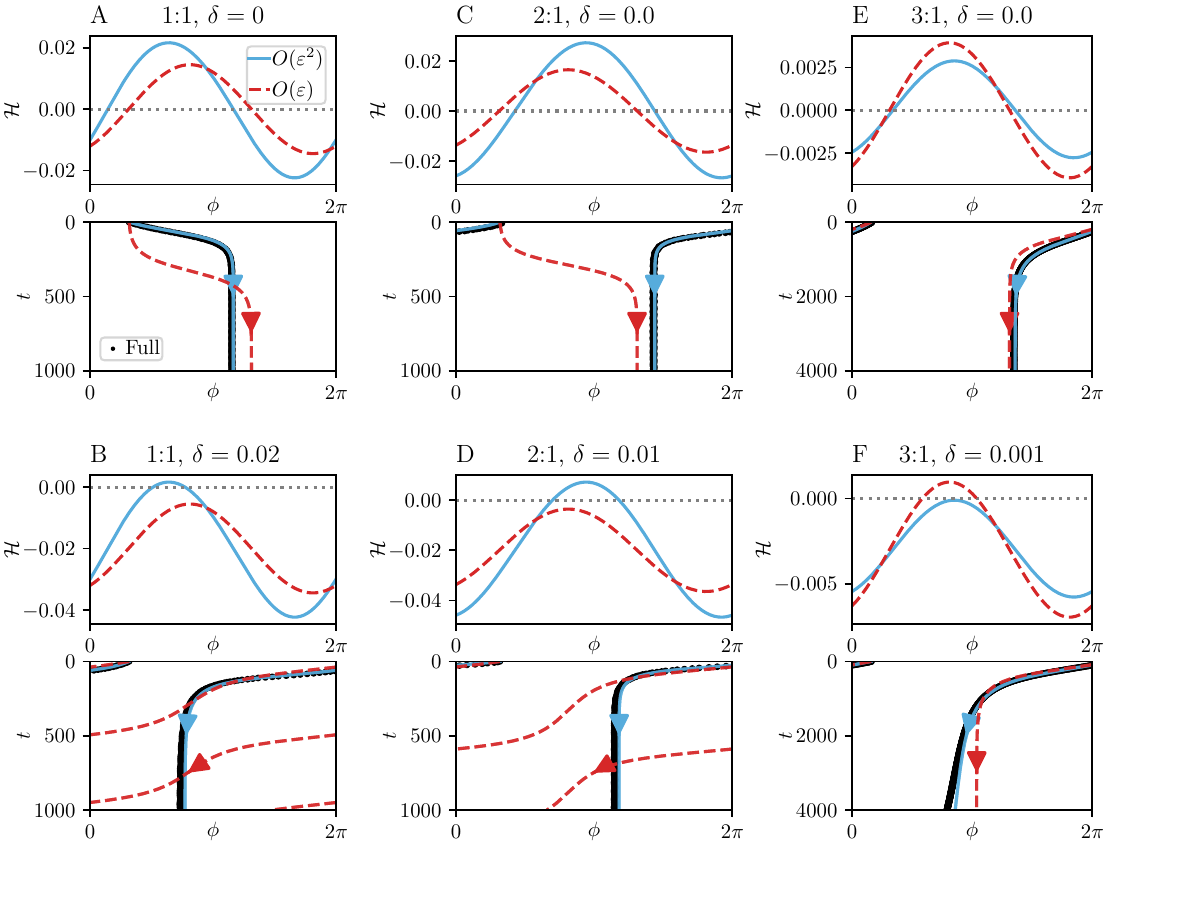}
	\caption{\yp{Frequency-locking and drift in a forced nonradial isochron clock. All panels use the same horizontal axis, $\phi\in[0,2\pi)$. Each label (A-F) corresponds to particular values of $\ve$ and $\delta$, and each labeled panel has two sub-panels, top and bottom. Each top panel shows the right-hand sides of the first-order (dashed red) and second-order (solid blue) scalar reductions. The zeros of these functions (where they intersect the dotted gray line) correspond to phase-locked states in the full model. Each bottom panel shows the solutions of the first (dashed red) and second-order scalar reductions (solid blue) and full model (black) over time. A,B: $1{:}1$ phase-locking for $\delta=0,0.02$ respectively, with initial condition $\phi(0)=1$, and $\omega_X = 1$, $\omega_Y = 1$. C, D: $2{:}1$ phase-locking for $\delta=0, 0.01$, respectively, with $\phi(0)=1$, $\omega_X=2$, and $\omega_Y = 1$. E, F: $3{:}1$ phase-locking for $\delta=0, 0.001$, respectively, with $\phi(0)=0.5$, $\omega_X=3$, and $\omega_Y = 1$. To reduce lag when viewing the figure, we plot every two-hundredth point in the full model's phase estimate. $\ve=0.04$ in all panels except $E,F$, where $\ve=0.01$.}}\label{fig_nric_force0}
\end{figure}

The nonradial isochron clock \cite{wilson2019phase,park2024body} is defined as
\begin{equation}\label{eq:nric}
    \begin{split}
        \frac{1}{\omega_X} \frac{\dd X}{\dd t} &= F_\text{NR}(X) + \ve[G(\theta_Y),0]^\tr,\\
        \frac{1}{\omega_Y}\frac{\dd\theta_Y}{\dd t} &= 1 + \delta/\omega_Y,
    \end{split}
\end{equation}
where $X = [x,y]^\tr$,
\begin{equation*}
    F_\text{NR}(X) =\left(\begin{matrix}
        \sigma x(1-R^2) - y(1+\rho(R^2-1)\\
        \sigma y(1-R^2) + x(1+\rho(R^2-1))
    \end{matrix}\right),
\end{equation*}
$R = \sqrt{x^2 + y^2}$, $\sigma=0.08$, and $\rho=0.12$. The Floquet exponent of the isolated limit cycle (at $\ve=0$) is $\kappa_X \approx -0.16$, making it weakly attracting enough to suitably test the proposed method.

\begin{remark}\label{remark:nric}
    We are limited to testing $n{:}1$ phase-locking for the nonradial isochron clock because the Fourier modes of the model's response functions typically only contain the principal harmonic. This feature poses a minor problem: for example, if we choose a forcing function with half a period compared to the original oscillator ($\omega_Y = 2$ and $\omega_X = 1$), then the integrals involved in computing the $\mathcal{H}$-functions cancel due to orthogonality, and there is no phase-locking for the nonradial isochron clock. $\blacksquare$
\end{remark}

Trajectories of the first- and second-order reduction of the nonradial isochron clock are shown in Figure \ref{fig_nric_force0} (dashed red for order $O(\ve)$ and solid blue for order $O(\ve^2)$) along with the trajectories of the original system (black). Phase differences in the full model are estimated using the limit cycle computed at $\ve=0$ (see \ref{a:phase_estimate} for additional details on phase estimation). Each panel (A-F) corresponds to particular values of $\ve$ and $\delta$. We choose the coupling strength to be $\ve=0.04$ (except for Panels E and F, where we choose $\ve=0.01$). Across all panels, the second-order reduction consistently captures the existence of phase-locked states in the presence of heterogeneity (i.e., when $\delta > 0$) that the first-order reduction cannot.

While the integrated solutions of the second-order reduction capture phase-locked states more accurately relative to the first-order reduction, we also wish to predict where the original system loses the existence of phase-locked states due to \yp{nonlinear} heterogeneity $\delta$ and forcing strength $\ve$. To better quantify how and when the second-order reduction loses phase-locked states (and compare these predictions to the original model), we compute bifurcation diagrams.

\begin{figure}[ht!]
    \includegraphics[width=\textwidth]{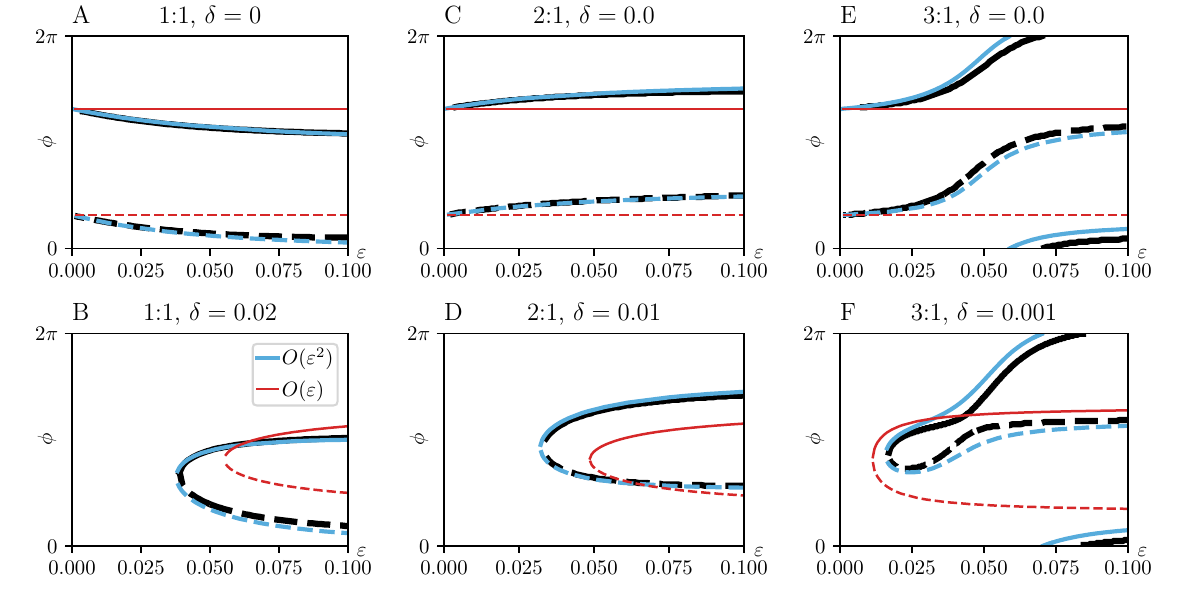}
    \caption{One-parameter bifurcation diagrams of the forced nonradial isochron clock as a function of coupling strength $\ve$. Each panel A-F corresponds to the same system in panels A-F of Figure \ref{fig_nric_force0}. The order $O(\ve)$ scalar reduction is shown in red, the order $O(\ve^2)$ scalar reduction is shown in blue, and the stable phase-locked states of the original system is shown in black. Solid (dashed) curves denote stable (unstable) phase-locked states.}\label{fig:bif1d_cgl1}
\end{figure}

One-parameter bifurcation diagrams of the reduced nonradial isochron clock are shown in Figure \ref{fig:bif1d_cgl1} (red for order $O(\ve)$ and blue for order $O(\ve^2)$) along with a superimposed one-parameter bifurcation diagram of the full model (black). \yp{Fixed points of the first- and second-order reduced models  are computed directly using \eqref{eq:phase_f}, while phase-locked states of the full model are found using Newton's method on a stroboscopic map. All panels A-F of Figure \ref{fig:bif1d_cgl1} correspond to the same underlying model in panels A-F in Figure \ref{fig_nric_force0}. We refer the reader to Appendix \ref{a:newton} for additional details on how we apply Newton's method for both forced and coupled oscillators. In brief, by assuming that the period of the phase-locked solution is equal to the forcing frequency, convergence of Newton's method implies the existence of a phase-locked state, whereas non-convergence implies phase drift. This convergence typically happens in under 10 iterations, assuming that the correction vector has a tolerance of \num{1e-5}. Non-convergence was assumed after 20 iterations.}

The second-order scalar reduction is significantly more capable of capturing the existence of frequency-locked states as a function of $\ve$ for representative values of $\delta$ compared to the first-order reduction. Note that the values of $\delta$ at which frequency-locking is lost differ by orders of magnitude between $1{:}1$ locking (Figure \ref{fig_nric_force0}A,B or Figure \ref{fig:bif1d_cgl1}A,B) and $3{:}1$ locking (Figure \ref{fig_nric_force0}E,F or Figure \ref{fig:bif1d_cgl1}E,F). This observation is consistent with the fact that Fourier coefficients become vanishingly small as $n\rightarrow\infty$ in $n{:}1$ locking, and that the bandwidth of $n{:}m$ locking decreases as a function of higher frequencies \cite{ermentrout1981}. 

\begin{figure}[ht!]
    \includegraphics[width=\textwidth]{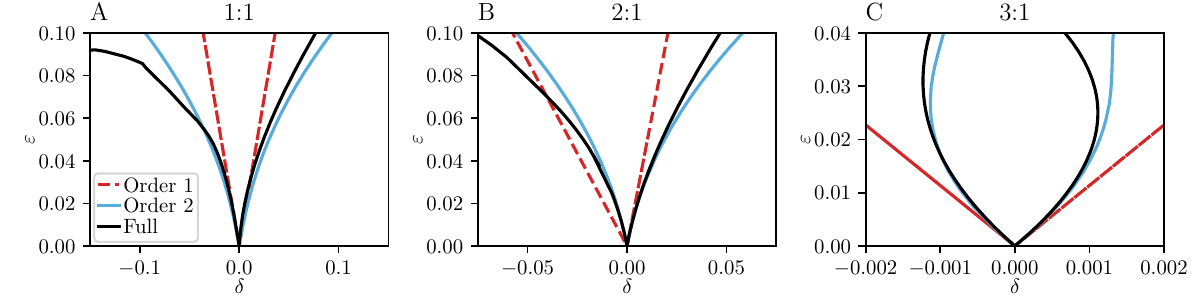}
    \caption{\yp{Two-parameter bifurcation diagrams (Arnold tongues) of the forced nonradial isochron clock. The vertical axis is the forcing amplitude $\ve$, and the horizontal axis is the forcing frequency change $\delta$. The black solid, red dashed, and blue solid curves denote boundaries of phase-locking regions for the full model, $O(\ve)$ phase reduction, and $O(\ve^2)$ phase reduction, respectively.}}\label{fig:nric_tongues}
\end{figure}

Two-parameter bifurcation diagrams of the reduced nonradial isochron clock are shown in Figure \ref{fig:nric_tongues}. The diagrams were computed numerically using XPPAUTO \cite{xpp} \yp{directly for the full model} and using Fourier approximations of the $\mathcal{H}$-functions in the scalar reduction up to $O(\ve^2)$. Coefficients of the Fourier approximation of each $\mathcal{H}$-function may be found in \ref{a:fourier}. The phase-locking regions in  Figure \ref{fig:nric_tongues} are equivalent to Arnold tongues, except that each horizontal axis shows the heterogeneity parameter $\delta$ as opposed to the more standard frequency ratio.

\subsubsection{Thalamic Neuron}\label{sec:thal1}

\begin{table}
    \caption{Parameter values of the nondimensionalized thalamic model \eqref{eq:thal_nondim_force} }\label{tab:thal_nondim}
    \centering
    \begin{tabular}{c|c|c|c|c|c|c|c|c|c|c|c|c}
        Parameter & $E_\text{K}$& $E_\text{Na}$ & $E_\text{T}$ & $E_\text{L}$ & $g_\text{L}$ & $g_\text{K}$ & $g_\text{Na}$ & $g_\text{syn}\equiv \ve$ & $I_\text{app}$ & $\alpha$ & $\beta$ \\
        \hline
        Value & -0.9 & 0.5 & 0 & -0.7 & 0.05 & 5 & 3 & 0-0.3 & 0.035 & 3 & 2\\
    \end{tabular}
\end{table}

\begin{figure}[ht]
    \includegraphics[width=\textwidth]{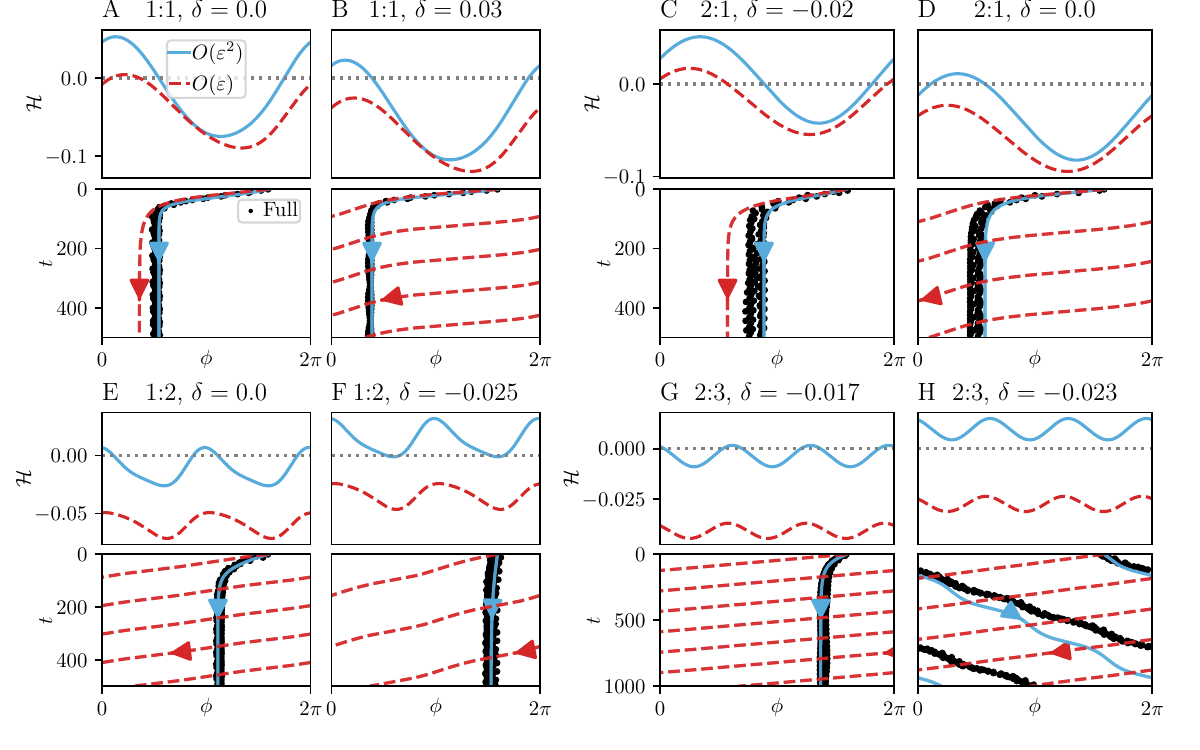}
    \caption{Frequency-locking and drift in a forced thalamic neuron. All panels use the same horizontal axis, $\phi\in[0,2\pi)$. Each label (A-H) corresponds to particular values of $\ve$ and $\delta$ and each labeled panel has two sub-panels top and bottom. Each top panel shows the right-hand sides of the first-order (dashed red) and second-order (solid blue) scalar reductions. The zeros of these functions (where they intersect the dotted gray line) correspond to phase-locked states in the full model. Each bottom panel shows the solutions of the first-order reduction (dashed red), second-order reductions (solid blue), and full model (black) over time. All panels use the initial condition $\phi(0)=5$. \yp{A,B: $1{:}1$ phase-locking for $\delta=0,0.03$ respectively, with $\ve=0.03$, $\omega_X=1$ and $\omega_Y=1$. C, D: $2{:}1$ phase-locking for $\delta=-0.02, 0$, respectively, with $\ve=0.02$, $\omega_X=2$ and $\omega_Y=1$. E, F: $1{:}2$ phase-locking for $\delta=0, -0.025$, respectively, with $\ve=0.04$, $\omega_X=1$ and $\omega_Y=2$. G, H: $2{:}3$ phase-locking for $\delta=-0.017, -0.023$, respectively, with $\ve=0.025$, $\omega_X=2$ and $\omega_Y=3$. We plot every two-hundredth point in the full model's phase estimate to reduce lag when viewing the figure.}}\label{fig:thal_force0}
\end{figure}

We now consider a nondimensionalized model of a thalamic neuron\footnote{To obtain the original thalamic model from \cite{rubi04}, multiply the voltage by $\SI{100}{mV}$, and divide the gating variable $r$ by $\SI{100}{mV}$. Conductances are scaled using $\SI{1}{nS}$. We choose the characteristic scale of $\SI{100}{mV}$ because it is similar in magnitude to the potassium reversal potential ($E_\text{K}=\SI{-90}{mV}$).} (adapted from \cite{rubi04}),
\begin{equation}\label{eq:thal_nondim_force}
    \begin{split}
        \frac{1}{\omega_X}\frac{\dd X}{\dd t} &= F_\text{Thal}(X) + \ve[G(\theta_Y),0,0]^\tr,\\
        \frac{1}{\omega_Y}\frac{\dd\theta_Y}{\dd t} &= 1 + \delta/\omega_Y,
    \end{split}
\end{equation}
where $X=[V,h,r]^\tr$,
\begin{equation*}
    F_\text{Thal}(X) = \left(\begin{matrix}
        -I_\text{L}(V) - I_{\text{Na}}(V) - I_\text{K}(V) - I_\text{T}(V) +I_\text{app}\\
        (h_\infty(V) - h)/\tau_h(V)\\
        (r_\infty(V) - r)/\tau_r(V)\\
    \end{matrix}\right).
\end{equation*}

\begin{figure}[ht!]
    \centering
    \includegraphics[width=\textwidth]{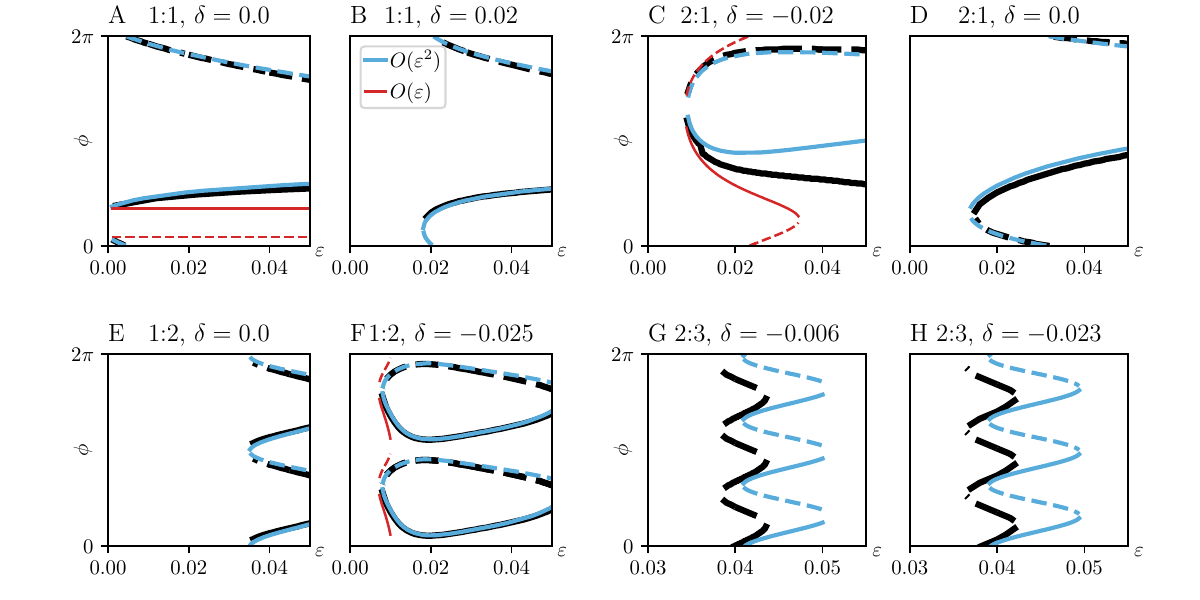}
    \caption{One-parameter bifurcation diagrams of a forced thalamic neuron as a function of coupling strength $\ve$. The order $O(\ve)$ scalar reduction is shown in red, the order $O(\ve^2)$ scalar reduction is shown in blue, and the stable phase-locked states of the original system are shown in black. Each panel A-H corresponds to the same system in panels A-H of Figure \ref{fig:thal_force0}. Solid (dashed) curves denote stable (unstable) fixed points of the scalar reduction. The vertical gray dashed lines correspond to the $\ve$ parameter value used in the corresponding panel in Figure \ref{fig:thal_force0}.}
    \label{fig:bif1d_thal1}
\end{figure}

We use the same forcing function introduced earlier \eqref{eq:force}, rewritten here for convenience:
\begin{equation}\tag{\ref{eq:force}}
    \yp{\ve G(X,\omega_Y t) \equiv \ve G(\omega_Y t) = -\ve p(\omega_Y t) + 20 \ve^2 p(\omega_Y (t+2)),}
\end{equation}
where \yp{$p(t) = [\exp(\cos t)/2.7]^3$} \eqref{eq:force0}. The remaining equations are given by
\begin{alignat*}{4}
    I_\text{L}(V) &= g_\text{L} (V-E_\text{L}), \quad&& I_{\text{Na}}(V) &&= g_\text{Na} h m_\infty^3(V)(V-E_\text{Na}),\\
    I_\text{K}(V) &= g_\text{K}[0.75 (1-h)]^4(V-E_\text{K}), \quad&& I_\text{T}(V) &&= g_\text{T} r p_\infty^2(V)  (V-E_\text{T}),
\end{alignat*}
and
\begin{equation}\label{eq:aux}
\begin{split}
    a_h(V) &= 0.128 \exp(-5.556(V+0.46)),\\
    b_h(V) &= 4/(1+\exp(-20(V+0.23))),\\
    m_\infty(V) &= 1/(1+\exp(-14.29(V+0.37))),\\
    h_\infty(V) &= 1/(1+\exp(25(V+0.41))),\\
    r_\infty(V) &= 100/(1+\exp(25(V+0.84))),\\
    p_\infty(V) &= 0.01/(1+\exp(-16.13(V+0.6))),\\
    \tau_h(V) &= 1/(a_h(V)+b_h(V)),\\
    \tau_r(V) &= 28+\exp(-9.52(V+0.25)),\\
    a_\infty(V) &= 1/(1+\exp(-125(V+0.2))).
\end{split}
\end{equation}
Nondimensional parameter values are listed in Table \ref{tab:thal_nondim}. The Floquet exponent of the isolated limit cycle (at $\ve=0$) is $\kappa_X \approx -0.024$, making it weakly attracting enough to suitably test the proposed method. 

\begin{figure}[ht]
    \includegraphics[width=\textwidth]{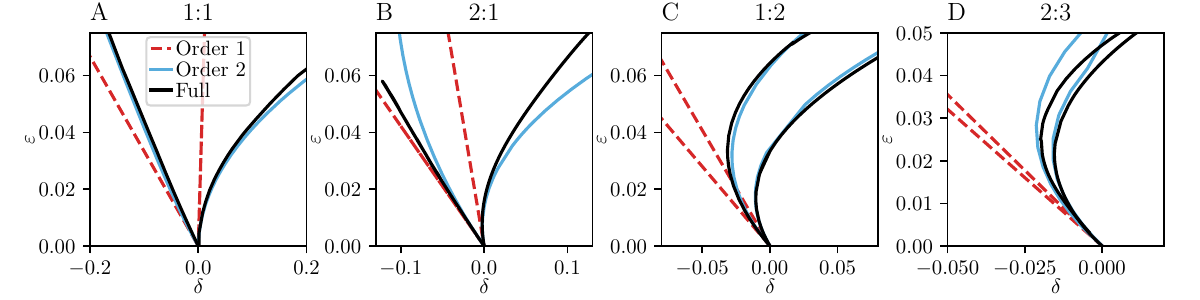}
    \caption{\yp{Two-parameter bifurcation diagrams (Arnold tongues) of the forced thalamic neuron. The vertical axis is the forcing amplitude $\ve$ and the horizontal axis is the forcing frequency change $\delta$. The black solid, red dashed, and blue solid curves denote boundaries of phase-locking regions for the full model, $O(\ve)$ phase reduction, and $O(\ve^2)$ phase reduction, respectively.}}\label{fig:thal_tongues}
\end{figure}

Trajectories of the first- and second-order scalar reductions are compared to the full model trajectories in Figure \ref{fig:thal_force0}. We now consider a different set of frequency ratios because $n{:}m$ locking is now possible for $n,m>1$. In particular, we choose $1{:}1$, $2{:}1$, $1{:}2$, and $2{:}3$ for the purpose of illustration.  In all panels, the second-order reduction is consistently better able to capture the existence or non-existence of phase-locked states relative to the first-order reduction. Phase differences in the full model are estimated using the limit cycle computed at $\ve=0$ (see \ref{a:phase_estimate} for additional details on phase estimation).

Just as in the previous example of the nonradial isochron clock, our goal is also to predict where the original system loses the existence of phase-locked states due to nonzero heterogeneity $\delta$ and non-weak forcing strength $\ve$, and thus compute one-parameter bifurcation diagrams.

One-parameter bifurcation diagrams of the reduced and full thalamic neuron models are shown in Figure \ref{fig:bif1d_thal1}. All panels A-H of Figure \ref{fig:bif1d_thal1} correspond to the same underlying model in panels A-H in Figure \ref{fig:thal_force0}. We observe that the second-order scalar reduction is significantly more capable of capturing the existence of frequency-locked states as a function of $\ve$ compared to the first-order reduction. \yp{The diagrams are computed numerically by directly computing fixed points in the order $O(\ve)$ and $O(\ve^2)$ reduced models, while we use Newton's method to find phase-locked states in the full model (see \ref{a:newton} for additional details on how we compute bifurcation diagrams using Newton's method). In brief, we use numerical integration to estimate the location of stable or unstable fixed points, then initialize Newton's method close to this state. Convergence to a phase-locked state is satisfied when the correction vector in Newton's method has a norm within a magnitude of \num{1e-5} achieved in under 20 Newton steps.}

Two-parameter bifurcation diagrams of the forced thalamic neuron are shown in Figure \ref{fig:thal_tongues}. The diagrams were computed numerically using XPPAUTO \cite{xpp} using Fourier approximations of the $\mathcal{H}$-functions in the scalar reduction up to $O(\ve^2)$. Coefficients of the Fourier approximation of each $\mathcal{H}$-function may be found in \ref{a:fourier}. The phase-locking regions in  Figure \ref{fig:thal_tongues} are equivalent to Arnold tongues, except that each horizontal axis shows the heterogeneity parameter $\delta$ as opposed to the more standard frequency ratio.

\subsection{Coupling}\label{sec:coupling}

\subsubsection{A Pair of Thalamic Neurons with Heterogeneity}\label{sec:thal_coupling}

We now consider a pair of synaptically coupled thalamic neurons (adapted from \cite{rubi04}). The model is identical to the forced thalamic model \eqref{eq:thal_nondim_force} (including the same parameters  as in Table \ref{tab:thal_nondim}), but we augment the system with a chemical synapse denoted by the variables $w_i$ for $i=1,2$:
\begin{equation}\label{eq:thalamic2}
\begin{split}
    \hat F_\text{Thal}(X) &= \left(\begin{matrix}
        F_\text{Thal}(X)\\
        \alpha (1-w_1) a_\infty(V_1) - \beta w_1
    \end{matrix}\right),\\
    \hat F_\text{Thal}(Y) &= \left(\begin{matrix}
        F_\text{Thal}(Y)\\
        \alpha (1-w_2) a_\infty(V_2) - \beta w_2
    \end{matrix}\right),
\end{split}
\end{equation}
where $X=[V_1,h_1,r_1,w_1]^\tr$ and $Y=[V_2,h_2,r_2,w_2]^\tr$. The coupled equations are given by,
\begin{equation}\label{eq:thal_nondim_coupling}
    \begin{split}
        \frac{1}{\omega_X}\frac{\dd X}{\dd t} &= \hat F_\text{Thal}(X) + \delta_1 + \delta_2^2 + \ve[G_\text{Thal}(X,Y)]^\tr,\\
        \frac{1}{\omega_Y}\frac{\dd Y}{\dd t} &= \hat F_\text{Thal}(Y) + \ve[G_\text{Thal}(Y,X)]^\tr,
    \end{split}
\end{equation}
where $\ve\equiv g_\text{syn}$ is the synaptic conductance, and
\begin{equation*}
    G_\text{Thal}(X,Y) = [w_2(V_1-E_\text{syn}),0,0,0]^\tr, \quad G_\text{Thal}(Y,X) = [w_1(V_2-E_\text{syn}),0,0,0]^\tr.
\end{equation*}
We choose an excitatory synapse with $E_\text{syn} = 0$.

The first- and second-order heterogeneity terms are given by
\begin{align*}
    \delta_1 \equiv \ve b c_1,\quad \delta_2 \equiv \ve b c_2,
\end{align*}
where $c_1$ and $c_2$ represent arbitrary coefficients of a Taylor expansion, and $b$ is a fixed heterogeneity parameter. In principle, there can be an infinite number of Taylor coefficients $c_i$, but we are only interested in testing whether the scalar reduction properly captures nonlinear effects of heterogeneity. Thus, choosing only $c_1$ and $c_2$ is equivalent to truncating a Taylor expansion of the heterogeneity function $J_X$.

\begin{figure}[ht!]
    \includegraphics[width=\textwidth]{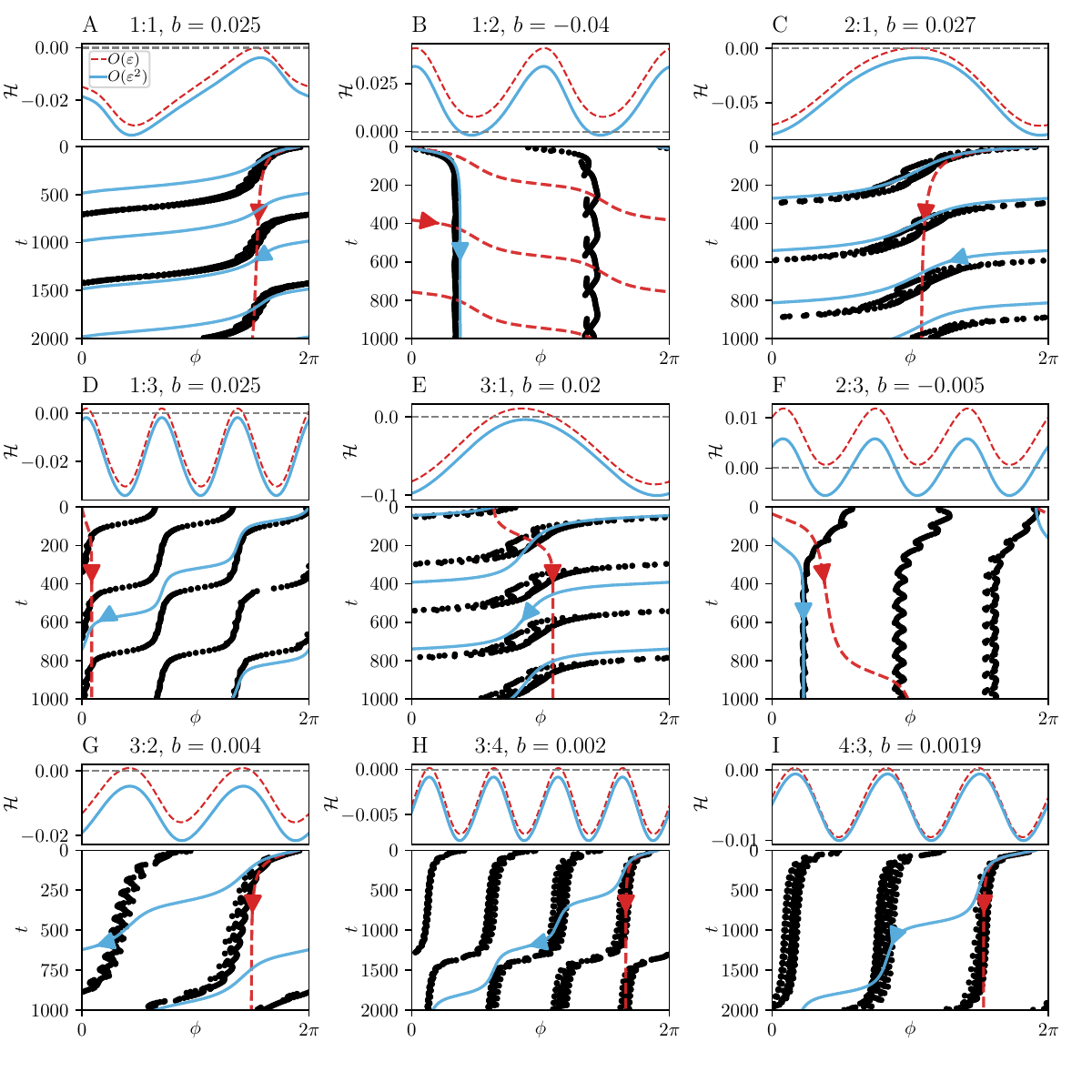}
    \caption{Representative trajectories of the scalar reduction (red dashed for order $O(\ve)$ and blue for order $O(\ve^2)$) compared to the original coupled thalamic models (black) for various degrees of heterogeneity $b$. The coupling strength $\ve$ is chosen to be $\ve=0.1$ for panels. We choose $c_1 = 1$ for all panels and $c_2 = [100,100,100,100,200,2000,2000,1000,1000]$ for respective panels A through I. To reduce lag when loading the figure, only one every two hundredth point of the full model's phase difference estimate is displayed. }\label{fig:traj_thal2}
\end{figure}

We choose $c_2 \gg c_1$ because we generally have that $\ve b \ll 1$. For instance, for $3{:}4$ phase-locking, we choose $c_1=1$ and $c_2=\num{1e3}$. Note that $c_2$ is not as excessively large as it may seem at first glance. First, $c_2$ is the coefficient for a term that is $\ve^2 b^2\ll \ve b \ll 1$, so $c_2$ must be exaggerated for non-linear effects to be noticeable. Second, for even modestly greater values of $n$ and $m$, such as $n=3$ and $m=4$, only small changes to $b$ lead to noticeable changes in phase-locked states, often on the order of $b=\num{1e-3}$. Then, $b^2$ is on the order of $\num{1e-5}$, and the second-order coefficient, $c_2 b^2$ (as it appears in the model equations), is on the order of $\num{1e-3}$ for $3{:}4$ phase-locking, which is reasonable. So the exaggerated choices for $c_2$ ultimately lead to more natural second-order coefficients.

\begin{figure}[ht]
    \includegraphics[width=\textwidth]{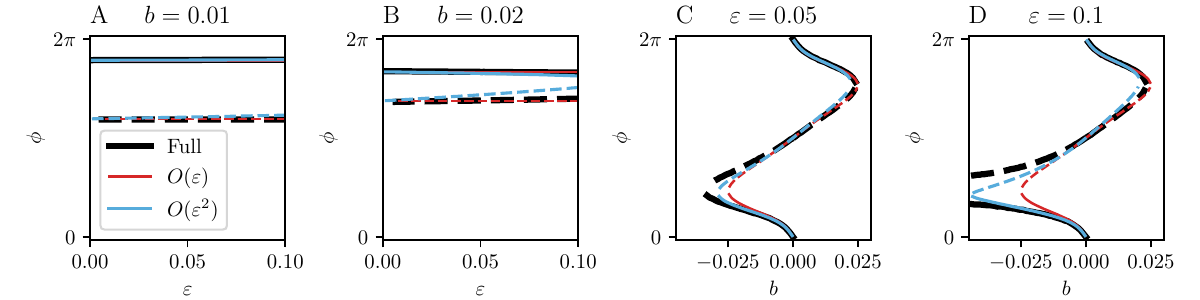}
    \caption{One-parameter bifurcation diagrams for $1{:}1$ phase-locking in a pair of thalamic neurons given varying levels of heterogeneity. All panels show the existence of fixed points in the $O(\ve)$ reduction (red), $O(\ve^2)$ reduction (blue), and full model (black). Solid (dashed) curves correspond to stable (unstable) fixed points. Panels A, B show bifurcation diagrams as a function of coupling strength $\ve$ given heterogeneity $b=0.01$ (A) and $b=0.02$ (B). Panels C, D show bifurcation diagrams as a function of heterogeneity $b$ given coupling strenghts $\ve=0.05$ (C) and $\ve=0.1$ (D). One-parameter bifurcations for remaining phase-locking combinations in Figure \ref{fig:traj_thal2} are shown in Figures \ref{fig:bif_thal2a} and \ref{fig:bif_thal2b}.}\label{fig:bif_thal2_11}
\end{figure}

\begin{remark}
    The simple choice of the heterogeneity terms leads to cases where phase-locking may not only be lost by increasing the magnitude of heterogeneity $b$, but also regained. Consider the heterogeneity terms with $b<0$:
    \begin{align*}
        \delta_1 + \delta_2^2 &= \ve b c_1 + \ve^2 b^2 c_2\\
        &= -\ve |b| c_1 + \ve^2 |b|^2 c_2.
    \end{align*}
    Suppose, for concreteness, that $c_1 = 1$, $c_2 = 1000$, and $\ve=0.1$. Then,
    \begin{equation*}
        \delta_1 + \delta_2^2 = -0.1|b| + 10 |b|^2.
    \end{equation*}
    If we set this equation to zero, we find two distinct values of $|b|$ that result in zero total heterogeneity:
    \begin{equation}\label{eq:b_cancel}
         |b|(-0.1 + 10 |b|) = 0,
    \end{equation}
    where $|b| = 0$ or $|b| = 0.1$. Since we assumed $b<0$, the only possible non-trivial root is $b = -0.1$. This observation shows that the resumption of phase-locking with this choice of heterogeneity is due to a cancelation of \yp{first- and second-order} heterogeneity terms (note that the same cancelation \yp{in \eqref{eq:b_cancel}} is possible for $b>0$ if $c_1 = -1$). $\blacksquare$
\end{remark}


Trajectories of the order $O(\ve)$ (red dashed) and order $O(\ve^2)$ (blue) scalar reductions, as well as the full model (black), are shown in Figure \ref{fig:traj_thal2}. Phase differences in the full model are estimated using the limit cycle computed at $\ve=0$ (see \ref{a:phase_estimate} for additional details about the phase estimation procedure). The second-order scalar reduction (blue) better reproduces phase-locked states than the first-order scalar reduction (red dashed) for all considered combinations of $n{:}m$ phase-locking and heterogeneity, including $1{:}1$ (Figure \ref{fig:traj_thal2}A), $1{:}2$ (Figure \ref{fig:traj_thal2}B), $2{:}1$ (Figure \ref{fig:traj_thal2}C), $1{:}3$ (Figure \ref{fig:traj_thal2}D), $3{:}1$ (Figure \ref{fig:traj_thal2}E), $2{:}3$ (Figure \ref{fig:traj_thal2}F), $3{:}2$ (Figure \ref{fig:traj_thal2}G), $3{:}4$ (Figure \ref{fig:traj_thal2}H), and $4{:}3$ (Figure \ref{fig:traj_thal2}I).

\begin{figure}[ht!]
    \includegraphics[width=\textwidth]{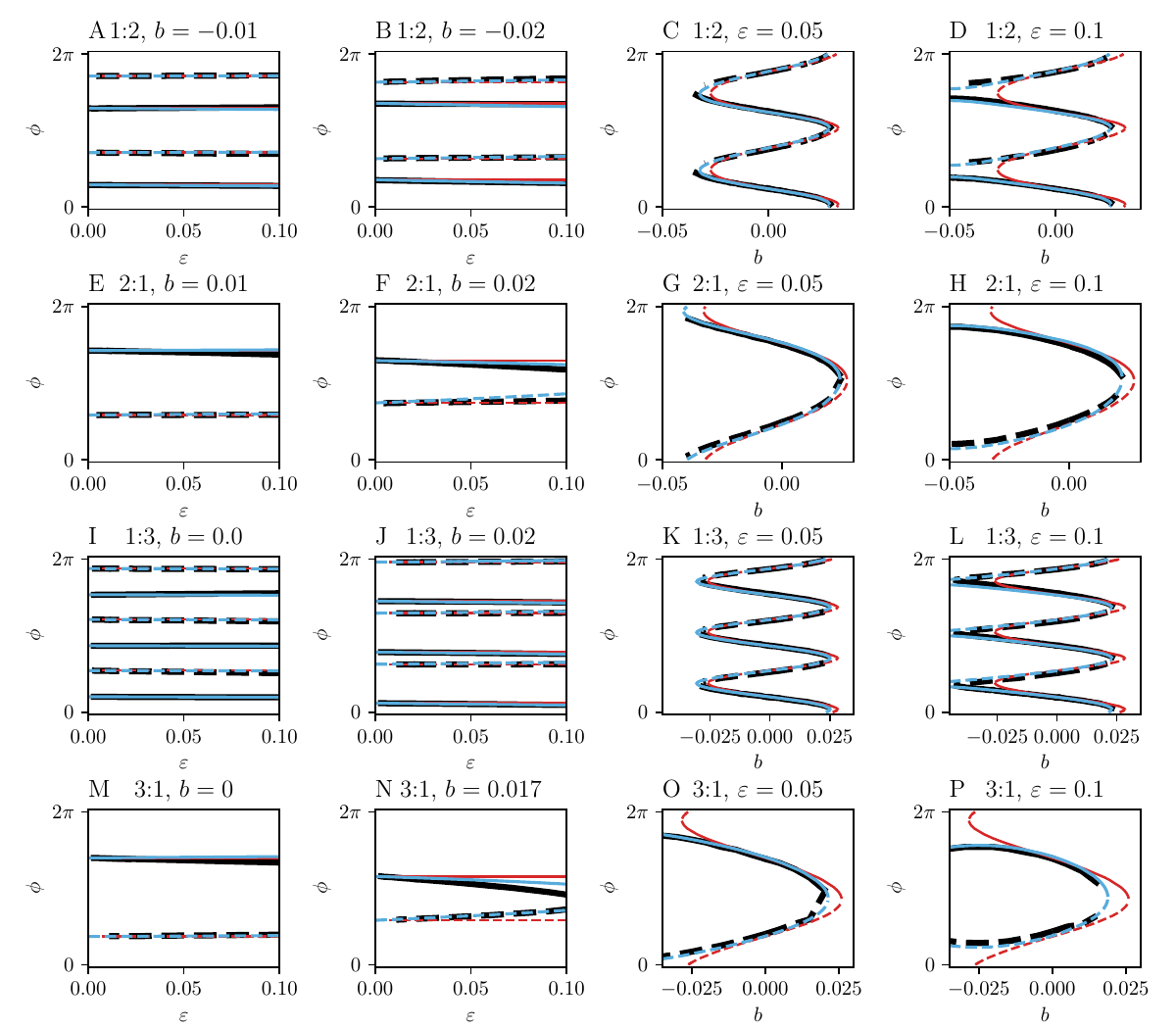}
    \caption{One-parameter bifurcation diagrams for $n{:}m$ phase-locking in a pair of thalamic neurons given varying levels of heterogeneity. All panels show the existence of fixed points in the $O(\ve)$ reduction (red), $O(\ve^2)$ reduction (blue), and full model (black). Solid (dashed) curves correspond to stable (unstable) fixed points.}\label{fig:bif_thal2a}
\end{figure}

\begin{figure}[ht!]
    \includegraphics[width=\textwidth]{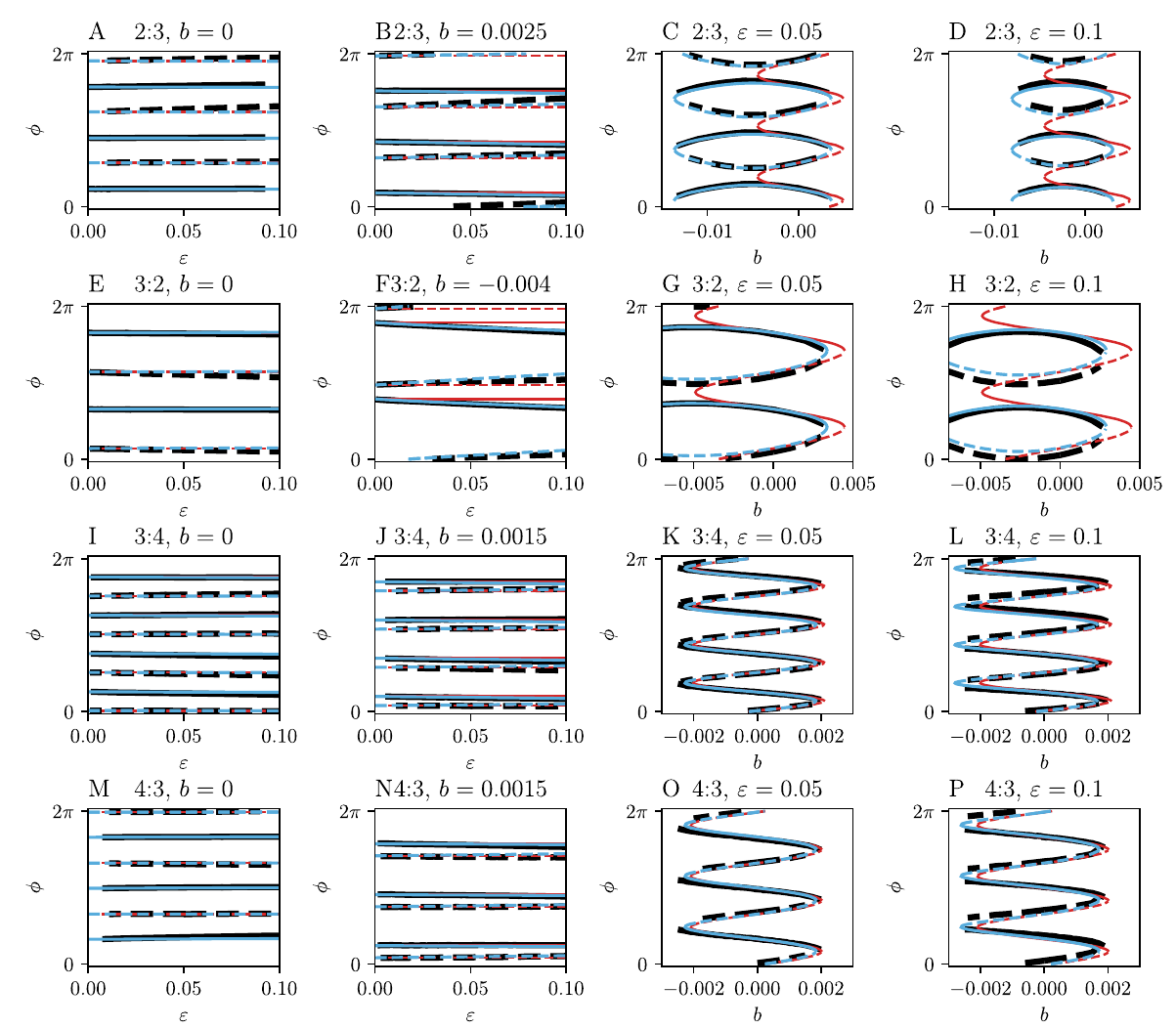}
    \caption{One-parameter bifurcation diagrams for $n{:}m$ phase-locking in a pair of thalamic neurons given varying levels of heterogeneity. All panels show the existence of fixed points in the $O(\ve)$ reduction (red), $O(\ve^2)$ reduction (blue), and full model (black). Solid (dashed) curves correspond to stable (unstable) fixed points.}\label{fig:bif_thal2b}
\end{figure}

\begin{figure}[ht!]
    \includegraphics[width=\textwidth]{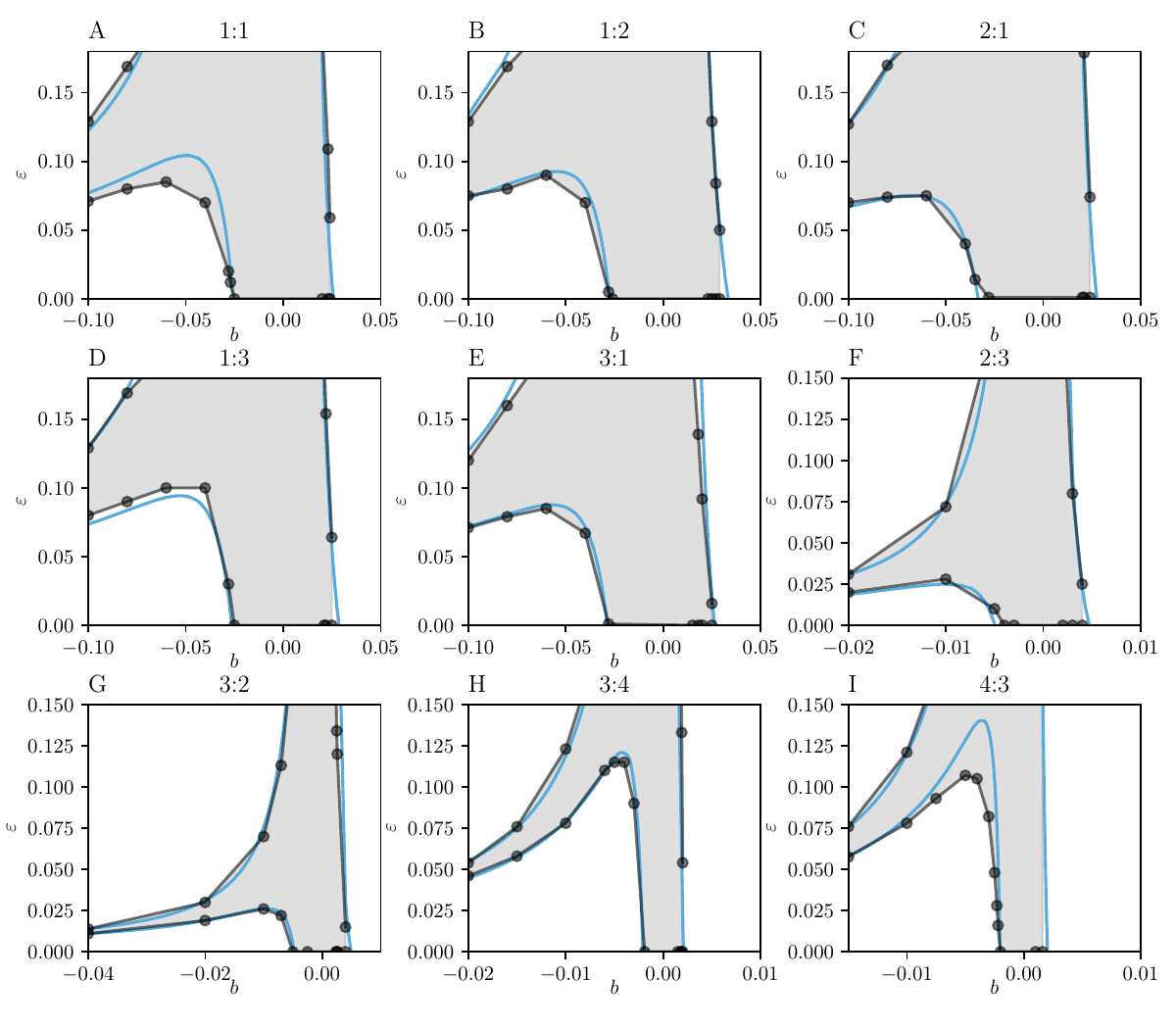}
    \caption{\yp{Two-parameter bifurcation diagrams of the reduced coupled thalamic model for $n{:}m$ phase-locking. The horizontal axis is the heterogeneity $b$ and the vertical axis is the coupling strength $\ve$. Black dots correspond to estimated fold bifurcations of periodic orbits in the full model. Blue curves denote fold bifurcations in the order $O(\ve^2)$ reduced system. Gray regions indicate phase-locking regions in the full model.}}\label{fig:bif2_thal2}
\end{figure}

A one-parameter bifurcation diagram for $1{:}1$ phase-locking is shown in Figure \ref{fig:bif_thal2_11} for the first-order reduction (red), second-order reduction (blue), and the full model (black). Fixed points of the first- and second-order reduced models  are computed directly using the phase equation \eqref{eq:dphi}, while phase-locked states of the full model are found using Newton's method on a stroboscopic map. All panels A-D of Figure \ref{fig:bif_thal2_11} correspond to the same underlying model in panel A of Figure \ref{fig:traj_thal2}. We refer the reader to Appendix \ref{a:newton} for additional details on how we apply Newton's method for coupled oscillators.

Because the coupling strength term $\ve$ always multiplies the heterogeneity term $b$, the system tends to be much less sensitive to changes in $\ve$ for a fixed value of $b$, as shown in Figure \ref{fig:bif_thal2_11} Panels A and B (intuitively, for our choice of the thalamic model, there happen to be fewer \yp{second}-order effects due to coupling strength). Thus, changes to coupling strength alone will not necessarily reveal significant differences between the first- and second-order reductions. In contrast, more interesting changes are visible in Figure \ref{fig:bif_thal2_11} Panels C and D, where we fix the coupling strength $\ve$ and alter the heterogeneity $b$. The \yp{second}-order effects of heterogeneity are especially visible in Figure \ref{fig:bif_thal2_11} Panel D due to the increased coupling strength $\ve$, which also exaggerates the \yp{second}-order effects of heterogeneity.

One-parameter diagrams for the remaining combinations of phase-locking in Figure \ref{fig:traj_thal2} are shown in  Figure \ref{fig:bif_thal2a} for $1{:}2$, $2{:}1$, $1{:}3$, $3{:}1$ locking and Figure \ref{fig:bif_thal2b} for $2{:}3$, $3{:}2$, $3{:}4$, $4{:}3$ locking, with the same corresponding values of $c_i$ as in Figure \ref{fig:traj_thal2}. The color scheme is the same as in Figure \ref{fig:bif_thal2_11}: red corresponds to the first-order reduction, blue to the second-order reduction, and black to the full model.

Two-parameter bifurcation diagrams of the second-order scalar reduction, along with estimated phase-locking boundaries of the full model, are shown in Figure \ref{fig:bif2_thal2}. In all cases, the second-order reduction accurately captures nonlinear phase-locked regions present in the original model.

\begin{remark}\label{remark:first_order_twopar}
    We exclude the first-order reduction in Figure \ref{fig:bif2_thal2} (and later in Figure \ref{fig:bif2_vdp_thal}) because the \yp{first}-order reduction has only vertical boundaries that separate the stable phase-locking regions from phase drift, i.e., the phase-locking regions are independent of the coupling strength. This observation can be proven simply by considering an equivalent form of the first-order reduction:
    \begin{equation*}
        x' = \ve(h_1(x) + b h_2(x)),
    \end{equation*}
    where $x$ is a scalar and $h_i: \mathbb{R} \rightarrow \mathbb{R}$ represents a $\mathcal{H}$ or $\mathcal{J}$ function. If $x=x^*$ is the coordinate of a saddle-node given $b=b^*$ and $\ve=\ve^*$, then by definition, we also have the following conditions satisfied:
    \begin{align*}
        0 &= \ve^*(h_1(x^*) + b^* h_2(x^*)),\\
        0 &= \ve^*(h_1'(x^*) + b^* h_2'(x^*)).
    \end{align*}
    Note, then, that so long as $x^*$ and $b^*$ are held constant, the conditions for the saddle-node will be satisfied for any $\ve^*$, thus proving that the first-order scalar reduction will not capture meaningful two-parameter bifurcation diagrams. $\blacksquare$
\end{remark}

\subsubsection{A Pair of Mixed Model Types with Heterogeneity}\label{sec:vdp_thal_coupling}

\yp{To further demonstrate the utility of the scalar reduction \eqref{eq:dphi}, consider the thalamic model synaptically coupled to the Van der Pol oscillator,}
\begin{equation}
    \begin{split}
        \frac{1}{\omega_X}\frac{\dd X}{\dd t} &= \hat F_\text{Thal}(X) + \delta_1 + \delta_2^2 + \ve G_\text{Thal}(X,Y),\\
        \frac{1}{\omega_Y}\frac{\dd Y}{\dd t} &= \hat F_\text{VDP}(Y) + \ve G_\text{VDP}(X,Y),
    \end{split}
\end{equation}
where $X=[V_1,h_1,r_1,w_1]^\tr$ and $Y=[V_2,h_2,w_2]^\tr$. The vector field $F_\text{VDP}$ is given by,
\begin{align}\label{eq:vdp}
    F_\text{VDP}(Y) = 
    \left(\begin{matrix}
        \mu (V_2-V_2^3-h_2)\\
        V_2/\mu\\
        \alpha (1-w_2) a_\infty(V_2) - \beta w_2
    \end{matrix}\right),
\end{align}
and the coupling functions $G_\text{Thal}$ and $G_\text{VDP}$ are defined as
\begin{equation*}
    G_\text{Thal}(X,Y) = [w_2(V_1-E_\text{syn,1}),0,0,0]^\tr, \quad G_\text{VDP}(X,Y) = [w_1(V_2-E_\text{syn,2}),0,0,0]^\tr.
\end{equation*}

\begin{figure}[ht!]
	\includegraphics[width=\textwidth]{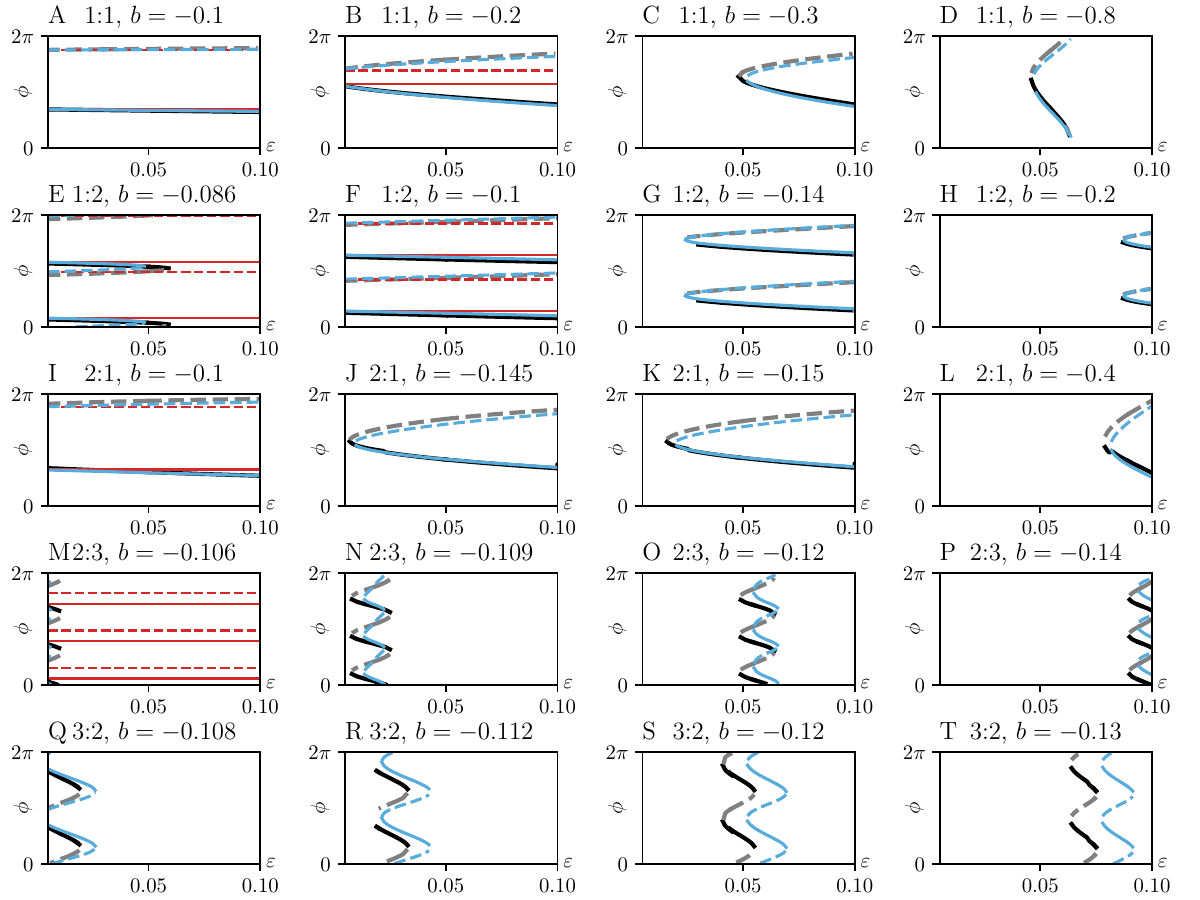}
	\caption{\yp{One-parameter bifurcation diagrams of the Van der Pol oscillator coupled to the thalamic model for $n{:}m$ phase-locking. The horizontal axis is the coupling strength $\ve$ and the vertical axis is the phase difference $\phi$.  Each panel shows the existence of fixed points in the first-order $O(\ve)$ reduction (red), second-order $O(\ve^2)$ reduction (blue), and full model (black). Solid (dashed) curves correspond to stable (unstable) fixed points for all models. The unstable fixed points of the full model have been shaded gray to assist in visualization. Subplots without red curves imply that the first-order $O(\ve)$ reduction predicts no phase-locked states.}}\label{fig:bif1_vdp_thal}
\end{figure}

We choose the Van der Pol equation because of its biological relevance, its comparable timescale to the thalamic model, and its easily tunable limit cycle, which can be made weakly attracting through the parameter $\mu$. We choose $\mu=0.04$ so that the slowest decaying Floquet exponent is $\kappa_\text{VDP}\approx-0.04$. The Van der Pol model is also of different dimension than the thalamic model and obeys very different rules such that no symmetry results can be used to examine phase-locked states. This example provides a starting point for determining the robustness of phase-locked states in biologically realistic settings. The thalamic neuron uses the same equations and parameters as in the coupling case (see Table \ref{tab:thal_nondim} in Section \ref{sec:thal_coupling}).

The synaptic equations in \eqref{eq:vdp} use the same equations as the coupled thalamic models \eqref{eq:thalamic2}, but with parameters modified so that the synaptic variable $w_2$ becomes nonzero as the $V_2$ variable of the Van der Pol oscillator \eqref{eq:vdp} reaches its peak. In particular, for the $w_2$ equation, we choose $\sigma_T = 0.1$, $V_T = 1$, $\alpha = 3$, and $\beta = 2$. Parameters of the Van der Pol oscillator are summarized in Table \ref{tab:vdp}. Further, we choose $E_\text{syn,1}=0$ so that the Van der Pol oscillator imparts an excitatory effect on the thalamic neuron, and we choose $E_\text{syn,2}=-2$ so that the Van der Pol oscillator receives an inhibitory effect from the thalamic neuron. These asymmetric reversal potentials further reduce symmetry.

\begin{table}
    \caption{\yb{Parameter values of the Van der Pol oscillator in \eqref{eq:vdp}. All values are nondimensional}.}\label{tab:vdp}
    \centering
    \begin{tabular}{c|c|c|c|c|c}
        Parameter & $\mu$ & $\sigma_T$& $V_T$ & $\alpha$ & $\beta$ \\
        \hline
        Value & 0.04 & 0.1 & 1 & 3 & 2 \\
    \end{tabular}
\end{table}

\begin{figure}[ht!]
	\includegraphics[width=\textwidth]{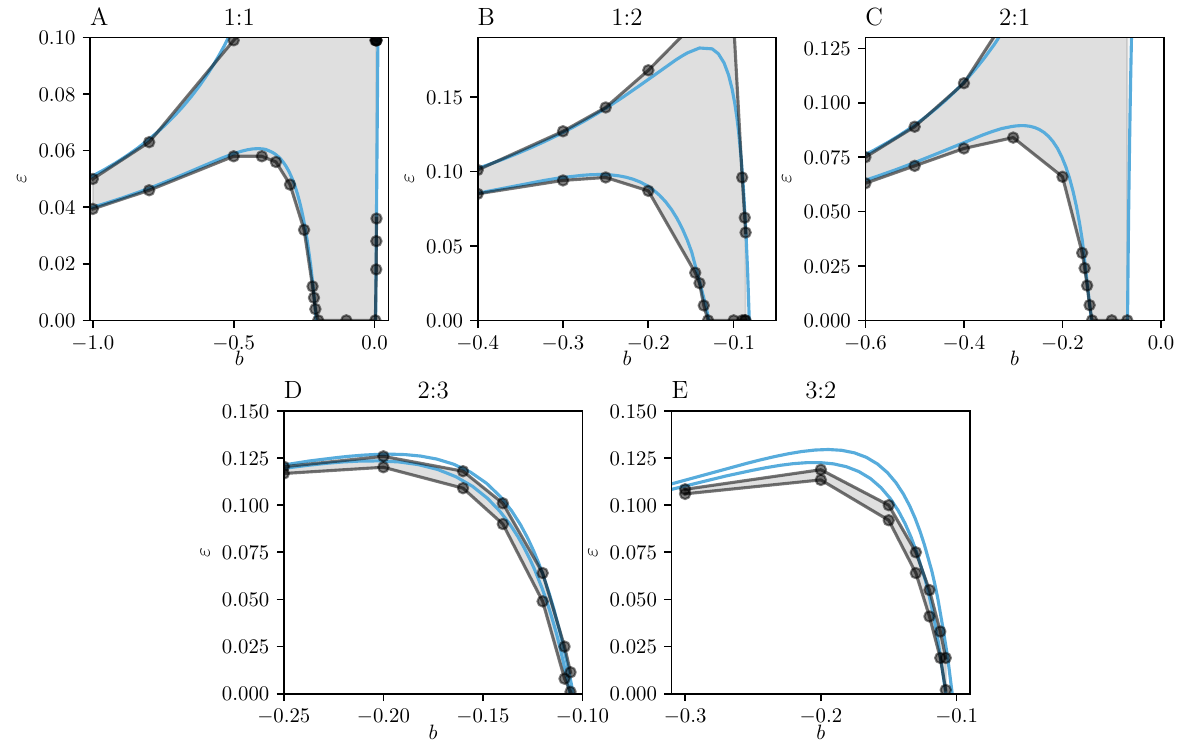}
	\caption{\yp{Two-parameter bifurcation diagrams of the Van der Pol oscillator coupled to the thalamic model for $n{:}m$ phase-locking. The horizontal axis is the heterogeneity $b$ and the vertical axis is the coupling strength $\ve$. Black dots correspond to estimated fold bifurcations of periodic orbits in the full model. Blue curves denote fold bifurcations in the order $O(\ve^2)$ reduced system. Gray regions indicate phase-locking regions in the full model.}}\label{fig:bif2_vdp_thal}
\end{figure}

\yp{One-parameter bifurcation diagrams of the mixed model types are shown in Figure \ref{fig:bif1_vdp_thal} for various combinations of $n{:}m$ phase-locking including $1{:}1$ (Figure \ref{fig:bif1_vdp_thal}A-D, $1{:}2$ (Figure \ref{fig:bif1_vdp_thal}E-H), $2{:}1$ (Figure \ref{fig:bif1_vdp_thal}I-L), $2{:}3$ (Figure \ref{fig:bif1_vdp_thal}M-P), and $3{:}2$ (Figure \ref{fig:bif1_vdp_thal}Q-T). The bifurcation diagram generated using the first-order reduction ($O(\ve)$) is shown in red, the bifurcation diagram using the second-order reduction ($O(\ve^2)$) is in blue, and the bifurcation diagram using the full model is shown in black. Note that several panels are missing the first-order $O(\ve)$ reduction (red curves) because the first-order reduction cannot capture nonlinear effects due to coupling strength $\ve$ or heterogeneity $\delta$.}

\yp{Both stable and unstable phase-locked states are computed using Newton's method, identical to the procedure used to calculate phase-locked states of the coupled thalamic neurons (Section \ref{sec:thal_coupling}). Details about this procedure may be found in Section \ref{a:newton}. In brief, we use Newton's method to find stable or unstable phase-locked states. Convergence to a phase-locked state is satisfied when the correction vector in Newton's method has a norm within \num{1e-5} achieved in under 20 Newton steps.}

\yp{Two-parameter bifurcation diagrams of the mixed model types are shown in Figure \ref{fig:bif2_vdp_thal} for the same combinations of $n{:}m$ phase-locking in the corresponding one-parameter diagram (Figure \ref{fig:bif1_vdp_thal}). The first-order $O(\ve)$ reduction is not shown because it cannot capture nonlinearities as a function of $\ve$; its phase-locking regions are bounded only by vertical lines independent of $\ve$ (see Remark \ref{remark:first_order_twopar}). }

\yp{These results -- including the coupled thalamic neurons in Section \ref{sec:thal_coupling} and the Van der Pol oscillator coupled to a thalamic neuron model in Section \ref{sec:vdp_thal_coupling} -- demonstrate that the proposed phase-reduction to second order \eqref{eq:dphi} robustly and reliably reproduces $n{:}m$ phase-locked behaviors of realistic conductance-based neural models in a manner that the first-order reduction cannot.}

\section{Discussion}\label{sec:discussion}

We have demonstrated that the proposed scalar reduction captures the same existence and non-existence of phase-locked states as the corresponding full model for various combinations of $n{:}m$ phase-locking -- in particular where there exist nonlinearities in the coupling strength $\ve$ and heterogeneity $\delta$.

Across virtually all results, the heterogeneity parameter $b$ may be orders of magnitude smaller than $\ve$ yet yield significant changes in the existence and stability of phase-locked states. This observation suggests that methods relying on symmetry, which are capable of showing the existence and stability of fixed points even with very strong coupling \cite{golubitsky2006symmetry}, may ultimately find limited use in biologically realistic contexts where heterogeneity is an important (if not necessary) part of function.

\subsection{Limitations}

The proposed scalar reduction relies on having a reasonable separation of timescales between the phase difference dynamics and the limit cycle period so that first-order averaging can be used to turn non-autonomous equations into relatively tractable autonomous equations. \yb{We have observed a similar separation of timescales in prior work for $1{:}1$ phase-locking \cite{park2021high,park2024body}. Establishing precise conditions for this separation of timescales for $n{:}m$  phase-locking is beyond the scope of this work, but our results show that first-order averaging is sufficient to capture nonlinear contributions of heterogeneity and coupling strength in phase-locked behaviors, provided that $\ve$ and $\delta$ are sufficiently small.}


Another possible limitation lies in the types of bifurcations that the proposed reduction can capture. While the scalar reduction naturally captures fold bifurcations of cycles via saddle-node bifurcations, it is unable to capture Hopf bifurcations in phase-locked states for a pair of oscillators. The scalar model also cannot capture transient oscillations toward phase-locked states that happen to be stable foci, although it can capture the existence and asymptotic stability of stable foci. Note, however, that for three or more oscillators, the scalar model can capture Hopf bifurcations in the phase difference dynamics, as demonstrated in \cite{park2024body} for $1{:}1$ phase-locking.

\subsection{Utility of the Proposed Method}

The proposed scalar reduction retains significant utility despite potential limitations. Multiple authors have developed and demonstrated scalar reductions beyond the weak coupling regime (i.e., beyond the linear regime) using a similar phase-isostable approach \cite{wilson2019phase,park2021high,park2024body,mau2023high}, but their results are restricted to $1{:}1$ phase-locking and typically without heterogeneity. While there are existing works on scalar reductions with heterogeneity, such as \cite{mau2023high}, their results apply only to planar systems. In contrast, this paper establishes a natural extension to arbitrarily high-dimensional systems for $n{:}m$ phase-locking with heterogeneity that is present in biological systems. We have also considered two coupling examples that are more biologically realistic. The proposed formulation for heterogeneity is not just restricted to the case of $N=2$ oscillators, as in \cite{mau2023high}. The generalization to $n{:}m$ phase-locking for $N \geq 2$ oscillators will be the topic of future work.

The proposed reduction is efficient to integrate numerically because the integrator only references an interpolating function for each $\mathcal{H}_i^{(\ell)}$ function. Integrating the scalar reduction often takes orders of magnitude less time compared to integrating the full system while being much easier to analyze. Therefore, the proposed reduction may be used with biologically realistic models in more biologically realistic settings to understand changes in phase-locked states as a function of key model parameters. We have primarily explored changes in phase-locked states as a function of coupling strength $\ve$ and heterogeneity $\delta$, but it is straightforward to apply the proposed method to understand how phase-locked states change as a function of any other model parameter -- this type of analysis will also be the topic of future work.

The proposed reduction may be used to introduce more realistic forms of coupling functions to idealized models such as the Kuramoto model, \yp{which uses a single sinusoid to describe oscillator interactions. Realistic interaction functions often involve more Fourier modes, as demonstrated in the present study (see, e.g., Figures \ref{fig:a:11} - \ref{fig:a:43}) and Figures \ref{fig:a:vdp11} - \ref{fig:a:vdp32})}. This idea may be applied naturally beyond the Kuramoto model, e.g., to the \yp{second}-order reduction of the Haken-Kelso-Bunz (HKB) equation in \cite{mckinley2021third}, which considers the \yp{first}-order Fourier coefficients of the $\mathcal{H}$-functions of the proposed method.

\section*{Acknowledgments}
The author acknowledges Bard Ermentrout for helpful discussions regarding the choice of forcing function in Section \ref{sec:forcing} and acknowledges support under the National Science Foundation Grant No. NSF-2512988. \yb{In addition, the author thanks both reviewers for greatly strengthening the paper -- Reviewer 1 for their tireless efforts to improve the writing and verify calculations across multiple iterations, and Reviewer 2 for their exceptionally thoughtful questions and comments.}

\appendix

\section{Elimination of Isostable Coordinates}\label{sec:elimination}

We solve the isostable coordinates to further reduce \eqref{eq:phase_isostable_na} from 4 dimensions to 2 dimensions. To this end, we will perform the following steps:
\begin{enumerate}
    \item Convert all expansions in $\psi_i$ to expansions in $\ve$ using the ansatz
    \begin{align}
        \psi_i(s) &\approx \ve p_i^{(1)}(s) + \ve^2 p_i^{(2)}(s) + \ve^3 p_i^{(3)}(s) + \cdots\tag{\ref{eq:psi_exp}}, \quad i\in\{X,Y\}.
    \end{align}
    \item Obtain inhomogeneous linear equations for $p_i^{(k)}$ in a hierarchy of equations in powers of $\ve$.
    \item Solve for each $p_i^{(k)}$ to the desired order in $\ve$ (this step is handled by a symbolic \yp{algebra} solver).
    \item Plug in each equation for $p_i^{(k)}$ into the phase equation, thus effectively eliminating the isostable variable (note, however, that the isostable dynamics are retained, and perturbed trajectories need not be near the unperturbed limit cycle).
\end{enumerate}

\paragraph{Step 1: Convert all expansions in $\psi_i$ to expansions in $\ve$} We plug in the expansion \eqref{eq:psi_exp} into all terms containing $\psi$ in \eqref{eq:phase_isostable_na}, and collect terms in powers of $\ve$. We then obtain the $\ve$-expansion for terms in $\hat{G}_i$ (recall that $\hat G_X(X,Y) = bJ_X(X) + G_X(X,Y)$, $\hat G_Y(X,Y) = bJ_Y(Y) + G_Y(X,Y)$, and $b = \delta/\ve$), such as the heterogeneity functions $J_i$:
\begin{align*}
    J_i(\theta_X,\psi_X) &= \yp{J_i(\theta_X,\ve p_X^{(1)} + \ve^2 p_X^{(2)} + \cdots)}\\
    &\approx \yp{J_i^{(0)}(\theta_X) + \ve J_i^{(1)}(\theta_X)p_X^{(1)} + \ve^2 J_i^{(2)}(\theta_X)p_X^{(2)} + \cdots},
\end{align*}
for $i\in\{X,Y\}$ and the $\ve$-expansion of $G_X$
\begin{equation}\label{eq:G_eps_body}
    \begin{split}
        G_{X}(\theta_X,\theta_Y,\psi_X,\psi_Y)= &\, K_{X}^{(0)}(\theta_X,\theta_Y)+ \ve K_{X}^{(1)}\left(\theta_X,\theta_Y,p_X^{(1)},p_Y^{(1)}\right)\\
        &+ \ve^2 K_{X}^{(2)}\left(\theta_X,\theta_Y,p_X^{(1)},p_X^{(2)},p_Y^{(1)},p_Y^{(2)}\right)\\
        &+ \cdots,
    \end{split}
\end{equation}
where \yp{$\theta_i\in[0,2\pi)$}. The $\ve$-expansion of $G_Y$ is virtually identical in form to \eqref{eq:G_eps_body} with parity in the indices. Some details of how these terms are computed and collected are discussed in \ref{a:g}. In brief, we use a symbolic \yp{algebra} package to handle the collection of the Taylor expansion of $G_X$ and $G_Y$, due to the \yp{large} number of terms for even small orders of $\ve$ \yp{including order $O(\ve)$ and order $O(\ve^2)$}.

The $\ve$-expansions of $\mathcal{Z}_i$ and $\mathcal{I}_i$ are straightforward to obtain by plugging in the $\ve$-expansion of $\psi_i$ \eqref{eq:psi_exp} into the $\psi$-expansion of $\mathcal{Z}_i$ \eqref{eq:z_exp} and $\mathcal{I}_i$ \eqref{eq:z_exp}, then collecting in powers of $\ve$ using a symbolic \yp{algebra} package. We then have all terms $\mathcal{Z}_i$, $\mathcal{I}_i$, and $\hat G_i$ expanded in $\ve$, and can proceed to rewrite the phase-isostable equations \eqref{eq:phase_isostable_na} in terms of $\ve$. In turn, we obtain a hierarchy of ODEs in powers of $\ve$.

\paragraph{Step 2: Obtain inhomogeneous linear equations for $p_i^{(k)}$ in a hierarchy of equations in powers of $\ve$} The left-hand side consists of straightforward time derivatives:
\begin{align*}
    \frac{1}{\omega}\frac{d}{ds}\psi_X &=  \frac{\ve}{\omega}\frac{d}{ds}p_X^{(1)} +  \frac{\ve^2}{\omega}\frac{d}{ds}p_X^{(2)}+  \frac{\ve^3}{\omega}\frac{d}{ds}p_X^{(3)} +\cdots,\\
    \frac{d}{ds}\psi_Y &= \ve \frac{d}{ds}p_Y^{(1)} + \ve^2 \frac{d}{ds}p_Y^{(2)}+ \ve^3 \frac{d}{ds}p_Y^{(3)} +\cdots.
\end{align*}
For the right-hand side of $\psi_X$ we have,
\begin{align*}
    &\kappa_X \psi_X +\ve\mathcal{I}_X(\hat \theta_X+\omega s,\psi_X)\cdot \hat G_X(\hat\theta_X+\omega s,\hat\theta_Y+s,\psi_X,\psi_Y)\\
    &= \kappa_X\left(\ve p_X^{(1)} + \ve^2 p_X^{(2)}+ \ve^3 p_X^{(3)}+\cdots\right)\\
    &\quad+ \left[ \ve\left\{I_X^{(0)}\right\} + \ve^2 \left\{p_X^{(1)}I_X^{(1)}\right\} + \ve^3 \left\{p_X^{(2)} I_X^{(1)} + (p_X^{(1)})^2 I_X^{(2)}\right\} +\cdots\right]\\
    &\quad\quad\quad  \cdot\Bigl[ \left\{b J_X^{(0)} + K_X^{(0)}\right\}+ \ve \left\{ b J_X^{(1)} + K_X^{(1)}\right\}+ \ve^2 \left\{ b J_X^{(2)}+  K_{X}^{(2)} \right\} + \cdots\Bigr]\\
    &= \ve \left\{ \kappa_X p_X^{(1)} + I_X^{(0)}\cdot[bJ_X^{(0)} + K_X^{(0)}] \right\}\\
    &\quad+ \ve^2 \left \{ \kappa_X p_X^{(2)} + I_X^{(0)} \cdot [bJ_X^{(1)} + K_X^{(1)}] + p_X^{(1)} I_X^{(1)} \cdot[b J_X^{(0)} + K_X^{(0)}]\right\}\\
    &\quad +\ve^3 \Bigl\{\kappa_X p_X^{(3)} + I_X^{(0)} \cdot [b J_X^{(2)} + K_X^{(2)}] + p_X^{(1)} I_X^{(1)}\cdot[b J_X^{(1)} + K_X^{(1)}]\\
    &\quad\quad\quad\quad\quad\quad\quad + [p_X^{(2)}I_X^{(1)} + (p_X^{(1)})^2 I_X^{(2)}]\cdot[bJ_X^{(0)} + K_X^{(0)}]\Bigr\},
\end{align*}
where we have suppressed the following function dependencies to keep the notation less cluttered: $p_i^{(\ell)}(s)$, $I_X^{(\ell)}(\hat\theta_X + \omega s)$, $J_X^{(0)}(\hat\theta_X+\omega s)$, $J_X^{(1)}(\hat\theta_X+\omega s,p_X^{(1)})$, $J_X^{(2)}(\hat\theta_X+\omega s,p_X^{(1)},p_X^{(2)})$, $K_X^{(0)}(\hat\theta_X+\omega s,\hat\theta_Y+s)$, $K_X^{(1)}(\hat\theta_X+\omega s,\hat\theta_Y+s,p_X^{(1)},p_Y^{(1)})$, and $K_{X}^{(2)}(\hat\theta_X+\omega s,\hat\theta_Y+s,p_X^{(1)},p_X^{(2)},p_Y^{(1)},p_Y^{(2)})$. The right-hand side of $\psi_Y$ follows similarly. 

Consolidating terms for each order in $\ve$ yields the following hierarchy of ODEs (starting with order $O(\ve)$),
\begin{equation}\label{eq:p_i_odes_X}
    \begin{split}
        \frac{1}{\omega}\frac{\dd p_X^{(1)}}{\dd s} &= \kappa_X p_X^{(1)} + I_X^{(0)} \cdot [bJ_X^{(0)} + K_X^{(0)}],\\
        \frac{1}{\omega}\frac{\dd p_X^{(2)}}{\dd s} &= \kappa_X p_X^{(2)} + I_X^{(0)} \cdot [bJ_X^{(1)} + K_X^{(1)}] + p_X^{(1)} I_X^{(1)} \cdot[b J_X^{(0)} + K_X^{(0)}],\\
        \frac{1}{\omega}\frac{\dd p_X^{(3)}}{\dd s} &= \kappa_X p_X^{(3)} + I_X^{(0)} \cdot [b J_X^{(2)} + K_X^{(2)}] + p_X^{(1)} I_X^{(1)}\cdot[b J_X^{(1)} + K_X^{(1)}]\\
    &\quad\quad\quad\quad\quad\quad\quad + [p_X^{(2)}I_X^{(1)} + (p_X^{(1)})^2 I_X^{(2)}]\cdot[bJ_X^{(0)} + K_X^{(0)}],\\
        &\,\,\,\vdots
    \end{split}
\end{equation}
The hierarchy of ODEs for $p_Y^{(\ell)}$ follows similarly. Here, we notice that all ODEs are first-order inhomogeneous differential equations with forcing terms that depend on previous-order solutions. As such, we can solve each ODE explicitly and pre-compute them.

\paragraph{Step 3: Solve for each $p_i^{(k)}$ to the desired order in $\ve$} It is straightforward to solve \eqref{eq:p_i_odes_X} with a combination of symbolic and numerical methods, but we show some explicit calculations for concreteness. We begin with the \yp{first}-order term, rewritten here with explicit dependencies for convenience:
\begin{equation*}
    \frac{1}{\omega}\frac{\dd p_X^{(1)}}{\dd s} = \kappa_X p_X^{(1)}(s) + I_X^{(0)}(\hat\theta_X+\omega s) \cdot [bJ_X^{(0)}(\hat\theta_X+\omega s) + K_X^{(0)}(\hat\theta_X+\omega s,\hat\theta_Y+s)]
\end{equation*}
The solution may be found using standard methods:
\begin{align*}
    p_X^{(1)}(s) &= \omega \int_0^s e^{\omega \kappa_X (s-r)} I_X^{(0)}(\hat\theta_X+\omega r) \cdot K_X^{(0)}(\hat\theta_X + \omega r,\hat\theta_Y+r) \,\mathrm{d} r\\
    &\quad+  \omega b\int_0^s e^{\omega \kappa_X (s-r)} I_X^{(0)}(\hat\theta_X+\omega r) \cdot J_X^{(0)}(\hat\theta_X + \omega r) \,\mathrm{d} r\\
    &\quad+p_X^{(1)}(0) e^{\omega \kappa_X s}.
\end{align*}
\begin{assumption}\label{as:transients}
    For each fixed $\hat\theta_X,\hat\theta_Y$, each $p_i^{(\ell)}$ term has transients that may be safely ignored.
\end{assumption}
This assumption allows us to rewrite the domain of integration from $[0,s]$ to $(-\infty,s]$ and drop the initial condition:
\begin{align*}
    p_X^{(1)}(s) &= \omega \int_{-\infty}^s e^{\omega \kappa_X (s-r)} I_X^{(0)}(\hat\theta_X+\omega r) \cdot K_X^{(0)}(\hat\theta_X + \omega r,\hat\theta_Y+r) \,\mathrm{d} r\\
    &+ \omega b \int_{-\infty}^s e^{\omega \kappa_X (s-r)} I_X^{(0)}(\hat\theta_X+\omega r) \cdot J_X^{(0)}(\hat\theta_X + \omega r) \,\mathrm{d} r.
\end{align*}
With the change of variables $r'=s-r$ \yp{followed by a trivial substitution in the dummy variable $r'\rightarrow r$}, the above equation becomes
\begin{align*}
    p_X^{(1)}(s) &= \omega \int_{0}^\infty e^{\omega \kappa_X r} I_X^{(0)}(\hat\theta_X+\omega (s-r)) \cdot \hat K_X^{(0)}(\hat\theta_X + \omega (s-r),\hat\theta_Y+s-r) \,\mathrm{d} r\\
    &\quad+\omega b \int_{0}^\infty e^{\omega \kappa_X r} I_X^{(0)}(\hat\theta_X+\omega (s-r)) \cdot J_X^{(0)}(\hat\theta_X + \omega (s-r)) \,\mathrm{d} r.
\end{align*}
For convenience and by definition, this function may be viewed purely as a function of the original fast phase variables $\theta_X$,$\theta_Y$:
\begin{align*}
    p_X^{(1)}(\theta_X,\theta_Y) &= \omega \int_{0}^\infty e^{\omega \kappa_X r} I_X^{(0)}(\theta_X-\omega r) \cdot \hat K_X^{(0)}(\theta_X -r\omega,\theta_Y-r) \,\mathrm{d} r\\
    &\quad+\omega b \int_{0}^\infty e^{\omega \kappa_X r} I_X^{(0)}(\theta_X-\omega r) \cdot J_X^{(0)}(\theta_X -r\omega) \,\mathrm{d} r.
\end{align*}
By an identical argument,
\begin{align*}
    p_Y^{(1)}(\theta_X,\theta_Y) &= \int_{0}^\infty e^{\kappa_Y r} I_Y^{(0)}(\theta_Y- r) \cdot K_Y^{(0)}(\theta_X -r\omega,\theta_Y-r) \,\mathrm{d} r\\
    &\quad+ b\int_{0}^\infty e^{\kappa_Y r} I_Y^{(0)}(\theta_Y- r) \cdot J_Y^{(0)}(\theta_Y-r) \,\mathrm{d} r.
\end{align*}
Because $p_X^{(1)}$ and $p_Y^{(1)}$ only depend on phase variables, these functions may be pre-computed numerically. The numerical computation can be made especially efficient by rewriting the above as a 1D convolution (see \ref{a:conv}). Note that while $p_Y^{(1)}(\theta_X,\theta_Y)$ is written in terms of fast phase variables, precomputing this quantity does not violate any separation of timescales. The dependence of $p_Y^{(1)}$ on $(\theta_X,\theta_Y)$ is simply a shorthand -- what we have really computed is $p_X^{(1)}(s)\equiv p_X^{(1)}(\hat\theta_X + \omega s,\hat \theta_Y + s)$. That is, we've explicitly solved for $p_X^{(1)}$ in terms of $s$ as desired, given the relatively constant phase variables $\hat\theta_X$ and $\hat\theta_Y$. Because $p_X^{(1)}$ is periodic in each coordinate\footnote{Periodicity of $p_X^{(1)}(s)$ follows from periodicity of its forcing functions, e.g., $I_X^{(0)}(\hat\theta_X+\omega (s + 2\pi -r)) = I_X^{(0)}(\hat\theta_X+\omega (s-r))$.}, we only need to compute $p_X^{(1)}(s)$ for one period in $s$ for each pair of slow variables $(\hat\theta_X,\hat\theta_Y) \in [0,2\pi]\times[0,2\pi]$.

We next solve $p_X^{(2)}$ ($p_Y^{(2)}$) for order $k=2$. The forcing function for $p_X^{(2)}$ ($p_Y^{(2)}$) includes the term $p_X^{(1)}(\theta_X,\theta_Y)$ ($p_Y^{(1)}(\theta_X,\theta_Y)$), or equivalently, $p_X^{(1)}(\hat\theta_X + \omega s,\theta_Y + s)$ ($p_Y^{(1)}(\hat\theta_X + \omega s,\theta_Y + s)$). These functions are integrated in the fast variable $s$ in the same manner as above to solve for $p_X^{(2)}(\theta_X,\theta_Y)$ ($p_Y^{(2)}(\theta_X,\theta_Y)$).

We may continue this process to include as many higher-order terms $p_X^{(\ell)}(\theta_X,\theta_Y)$, $p_Y^{(\ell)}(\theta_X,\theta_Y)$ as needed, thus obtaining explicit terms for the coefficients of the $\psi_i$ $\ve$-expansion $\psi_X = \ve p_X^{(1)} + \ve^2 p_X^{(2)} + \cdots$ ($\psi_Y = \ve p_Y^{(1)} + \ve^2 p_Y^{(2)} + \cdots$), effectively eliminating the $\psi_X$ ($\psi_Y$) equations (without requiring that solutions remain near the unperturbed limit cycle).

\begin{remark}\label{rem:b}
    The heterogeneity parameter $b$ appears in each solution $p_X^{(\ell)}$ (or $p_Y^{(\ell)}$) up to the power $b^\ell$. 
\end{remark}
The intuition behind this remark is straightforward to confirm by taking $J_X = (1,0,\ldots,0)^\tr$, and $J_Y = 0$ so that the heterogeneity parameter $b$ only appears in the first coordinate of oscillator $X$. Then, $\hat K_X^{(0)} = K_X^{(0)} + (b,0,\ldots,0)^\tr$ and $\hat K_X^{(\ell)} = K_X^{(\ell)}$ for $\ell\geq 1$. With this choice of heterogeneity, the solution of $p_X^{(1)}$ will have an additive term scaled by $b$. It then follows that the right-hand side of $p_X^{(2)}$, which contains the term $p_X^{(1)} I_X^{(1)} \cdot (K_X^{(0)} + (b,0,\ldots,0)^\tr)$, will contain a $b^2$ term. It is possible to argue inductively to higher powers. We rely on automated symbolic \yp{algebra package} to handle these calculations for us, but we refer the reader to the derivation in \cite{park2024body} for additional details in the case of $1{:}1$ phase-locking with $N\geq 2$ oscillators.


Now that we have solved for each $p_i^{(\ell)}$, we turn to writing down an explicit phase difference equation.

\paragraph{Step 4: Plug in the $\ve$-expansions of $\psi_i$ into the phase equations, thus effectively eliminating the isostable variable} 

To provide a bird's-eye view, we perform the average,
\begin{align*}
    \frac{1}{\omega}{\hat \theta}_X' &= \frac{1}{2\pi m} \int_0^{2\pi m} \mathcal{Z}_X(\hat \theta_X+ \omega s,\hat \theta_Y+s) \cdot \hat G_{X}(\hat \theta_X+ \omega s,\hat \theta_Y+s) \dd s,\\
    {\hat \theta}_Y' &= \frac{1}{2\pi m} \int_0^{2\pi m} \mathcal{Z}_Y(\hat \theta_X+ \omega s,\hat \theta_Y+s) \cdot \hat G_{Y}(\hat \theta_X+\omega s,\hat \theta_Y+s)\dd s,
\end{align*}
where we momentarily abuse notation and let $\hat \theta_i$ denote the averaged version of the variables $\hat\theta_i$ and do not explicitly include the $\psi_i$ or $p_i^{(\ell)}$ terms because they are included implicitly. By applying the transformation $s \rightarrow \theta_Y + s$, we arrive at a more convenient form,
\begin{align*}
    \frac{1}{\omega}{\hat \theta}_X' &= \frac{1}{2\pi m} \int_0^{2\pi m} \mathcal{Z}_X(\hat \theta_X - \omega \hat \theta_Y + \omega s,s) \cdot \hat G_{X}(\hat \theta_X- \omega \hat \theta_Y+ \omega s,s) \dd s,\\
    {\hat \theta}_Y' &= \frac{1}{2\pi m} \int_0^{2\pi m} \mathcal{Z}_Y(\hat \theta_X- \omega \hat \theta_Y+ \omega s,s) \cdot \hat G_{Y}(\hat \theta_X- \omega \hat \theta_Y+\omega s,s)\dd s.
\end{align*}
These equations are useful because the integrals can be expressed as functions of the phase difference $\phi=\hat \theta_X - \omega \hat \theta_Y$, which will further reduce the dimension of the system into a scalar equation.  For the following calculations, we resume working explicitly with $bJ_X(X) + G_X(X,Y)$ ($bJ_Y(Y) + G_Y(X,Y)$) as opposed to $\hat G_X(X,Y)$ ($\hat G_Y(X,Y)$). 

Substituting the expansion for $\psi_X$ and collecting in powers of $\ve$ yields
\begin{equation}\label{eq:th_expansion_X}
    \begin{split}
        \frac{1}{\omega}\hat\theta_X' &=  \ve Z_X^{(0)}\cdot K_{X}^{(0)}  + \ve^2 \left( Z_X^{(0)}\cdot K_X^{(1)} + p_X^{(1)} Z_X^{(1)}\cdot K_X^{(0)}\right) \\
        &\quad + \ve^3 \left( Z_X^{(0)}\cdot K_X^{(2)} + p_X^{(1)} Z_X^{(1)}\cdot K_X^{(1)}  +  p_X^{(2)}Z_X^{(1)} \cdot K_X^{(0)}   + \left( p_X^{(1)}\right)^2 Z_X^{(2)} \cdot K_X^{(0)} \right)\\
        &\quad + \cdots\\
        &\quad+ \ve b Z_X^{(0)}\cdot J_X^{(0)}  + \ve^2 b \left(Z_X^{(1)}J_X^{(0)} + Z_X^{(0)}J_X^{(1)}\right)\\
        &\quad+\ve^3 b\left(Z_X^{(2)}J_X^{(0)} +  Z_X^{(1)}J_X^{(1)} +Z_X^{(0)}J_X^{(2)}\right)\\
        &\quad + \cdots,
    \end{split}
\end{equation}
Similarly,
\begin{equation}\label{eq:th_expansion_Y}
    \begin{split}
        {\hat\theta}_Y' &=  \ve Z_Y^{(0)}\cdot K_Y^{(0)} + \ve^2 \left( Z_Y^{(0)}\cdot K_Y^{(1)} + p_Y^{(1)} Z_Y^{(1)}\cdot K_Y^{(0)}\right) \\
        &\quad +\ve^3 \left( Z_Y^{(0)}\cdot K_Y^{(2)} + p_Y^{(1)} Z_Y^{(1)}\cdot K_Y^{(1)}  +  p_Y^{(2)}Z_Y^{(1)} \cdot K_Y^{(0)}   + \left( p_Y^{(1)}\right)^2 Z_Y^{(2)} \cdot K_Y^{(0)} \right)\\
        &\quad + \cdots\\
        &\quad + \ve b Z_Y^{(0)}\cdot J_Y^{(0)}  + \ve^2 b \left(Z_Y^{(1)}J_Y^{(0)} + Z_Y^{(0)}J_Y^{(1)}\right)\\
        &\quad +\ve^3 b\left(Z_Y^{(2)}J_Y^{(0)} +  Z_Y^{(1)}J_Y^{(1)} +Z_Y^{(0)}J_Y^{(2)}\right)\\
        &\quad + \cdots.
    \end{split}
\end{equation}
Note that the heterogeneity parameter $b$ is implicitly taken to higher orders as argued in Remark \ref{rem:b}. 

We average each $O(\ve^\ell)$ term in \eqref{eq:th_expansion_X} and \eqref{eq:th_expansion_Y}, yielding a system of autonomous phase equations
\begin{align}
    \frac{1}{\omega}\theta_X' &= b \sum_{\ell=1}^M \ve^{\ell}\mathcal{J}_X^{(\ell)} + \sum_{\ell=1}^M \ve^{\ell} \left[\mathcal{H}_X^{(\ell)}(\theta_X - \omega \theta_Y)\right] , \tag{\ref{eq:th_avg_X}}\\
    \theta_Y' &= b \sum_{\ell=1}^M \ve^{\ell}\mathcal{J}_Y^{(\ell)} + \sum_{\ell=1}^M \ve^{\ell} \left[ \mathcal{H}_Y^{(\ell)}(\theta_X - \omega \theta_Y)\right], \tag{\ref{eq:th_avg_Y}}
\end{align}
where
\begin{align*}
    \mathcal{J}_i^{(1)} &= \frac{1}{2\pi}\int_0^{2\pi} Z_i^{(0)}(s)\cdot J_i^{(0)}(s) \dd s,\\
    \mathcal{J}_i^{(2)} &= \frac{1}{2\pi}\int_0^{2\pi} Z_i^{(1)}(s)\cdot J_i^{(0)}(s) + Z_i^{(0)}(s)\cdot J_i^{(1)}(s) \dd s,\\
    \mathcal{J}_i^{(3)} &= \frac{1}{2\pi}\int_0^{2\pi} Z_i^{(2)}(s)\cdot J_i^{(0)}(s) +  Z_i^{(1)}(s)\cdot J_i^{(1)}(s) +Z_i^{(0)}(s)\cdot J_i^{(2)}(s) \dd s,\\
    &\quad \vdots
\end{align*}
and
\begin{align*}
    \mathcal{H}_X^{(1)}(\xi) &= \frac{1}{2\pi m}\int_0^{2\pi m}Z_X^{(0)}(\xi+\omega s)\cdot K_X^{(0)}(\xi+\omega s, s) \dd s,\\
    \mathcal{H}_X^{(2)}(\xi) &= \frac{1}{2\pi m}\int_0^{2\pi m} Z_X^{(0)}(\xi + \omega s)\cdot K_X^{(1)}(\xi + \omega s, s)\\
    &\quad\quad\quad\quad\quad\quad +p_X^{(1)}(\xi + \omega s, s) Z_X^{(1)}(\xi+\omega s)\cdot K_X^{(0)}(\xi+\omega s, s) \dd s,\\
    \mathcal{H}_X^{(3)}(\xi) &= \frac{1}{2\pi m}\int_0^{2\pi m} Z_X^{(0)}(\xi+\omega s)\cdot K_X^{(2)}(\xi+\omega s, s)\\
    &\quad\quad\quad\quad\quad\quad+ p_X^{(1)}(\xi+\omega s, s) Z_X^{(1)}(\xi+\omega s)\cdot K_X^{(1)}(\xi+\omega s, s)\\
    &\quad\quad\quad\quad\quad\quad+  p_X^{(2)}(\xi+\omega s, s)Z_X^{(1)}(\xi+\omega s) \cdot K_X^{(0)}(\xi+\omega s, s)\\
    &\quad\quad\quad\quad\quad\quad+ p_X^{(1)}(\xi+\omega s,s)^2 Z_X^{(2)}(\xi+\omega s) \cdot K_X^{(0)}(\xi+\omega s, s) \dd s,\\
    &\vdots.
\end{align*}
Again, the heterogeneity parameter $b$ is implicit.

Similarly,
\begin{align*}
    \mathcal{H}_Y^{(1)}(\xi) &= \frac{1}{2\pi m}\int_0^{2\pi m}Z_Y^{(0)}(\xi+\omega s)\cdot K_Y^{(0)}(\xi+\omega s, s) \dd s,\\
    \mathcal{H}_Y^{(2)}(\xi) &= \frac{1}{2\pi m}\int_0^{2\pi m} Z_Y^{(0)}(\xi+\omega s)\cdot K_Y^{(1)}(\xi+\omega s, s)\\
    &\quad\quad\quad\quad\quad\quad +p_Y^{(1)}(\xi+\omega s, s) Z_Y^{(1)}(\xi+\omega s)\cdot K_Y^{(0)}(\xi+\omega s, s) \dd s,\\
    \mathcal{H}_Y^{(3)}(\xi) &= \frac{1}{2\pi m}\int_0^{2\pi m} Z_Y^{(0)}(\xi+\omega s)\cdot K_Y^{(2)}(\xi+\omega s, s)\\
    &\quad\quad\quad\quad\quad\quad+ p_Y^{(1)}(\xi+\omega s, s) Z_Y^{(1)}(\xi+\omega s)\cdot K_Y^{(1)}(\xi+\omega s, s)\\
    &\quad\quad\quad\quad\quad\quad+  p_Y^{(2)}(\xi+\omega s, s)Z_Y^{(1)}(\xi+\omega s) \cdot K_Y^{(0)}(\xi+\omega s, s)\\
    &\quad\quad\quad\quad\quad\quad+  p_Y^{(1)}(\xi+\omega s,s)^2 Z_Y^{(2)}(\xi+\omega s) \cdot K_Y^{(0)}(\xi+\omega s, s) \dd s,\\
    &\vdots
\end{align*}

Defining $\phi = \theta_X - \omega \theta_Y$, the corresponding ordinary differential equation is given by
\begin{align*}
    \phi' &= \theta_X' - \omega \theta_Y'\\
    &= \omega \sum_{\ell=1}^M \ve^{\ell}\left(\mathcal{J}_X^{(\ell)} - \mathcal{J}_Y^{(\ell)}\right)b + \omega \sum_{\ell=1}^M \ve^{\ell} \left(\mathcal{H}_X^{(\ell)}(\phi) - \mathcal{H}_Y^{(\ell)}(\phi)\right).
\end{align*}
Finally, defining $\mathcal{H}_{n,m}^{(\ell)}(\phi) =  \mathcal{H}_X^{(\ell)}(\phi) - \mathcal{H}_Y^{(\ell)}(\phi)$, and $b^{(\ell)} = (\mathcal{J}_X^{(\ell)} - \mathcal{J}_Y^{(\ell)})b$, the phase difference dynamics is given by the scalar equation,
\begin{equation}\tag{\ref{eq:dphi}}
    \frac{1}{\omega}\phi' = \sum_{\ell=1}^M\ve^{\ell} [b^{(\ell)}  + \mathcal{H}_{n,m}^{(\ell)}(\phi)].
\end{equation}

\section{Notes on Numerical Continuation}\label{a:continuation}

\subsection{XPPAUTO Bifurcation Parameters}\label{a:xpp}

We used XPPAUTO to compute the two-parameter diagrams of the forced systems (Figure \ref{fig:nric_tongues} for the forced nonradial isochron clock and Figure \ref{fig:thal_tongues} for the forced thalamic neuron model). These instructions are provided because convergence can be difficult to obtain. Computing bifurcation diagrams of the nonradial isochron clock is more forgiving, so we focus on the forced thalamic oscillator (\texttt{./v2\_xpp/thal1f.ode} in the GitHub repository\footnote{https://doi.org/10.5281/zenodo.13824274}).

\begin{enumerate}[noitemsep]
    \item Set $\ve=0.03$ and $\delta=0.01$ (parameters \texttt{eps} and \texttt{d}, respectively). Set $\omega_X$ and $\omega_f$ (parameters \texttt{omx} and \texttt{omf}, respectively) to the desired natural number values.
    \item Set the AUTO parameters as shown in Table \ref{tab:auto}.
    \item Compute a periodic solution, import it into AUTO and run a one-parameter diagram (follow the instructions in Chapter 7.1.1 of \cite{xpp}). Use the Gears solver with a relative tolerance of \num{1e-5} to compute the periodic solution. Use a time step size of \texttt{dt=0.0001} so that the estimate of the period is accurate to three decimal places.
    \item In the one-parameter diagram (which should be a function of $\ve$), there should be a limit point, i.e., a fold bifurcation of periodic orbits. You may need to change \texttt{ds} to be negative to find it.
    \item Follow the instructions later in Chapter 7.1.1 of \cite{xpp} to compute a two-parameter diagram of the limit point as a function of $\ve$ and $\delta$ (parameters \texttt{eps} and \texttt{d}, respectively).
\end{enumerate}
The last step may require a significant amount of searching, particularly beyond $1{:}1$ phase-locking. Consider some broad guidelines:
\begin{itemize}[noitemsep]
    \item The two-parameter diagram typically follows one boundary of the Arnold tongue, so the above instructions should be repeated for $\delta=-0.01$.
    \item Based on the choice of $\omega_X$ and $\omega_f$, different values of $\delta$ may need to be chosen. The auxiliary \texttt{diff} variable in the ode file can be used to determine if there is phase-locking: a constant envelope typically indicates phase-locking. A low-frequency non-constant envelope of this auxiliary variable typically indicates phase drift.
    \item There are often multiple fold bifurcations of cycles close together in parameter space. If tolerances are not tight enough, AUTO will jump between them. Just keep grabbing points and switching signs of the step parameter \texttt{ds} until the phase-locking boundaries are found.
    \item Reducing EPSL, EPSU, and EPSS from \num{1e-5} to \num{1e-6} yields a more robust continuation of Arnold tongue boundaries of at the cost of increased computation time. If EPSL, EPSU, and EPSS are chosen to be greater than \num{1e-5}, the two-parameter continuation often follows limit points between boundaries, which is undesirable.
\end{itemize}

\begin{table}
    \caption{AUTO parameter values for two-parameter diagrams. Unlisted parameters are kept at default values. Parameter descriptions are from \myurla~and \myurlb.}\label{tab:auto}
    \centering
    \begin{tabular}{c|c|c}
        Parameter & Value & Description\\
        \hline
        EPSL & \num{1e-5} & Convergence criterion\\
        EPSU & \num{1e-5} & Convergence criterion\\
        EPSS & \num{1e-5} & Convergence criterion\\
        NWTN & 25 & Number of Newton iterations\\
        ITMX & 50 & Max iterations to find bifurcations\\
        MXBF & 0 & Number of additional branches to follow\\
        NTST & 50 & Number of mesh intervals for discretization
    \end{tabular}
\end{table}

\subsection{Newton's Method for Phase-Locking}\label{a:newton}

Let $f$ be the function that maps $g^\alpha(0)$ to $g^\alpha(T)$ for a particular $g^\alpha$  we are looking for. Assuming $g^*(0)$ is a close guess for the periodic solution, $f(g^*(0)) = g^*(T)$. We then look for a vector $\Delta g$ that solves for the periodic solution \cite{wilson_newton}, i.e.,
\begin{align*}
    &f(g^*(0) + \Delta g) = g^*(0) + \Delta g,\\
    \Rightarrow &f(g^*(0)) + J \Delta g = g^*(0) + \Delta g,\\
    \Rightarrow &(J-I)\Delta g = g^*(0)  f(g^*(0)),\\
    \Rightarrow &\Delta g = (J-I)^{-1} (g^*(0) - f(g^*(0)),
\end{align*}
where $J$ is the Jacobian matrix of $f$, and $I$ is the identity matrix. The function $f$ is the stroboscopic map after the times $T = nT_X$ or $T = mT_Y$, where the quantities $T$ or $T_i$ are estimated by simulating the coupled system for sufficiently long times until it converges to a phase-locked state. 

The Jacobian matrix J is estimated as follows. Take an initial condition, i.e., an initial guess $Y(0)$ on or near the limit cycle. Without loss of generality, consider two spatially-perturbed initial conditions, $Y_1(0) = Y(0)+[\ve,0,\ldots,0]$ and $Y_2(0) = Y(0)-[\ve,0,\ldots,0]$. We integrate each initial condition forward in time up to time $T$, where we obtain the estimate of the first column of the Jacobian matrix, $(Y_2(T)-Y_1(T))/(2\ve)$. Repeat this process for the remaining columns of $J$.

To include the time estimate $T$ in the Newton iteration, we  augment $J$ with a column given by two solutions $Z_1(T+\ve)$ and $Z_2(T-\ve)$ with initial conditions $Z_1(0) = Z_2(0) = Y(0)$. We then augment $J$ with a new row, with the last row and column set to zero, and the remaining elements of the last row given by the right-hand side of the coupled system evaluated on the initial condition $Y(0)$.

When starting the Newton iteration, the term $g^*(0)$ is given by the initial guess $Y(0)$ and $f(g^*(0))$ is given by the time integration of the coupled system given the initial condition $Y(0)$ up to time $T$. Once $\Delta g$ is computed, it is used to update the initial guess $Y(0)$ (simply, $Y(0)\rightarrow Y(0) + \Delta g$, which is used as the new initial condition for the next iteration.

For all Newton iterations, we use a perturbation size of $\ve=\num{1e-4}$. The condition to stop the iteration is determined by the magnitude of $\|Y(0)\|$: if $\|Y(0)\| \leq \num{1e-5}$, then we say the system has converged to a phase-locked state. If Newton's method exceeds 20 iterations without the magnitude reducing below the desired tolerance, we assume that there is no phase-locked solution.

In the case of forcing, we exclude the time condition because there is no stability information from the forcing function. Instead, we only compute the Jacobian given the spatially-perturbed solutions, and take the integration time $T$ to be the period of the forcing function. We then use the same convergence criteria as the coupling case.

\section{Coupling Term Expansions}\label{a:g}
To expand the coupling functions $G_{i}$ in powers of $\ve$, we use the Floquet eigenfunction approximation ($g_i^{(k)}$) for each oscillator,
\begin{equation}\label{eq:efuns}
    \begin{split}
        \Delta X &\approx \psi_X g_X^{(1)}(\theta_X) + \psi_X^2 g_X^{(2)}(\theta_X) + \cdots,\\
        \Delta Y &\approx \psi_Y g_Y^{(1)}(\theta_Y) + \psi_Y^2 g_Y^{(2)}(\theta_Y) + \cdots,
    \end{split}
\end{equation}
where $\Delta X \equiv X(t)-X_0(\theta_X(t))$ ($\Delta Y \equiv Y(t) - Y_0(\theta_Y(t))$) is the difference between the limit cycle solution $X_0$ ($Y_0$) and an arbitrary trajectory $X(t)$ ($Y(t)$) in the basin of attraction. We focus on $G_X$ without loss of generality because the calculations are identical for $G_Y$, $\hat G_i$, and $J_i$.

We view the coupling function as the map $G_{X}:\mathbb{R}^{n_X + n_Y}\rightarrow \mathbb{R}^{n_X}$, where
\begin{align*}
    G_{X}(\Xi) &= \left[G_{X,1}(\Xi), G_{X,2}(\Xi), \ldots G_{X,n_X}(\Xi)\right]^\tr \in\mathbb{R}^{n_X}, \\
    &G_{X,m}:\mathbb{R}^{n_X+n_Y}\rightarrow \mathbb{R},\quad m=1,2,\ldots,n_X,
\end{align*}
and $\Xi = [X^\tr,Y^\tr]^\tr \in \mathbb{R}^{n_X+n_Y}$ is an $(n_X+n_Y)\times 1 $ column vector. To make the expansion explicit, we write $\Xi = \Lambda + \Delta \Xi$, where
\begin{align*}
    \Lambda &= [X_0(\theta_X)^\tr,Y_0(\theta_Y)^\tr]^\tr\in \mathbb{R}^{n_X+n_Y},\\
    \Delta \Xi &= [\Delta X^\tr,\Delta Y^\tr]^\tr\in \mathbb{R}^{n_X+n_Y},
\end{align*}
so both are $(n_X + n_Y) \times 1$ column vectors. We now use the ``$\vc$'' operator from \cite{magnus2019matrix,wilson2020phase} to obtain the multivariate Taylor expansion of $G_{X,m}$ for each $m=1,2,\ldots,n_X$:
\begin{align}\label{eq:g_raw}
    G_{X,m}(\Lambda + \Delta \Xi)&= G_{X,m}(\Lambda)+ G_{X,m}^{(1)}(\Lambda)\Delta \Xi + \sum_{k=2}^\infty \frac{1}{k!}\left[ \stackrel{k}{\otimes} \Delta \Xi^\tr\right] \vc\left(G_{X,m}^{(k)}(\Lambda)\right),
\end{align}
where $\otimes$ is the Kronecker product, and, temporarily treating $\Lambda$ as a vector of dummy variables,
\begin{equation}\label{eq:g_deriv}
    G_{X,m}^{(k)}(\Lambda) = \frac{\pa \vc\left(G_{X,m}^{(k-1)}(\Lambda)\right)}{\pa \Lambda^\tr} \in \mathbb{R}^{(n_X+n_Y)^{(k-1)}\times (n_X+n_Y)}.
\end{equation}
The $\vc$ operator simply reshapes a matrix by stacking its columns, which allows us to avoid calculating high-dimensional tensors. For example, if an $n\times m$ matrix $A$ has columns $a_i$ for $i=1,\ldots,n$ for $a_i \in\mathbb{R}^m$, then $\vc(A)$ is the $mn\times 1$ column vector $(a_1^\tr,a_2^\tr,\ldots,a_n^\tr)^\tr$. If $A$ is a Jacobian matrix, taking partial derivatives yields a tensor, whereas taking partials of $\vc(A)$ yields a matrix.

\subsection{Concrete Calculations}

For concreteness, we show explicit expansion terms of $G_{X,m}$ up to order $\psi^2$ and follow up with the order $\ve^3$ expansion. The lowest-order term $G_{X,m}^{(0)}(\Lambda)$ is trivial, so we examine the next-order term, $G_{X,m}^{(1)}(\Lambda)\Delta \Xi $. We first compute the first derivative using the $\vc$ notation,
\begin{align*}
    G_{X,m}^{(1)}(\Lambda) &= \frac{\pa \vc\left(G_{X,m}^{(0)}(\Lambda)\right)}{\pa \Lambda^\tr}\\
    &\equiv \left(\pa_{X_1} G_{X,m}^{(0)}(\Lambda), \ldots, \pa_{X_{n_X}} G_{X,m}^{(0)}(\Lambda), \pa_{Y_1} G_{X,m}^{(0)}(\Lambda), \ldots, \pa_{Y_{n_Y}} G_{X,m}^{(0)}\right).
\end{align*}
which is a vector in $\mathbb{R}^{1\times (n_X+n_Y)}$. Note that because $G_{X,m}^{(0)}$ is scalar-valued, the $\vc$ operator does nothing, and the resulting derivative is the gradient of $G_{X,m}^{(0)}$. Recall that 
\begin{equation*}
    \Delta \Xi = \left([\psi_X g_X^{(1)} + \psi_X^2 g_X^{(2)} + \cdots]^\tr, [\psi_Y g_Y^{(1)} + \psi_Y^2 g_Y^{(2)} + \cdots]^\tr \right)^\tr
\end{equation*}
is a column vector. If we denote $g_i^{(\ell)}$ using the vector
\begin{equation*}
    g_i^{(\ell)} = \left(\begin{matrix}
        g_{i,1}^{(\ell)}\\
        g_{i,2}^{(\ell)}\\
        \vdots\\
        g_{i,n_i}^{(\ell)}
    \end{matrix}\right),
\end{equation*}
then $\Delta \Xi$ can be written,
\begin{equation*}
    \Delta \Xi = \left(\begin{matrix}
        \psi_X g_{X,1}^{(1)} + \psi_X^2 g_{X,1}^{(2)}+\cdots\\
        \psi_X g_{X,2}^{(1)} + \psi_X^2 g_{X,2}^{(2)}+\cdots\\
        \vdots\\
        \psi_X g_{X,n_X}^{(1)} + \psi_X^2 g_{X,n_X}^{(2)}+\cdots\\
        \psi_Y g_{Y,1}^{(1)} + \psi_Y^2 g_{Y,1}^{(2)}+\cdots\\
        \psi_Y g_{Y,2}^{(1)} + \psi_Y^2 g_{Y,2}^{(2)}+\cdots\\
        \vdots\\
        \psi_Y g_{Y,n_X}^{(1)} + \psi_Y^2 g_{Y,n_X}^{(2)}+\cdots
    \end{matrix}\right)
\end{equation*}
It then follows that
\begin{align*}
    G_{X,m}^{(1)}(\Lambda)\Delta \Xi &= \pa_{X_1} G_{X,m}^{(0)}(\Lambda) \left[\psi_X g_{X,1}^{(1)} + \psi_X^2 g_{X,1}^{(2)}+\cdots\right] +\cdots\\
    &\quad + \pa_{X_{n_X}} G_{X,m}^{(0)}(\Lambda) \left[\psi_X g_{X,n_X}^{(1)} + \psi_X^2 g_{X,n_X}^{(2)}+\cdots\right]\\
    &\quad + \pa_{Y_1} G_{X,m}^{(0)}(\Lambda)\left[\psi_Y g_{Y,1}^{(1)} + \psi_Y^2 g_{Y,1}^{(2)}+\cdots\right] +\cdots\\
    &\quad + \pa_{Y_{n_Y}} G_{X,m}^{(0)}(\Lambda)\left[\psi_Y g_{Y,n_X}^{(1)} + \psi_Y^2 g_{Y,n_X}^{(2)}+\cdots\right].
\end{align*}
To finish collecting all order $\psi^2$ terms, we examine the next term,
\begin{equation*}
    \frac{1}{2} \left[\stackrel{2}{\otimes} \Delta \Xi^\tr\right] \vc\left(G_{X,m}^{(2)}(\Lambda) \right).
\end{equation*}
The derivative $G_{X,m}^{(2)}(\Lambda)$, given by,
\begin{equation*}
    \frac{\pa \vc\left(G_{X,m}^{(1)}(\Lambda)\right)}{\pa \Lambda^\tr},
\end{equation*}
transposes the first derivative $G_{X_m}^{(1)}$ into a column vector (because it stacks the columns of the row vector) before taking the gradient of each element. We thus end up with the following matrix for $G_{X,m}^{(2)}(\Lambda)$:
\begin{equation*}
    \left(\begin{matrix}
        \pa_{X_1} \pa_{X_1}  G_{X,m}^{(1)},  \cdots, \pa_{X_n} \pa_{X_1}  G_{X,m}^{(1)}, \pa_{Y_1} \pa_{X_1}  G_{X,m}^{(1)}, \cdots,   \pa_{Y_{n_Y}} \pa_{X_1}  G_{X,m}^{(1)}\\
          \vdots  \\
        \pa_{X_{1}} \pa_{X_{n_X}} G_{X,m}^{(1)}(\Lambda), \cdots , \pa_{X_{n_X}} \pa_{X_{n_X}} G_{X,m}^{(1)},  \pa_{Y_{1}} \pa_{X_{n_X}} G_{X,m}^{(1)}, \cdots , \pa_{Y_{n_Y}} \pa_{X_{n_X}} G_{X,m}^{(1)}\\
        \pa_{X_{1}} \pa_{Y_1} G_{X,m}^{(1)}(\Lambda), \cdots , \pa_{X_{n_X}} \pa_{Y_1} G_{X,m}^{(1)},  \pa_{Y_{1}} \pa_{Y_1} G_{X,m}^{(1)}, \cdots , \pa_{Y_{n_Y}} \pa_{Y_{1}} G_{X,m}^{(1)}\\
        \vdots\\
        \pa_{X_{1}} \pa_{Y_{n_Y}} G_{X,m}^{(1)}(\Lambda), \cdots , \pa_{X_{n_X}} \pa_{Y_{n_Y}} G_{X,m}^{(1)},  \pa_{Y_{1}} \pa_{Y_{n_Y}} G_{X,m}^{(1)}, \cdots , \pa_{Y_{n_Y}} \pa_{Y_{n_Y}} G_{X,m}^{(1)},
    \end{matrix}\right)
\end{equation*}
which is $(n_X+n_Y)\times (n_X+n_Y)$. We then apply the $\vc$ operator to this matrix to transform it into a large column vector with $(n_X+n_Y)^2$ elements before taking its dot product with $\stackrel{2}{\otimes} \Delta \Xi^\tr$, which is given by the large row vector,
\begin{equation*}
    \stackrel{2}{\otimes} \Delta \Xi^\tr = \left(\begin{matrix}
        [\psi_X g_{X,1}^{(1)} +\cdots ]\Delta\Xi\\
        [\psi_X g_{X,2}^{(1)} +\cdots ]\Delta\Xi\\
        \vdots\\
        [\psi_X g_{X,n_X}^{(1)} +\cdots ]\Delta\Xi\\
        [\psi_Y g_{Y,1}^{(1)} +\cdots ]\Delta\Xi\\
        [\psi_Y g_{Y,2}^{(1)} +\cdots ]\Delta\Xi\\
        \vdots\\
        [\psi_Y g_{Y,n_Y}^{(1)} +\cdots ]\Delta\Xi
    \end{matrix}\right)^\tr,
\end{equation*}
which rightly has $(n_X+n_Y)^2$ elements. Note that each $\Delta \Xi$ vector is of order $O(\psi_i)$, therefore $\stackrel{2}{\otimes} \Delta \Xi^\tr$ is of order $O(\psi_i^2)$.

Due to the large number of terms, we rely on a symbolic \yp{algebra package} to handle the collection of terms. In the main text, we simply abbreviate plugging in the $\ve$-expansion of $\psi_i$ ($\psi_i = \ve p_i^{(1)} + \ve^2 p_i^{(2)} + \cdots$) by writing
\begin{equation*}
    \begin{split}
        G_{X}(\theta_X,\psi_X,\theta_Y,\psi_Y)= &\, K_{X}^{(0)}(\theta_X,\theta_Y)+ \ve K_{X}^{(1)}\left(\theta_X,\theta_Y,p_X^{(1)},p_Y^{(1)}\right)\\
        &+ \ve^2 K_{X}^{(2)}\left(\theta_X,\theta_Y,p_X^{(1)},p_X^{(2)},p_Y^{(1)},p_Y^{(2)}\right)\\
        &+ \ve^3 K_{X}^{(3)}\left(\theta_X,\theta_Y,p_X^{(1)},p_X^{(2)},p_X^{(3)},p_Y^{(1)},p_Y^{(2)},p_Y^{(3)}\right)\\
        &+ \cdots.
    \end{split}
\end{equation*}
Each $K_{X}^{(\ell)}$ function contains the Floquet eigenfunctions and the partials of $G_{X}$. It is straightforward to verify using a symbolic \yp{algebra package} that the function $K_{X}^{(\ell)}$ only depends on terms $p_X^{(\ell)}$, $p_Y^{(\ell)}$ for $\ell \leq k$.

\section{Phase Estimation}\label{a:phase_estimate}

    We briefly describe the \yp{two} phase estimation methods used in this paper. \yp{The first method} is similar to the estimation done in \cite{park2016weakly,park2024body} \yp{and is used in the plots of phase difference trajectories of the full model. The second method uses peak-to-peak time differences to estimate the phase, and is used in the bifurcation diagrams of the full model. Two different estimates are used because full model trajectories include transients where oscillator periods may not be well-defined. One the other hand, bifurcation diagrams of the full model do not involve transients. Thus a well-defined period exists and peak-to-peak time differences can provide accurate phase estimates.}
    
    \paragraph{Phase estimation of full model trajectories} Consider a model with state variables $x_1,\ldots,x_n$ and suppose that we have saved a $T$-periodic limit cycle trajectory (at $\ve=0$) to some array $[y_1,\ldots,y_n]$. Then, for a given simulation, we can define the phase to be a point $\theta\in[0,T)$ that minimizes
    \begin{equation*}
    	\text{dist}(x_1(t)-y_1(\theta),\ldots, x_n(t)-y_n(\theta)),
    \end{equation*}
    where
    \begin{equation*}
    	\text{dist}(\Delta x_1,\ldots, \Delta x_n) := \sqrt{(\Delta x_1)^2 + \cdots + (\Delta x_n)^2}.
    \end{equation*}
    By simulating nondimensionalized versions of the equations, we need not normalize by the variance as in \cite{park2016weakly}. This method is efficient with vectorization.

    \yp{\paragraph{Phase estimation of full model phase differences for bifurcation diagrams}
    Phase differences are estimated using peak-to-peak time differences in the full model. For simplicity, only the tallest peak in oscillator $X$ in a post-transient time interval is used to take time differences with the remaining $m$ peaks of oscillator $Y$. The period is estimated by adding $n$ consecutive peak-to-peak time differences of a single oscillator $X$ (if a phase-locked state exists, this estimated period will be the same as adding $m$ consecutive peak-to-peak time differences of oscillator $Y$).

    Note that this method uses only one peak of oscillator $X$ to estimate differences, even if $n>1$.}

\section{Convolution}\label{a:conv}

Recall the $p_X^{(1)}$ term,
\begin{equation*}
    p_X^{(1)}(\theta_X,\theta_Y) = \omega \int_{0}^\infty e^{\omega \kappa_X r} I_X^{(0)}(\theta_X-\omega r) \cdot \hat K_X^{(0)}(\theta_X -r\omega,\theta_Y-r) \,\mathrm{d} r.
\end{equation*}
The numerical computation of this term can be performed efficiently by utilizing a 1D convolution. The transformation is performed by fixing some $c\in[0,2\pi)$ such that $\theta_X = c+\theta_Y\omega$. Let $H$ be the Heaviside function. Then,
\begin{align*}
    p_X^{(1)}(c+\theta_Y\omega,\theta_Y) &= \omega \int_0^\infty e^{\omega \kappa_X r} I_X^{(0)}(c+\omega( \theta_Y- r))\cdot \hat K_X^{(0)}(c+\omega (\theta_Y-r), \theta_Y-r)\, \mathrm{d}r\\
    &= \omega \int_{-\infty}^\infty H(r) e^{\omega \kappa_X r} I_X^{(0)}(c+\omega( \theta_Y- r))\cdot \hat K_X^{(0)}(c+\omega (\theta_Y-r), \theta_Y-r)\, \mathrm{d}r\\
    &= \omega  [H(\theta_Y) e^{\omega \kappa_X \theta_Y}] * f_{X,c}(\theta_Y),
\end{align*}
where $f_{X,c}(\theta_Y) = I_X^{(0)}(c+\omega \theta_Y)\cdot \hat K_X^{(0)}(c+\omega \theta_Y, \theta_Y)$, and $*$ is the convolution operator. Similarly,
\begin{align*}
    p_Y^{(1)}(c+\theta_Y\omega,\theta_Y) = [H(\theta_Y) e^{\kappa_Y \theta_Y}] * f_{Y,c}(\theta_Y),
\end{align*}
where  $f_{Y,c}(\theta_Y) = I_Y^{(0)}(c+\omega \theta_Y)\cdot \hat K_Y^{(0)}(c+\omega \theta_Y, \theta_Y)$. Thus, obtaining $p_X^{(1)}(\theta_X,\theta_Y)$ is a matter of transforming the first coordinate.

\section{Fourier Series}\label{a:fourier}

For all functions in this section, assume $x\in[0,2\pi n)$.

\subsection{Nonradial Isochron Clock}
The forced nonradial isochron clock in Section \ref{sec:nric} has the following $\mathcal{H}$ functions. 
\begin{align*}
    \mathcal{H}_{1,1}^{(1)}(x) &= -0.3296\cos(x)+0.2196\sin(x),\\
    \mathcal{H}_{1,1}^{(2)}(x)&= 0.0272+7.2584\cos(x)+3.172\sin(x)+0.4926\sin(2x),\\
    \mathcal{H}_{2,1}^{(1)}(x)&=-0.1556 \cos(2 x) + 0.1038 \sin(2 x),\\
    \mathcal{H}_{2,1}^{(2)}(x)&=0.1052 + 0.5916 \cos(2 x) + 3.6946 \sin(2 x) + 0.1136 \sin(4 x),\\
    \mathcal{H}_{3,1}^{(1)}(x)&=-0.0446 \cos(3 x) + 0.0298 \sin(3 x),\\
    \mathcal{H}_{3,1}^{(2)}(x)&=0.0788-0.7992 \cos(3 x)+0.7146\sin(3x)+0.0094\sin(6x),\\
    \mathcal{H}_{4,1}^{(1)}(x)&=-0.0078\cos(4x)+0.0052\sin(4x),\\
    \mathcal{H}_{4,1}^{(2)}(x)&=0.0544 - 0.1796 \cos(4 x) - 0.0498 \sin(4 x).
\end{align*}

\subsection{Forced Thalamic Oscillator}

\begin{align*}
\mathcal{H}_{1,1}^{(1)}(x) &= 1.2816 \sin(x) + 0.0746 \sin(2 x) - 0.014 \sin(3 x) - 0.0018 \sin(4 x)\\
&\quad + 1.1556 \cos(1 x) + 0.1188 \cos(2 x) + 0.014 \cos(3 x) + 0.001 \cos(4 x)\\
\mathcal{H}_{1,1}^{(2)}(x) &= 
9.6616 \cos(1 x) + 1.8382 \cos(2 x) + 0.466 \cos(3 x) + 0.0608 \cos(4 x) \\
&\quad - 0.0046 \cos(5 x) - 0.0004 \cos(6 x) + 2.2889 -32.8754 \sin(x) \\
&\quad - 3.9426 \sin(2 x) - 0.8276 \sin(3 x) - 0.0028 \sin(4 x) - 0.0024 \sin(5 x)\\
&\quad - 0.0022 \sin(6 x) - 0.0004 \sin(7 x),\\
\mathcal{H}_{2,1}^{(1)}(x)&= 0.6054 \sin(2 x) + 0.0038 \sin(4 x) + 0.5458 \cos(2 x) + 0.006 \cos(4 x),\\
\mathcal{H}_{2,1}^{(2)}(x)&= -4.8458 \sin(2 x) - 0.3894 \sin(4 x) - 0.011 \sin(6 x) + 15.5844 \cos(2 x)\\
&\quad - 0.0916 \cos(4 x) + 0.005 \cos(6 x) + 0.533,\\
\mathcal{H}_{1,2}^{(1)}(x)&= 0.1578 \sin(2 x) - 0.038 \sin(4 x) - 0.0024 \sin(6 x) + 0.2514 \cos(2 x) \\
&\quad + 0.0194 \cos(4 x) - 0.0012 \cos(6 x) - 0.0002 \cos(8 x),\\
\mathcal{H}_{1,2}^{(2)}(x)&= -0.1508 \cos(2 x) - 0.2636 \cos(4 x) - 0.0728 \cos(6 x) + 0.0468 \\
&\quad -5.93 \sin(2 x) - 0.8184 \sin(4 x) - 0.1026 \sin(6 x)\\
&\quad - 0.01 \sin(8 x) - 0.0006 \sin(10 x) - 0.0062 \cos(8 x) - 0.001 \cos(10 x),\\
\mathcal{H}_{2,3}^{(1)}(x)&= -0.0486 \sin(6 x) - 0.0004 \sin(12 x) + 0.0488 \cos(6 x) - 0.0002 \cos(12 x),\\
\mathcal{H}_{2,3}^{(2)}(x)&=-1.2122 \sin(6 x) - 0.0224 \sin(12 x) - 0.0004 \sin(18 x) - 0.4998 \cos(6 x) \\
&\quad + 0.0154 \cos(12 x) - 0.0766.
\end{align*}

\subsection{Coupled Thalamic Oscillators}
In this section, denote the heterogeneous term of $O(\ve)$ by $\mathcal{J}_i^{(1)}$ (denoted $b\mathcal{J}_i$ in Figures \ref{fig:a:11} through \ref{fig:a:43}), and the heterogeneous term of $O(\ve^2)$ by $\mathcal{J}_i^{(2,\ell)}$ for $\ell=1,2$ (denoted $b^2\mathcal{J}_i$ in Figures \ref{fig:a:11} through \ref{fig:a:43}). 

\subsubsection{1:1}
\begin{figure}[ht!]
    \centering
    \includegraphics[width=\textwidth]{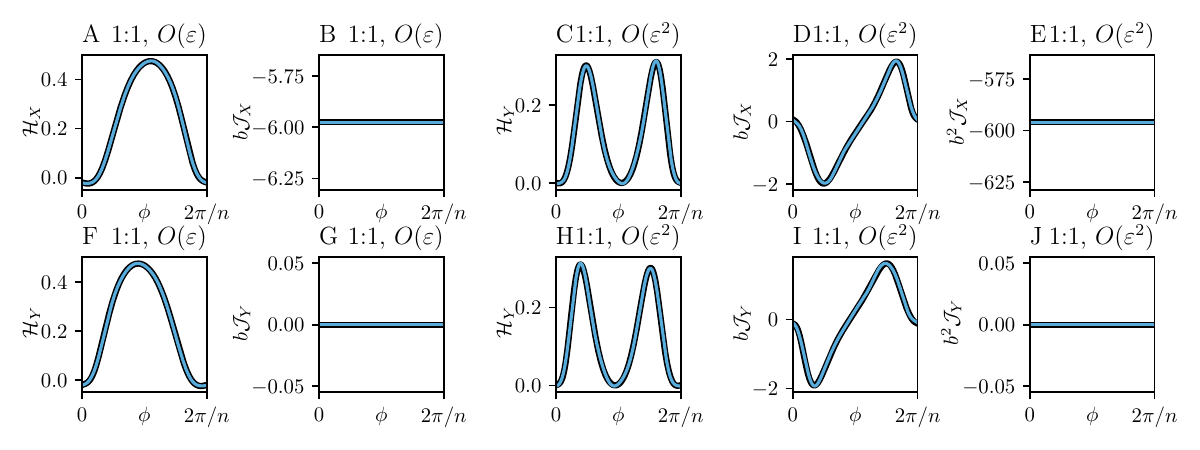}
    \caption{Original $\mathcal{H}$- and $\mathcal{J}$-functions (black) vs the corresponding Fourier approximation (blue) for $1{:}1$ phase-locking for the pair of coupled thalamic models.}
    \label{fig:a:11}
\end{figure}

Plots of the following $\mathcal{H}$ and $\mathcal{J}$ functions are shown in Figure \ref{fig:a:11}.

\begin{align*}
\mathcal{H}_X^{(1)} &= 0.2342-0.2549\cos(x) -0.0167\cos(2x) +0.0092\cos(3x)\\
&\quad +0.0050\cos(4x) +0.0020\cos(5x) -0.0646\sin(x) +0.0095\sin(3x) \\
&\quad +0.0038\sin(4x) \\
\mathcal{J}_X^{(1)}&= -5.9757\\
\mathcal{H}_X^{(2)}&= 0.1225+0.0198\cos(x) -0.1412\cos(2x) -0.0335\cos(3x) \\
&\quad +0.0161\cos(4x) +0.0114\cos(5x) +0.0035\cos(6x) +0.0044\sin(x) \\
&\quad -0.0393\sin(2x) -0.0098\sin(3x) +0.0130\sin(4x) +0.0089\sin(5x) \\
\mathcal{J}_X^{(2,1)}&= -0.1329+0.3773\cos(x) +0.0519\cos(2x) -0.1553\cos(3x) \\
&\quad -0.0872\cos(4x) -1.5871\sin(x) -0.2154\sin(2x) +0.1683\sin(3x) \\
&\quad +0.1167\sin(4x) +0.0569\sin(5x) +0.0276\sin(6x) \\
\mathcal{J}_X^{(2,2)} &=-595.9792\\
\mathcal{H}_Y^{(1)}&= 0.2342-0.2549\cos(x) -0.0167\cos(2x) +0.0092\cos(3x) \\
&\quad +0.0050\cos(4x) +0.0020\cos(5x) +0.0646\sin(x) -0.0095\sin(3x) \\
&\quad -0.0038\sin(4x) \\
\mathcal{J}_Y^{(1)}&=0\\
\mathcal{H}_Y^{(2)} &=0.1225+0.0198\cos(x) -0.1412\cos(2x) -0.0335\cos(3x) \\
&\quad +0.0161\cos(4x) +0.0114\cos(5x) +0.0035\cos(6x) -0.0044\sin(x) \\
&\quad +0.0393\sin(2x) +0.0098\sin(3x) -0.0130\sin(4x) -0.0089\sin(5x) \\
\mathcal{J}_Y^{(2,1)} &=-0.3363\cos(x) +0.1566\cos(3x) +0.0843\cos(4x) \\
&\quad -1.4361\sin(x) -0.1853\sin(2x) +0.1536\sin(3x) +0.1112\sin(4x) \\
&\quad +0.0559\sin(5x) +0.0274\sin(6x) \\
\mathcal{J}_Y^{(2,2)} &=0
\end{align*}

\subsubsection{1:2}
\begin{figure}[ht!]
    \centering
    \includegraphics[width=\textwidth]{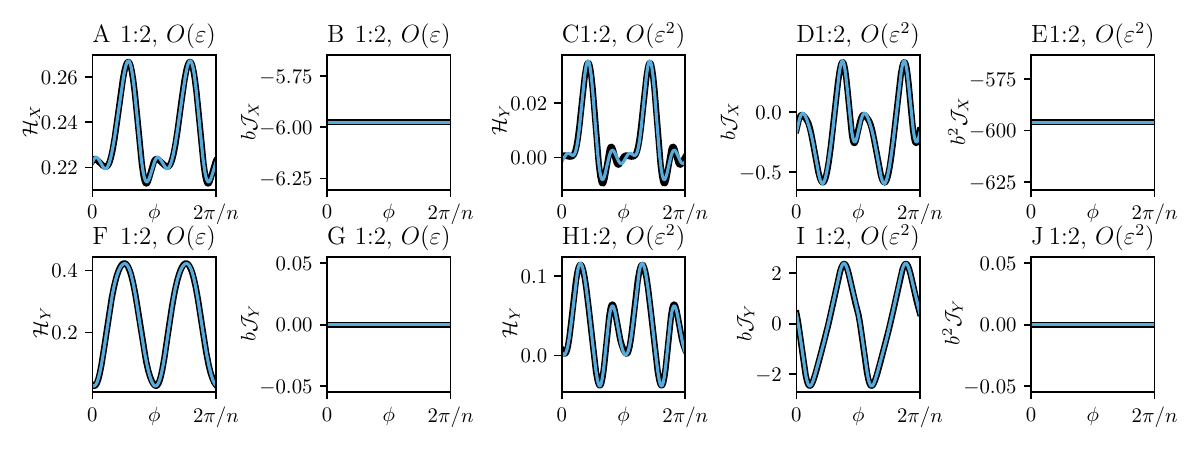}
    \caption{Original $\mathcal{H}$- and $\mathcal{J}$-functions (black) vs the corresponding Fourier approximation (blue) for $1{:}2$ phase-locking for the pair of coupled thalamic models.}
    \label{fig:a:12}
\end{figure}

Plots of the following $\mathcal{H}$ and $\mathcal{J}$ functions are shown in Figure \ref{fig:a:12}.

\begin{align*}
\mathcal{H}_X^{(1)} &= 0.2342-0.0208\cos(2x) +0.0067\cos(4x) +0.0021\cos(6x) \\
&\quad -0.0070\sin(2x) +0.0093\sin(4x) \\
\mathcal{J}_X^{(1)}&=-5.9757\\
\mathcal{H}_X^{(2)}&= 0.0068-0.0105\cos(2x) +0.0049\cos(4x) -0.0026\cos(6x)\\
&\quad +0.0081\sin(2x) -0.0091\sin(4x) +0.0051\sin(6x) \\
\mathcal{J}_X^{(2,1)}&= -0.1329+0.1519\cos(2x) -0.2134\cos(4x) +0.0251\cos(8x)\\
&\quad -0.2476\sin(2x) +0.1569\sin(4x) +0.0697\sin(6x) +0.0246\sin(8x) \\
\mathcal{J}_X^{(2,2)} &=-595.9792\\
\mathcal{H}_Y^{(1)}&= 0.2342-0.2001\cos(2x) -0.0088\cos(4x) +0.0045\cos(6x) \\
&\quad -0.0132\sin(2x) -0.0026\sin(4x) -0.0031\sin(6x) \\
\mathcal{J}_Y^{(1)} &=0\\
\mathcal{H}_Y^{(2)} &=0.0368-0.0303\cos(4x) -0.0059\cos(6x) +0.0035\cos(8x) \\
&\quad +0.0358\sin(2x) -0.0400\sin(4x) +0.0022\sin(6x) +0.0024\sin(8x) \\
\mathcal{J}_Y^{(2,1)} &=0.1786\cos(2x) +0.0576\cos(4x) +0.1024\cos(6x) \\
&\quad +0.0450\cos(8x) -2.2168\sin(2x) -0.1944\sin(4x) +0.1486\sin(6x) \\
&\quad +0.0670\sin(8x) \\
\mathcal{J}_Y^{(2,2)} &= 0
\end{align*}

\subsubsection{2:1}
\begin{figure}[ht!]
    \centering
    \includegraphics[width=\textwidth]{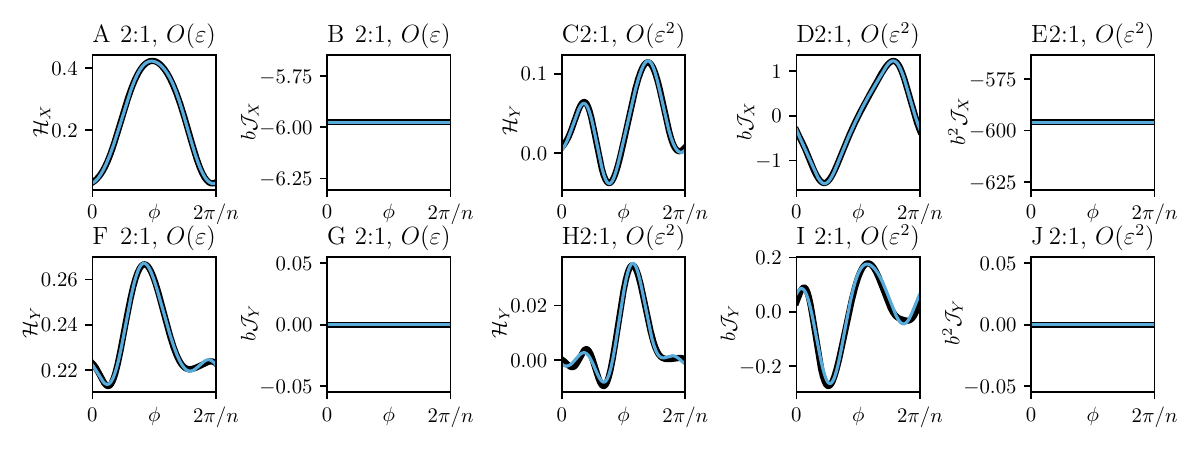}
    \caption{Original $\mathcal{H}$- and $\mathcal{J}$-functions (black) vs the corresponding Fourier approximation (blue) for $2{:}1$ phase-locking for the pair of coupled thalamic models.}
    \label{fig:a:21}
\end{figure}

Plots of the following $\mathcal{H}$ and $\mathcal{J}$ functions are shown in Figure \ref{fig:a:21}.

\begin{align*}
\mathcal{H}_X^{(1)} &=0.2342-0.2001\cos(2x) -0.0088\cos(4x) +0.0045\cos(6x) \\
&\quad +0.0132\sin(2x) +0.0026\sin(4x) +0.0031\sin(6x)\\ 
\mathcal{J}_X^{(1)} &=-5.9757\\
\mathcal{H}_X^{(2)}&= 0.0368-0.0303\cos(4x) -0.0059\cos(6x) +0.0035\cos(8x) \\
&\quad -0.0358\sin(2x) +0.0400\sin(4x) -0.0022\sin(6x) -0.0024\sin(8x) \\
\mathcal{J}_X^{(2,1)}&= -0.1346-0.1011\cos(2x) -0.0490\cos(6x) -0.0230\cos(8x) \\
&\quad -1.2527\sin(2x) -0.1229\sin(4x) +0.0814\sin(6x) +0.0356\sin(8x) \\
\mathcal{J}_X^{(2,2)}&= -595.9595\\
\mathcal{H}_Y^{(1)}&= 0.2342-0.0208\cos(2x) +0.0067\cos(4x) +0.0021\cos(6x) \\
&\quad +0.0070\sin(2x) -0.0093\sin(4x) \\
\mathcal{J}_Y^{(1)}&= 0\\
\mathcal{H}_Y^{(2)} &=0.0068-0.0105\cos(2x) +0.0049\cos(4x) -0.0026\cos(6x) \\
&\quad -0.0081\sin(2x) +0.0091\sin(4x) -0.0051\sin(6x) \\
\mathcal{J}_Y^{(2,1)} &=-0.0373\cos(2x) +0.1034\cos(4x) -0.1184\sin(2x) \\
&\quad +0.0769\sin(4x) +0.0346\sin(6x) \\
\mathcal{J}_Y^{(2,2)} &=0
\end{align*}

\subsubsection{1:3}
\begin{figure}[ht!]
    \centering
    \includegraphics[width=\textwidth]{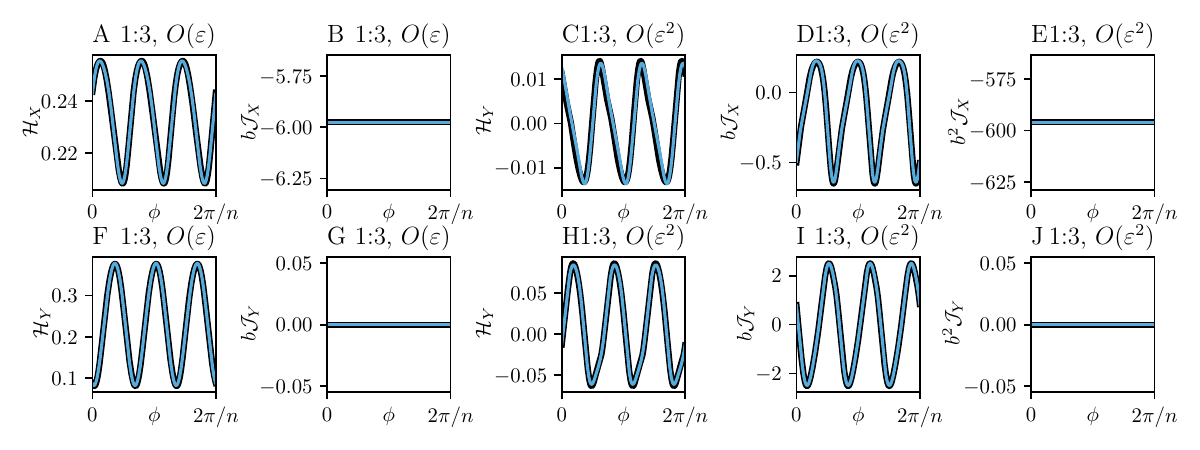}
    \caption{Original $\mathcal{H}$- and $\mathcal{J}$-functions (black) vs the corresponding Fourier approximation (blue) for $1{:}3$ phase-locking for the pair of coupled thalamic models.}
    \label{fig:a:13}
\end{figure}

Plots of the following $\mathcal{H}$ and $\mathcal{J}$ functions are shown in Figure \ref{fig:a:13}.

\begin{align*}
\mathcal{H}_X^{(1)}&= 0.2342+0.0062\cos(3x) +0.0028\cos(6x) +0.0226\sin(3x) \\
\mathcal{J}_X^{(1)}&= -5.9757\\
\mathcal{H}_X^{(2)}&= 0.0126\cos(3x) -0.0027\sin(6x) \\
\mathcal{J}_X^{(2,1)}&= -0.1329-0.3844\cos(3x) +0.0258\cos(9x) +0.1298\sin(3x) \\
&\quad +0.0938\sin(6x) +0.0245\sin(9x) \\
\mathcal{J}_X^{(2,2)} &=-595.9792\\
\mathcal{H}_Y^{(1)}&= 0.2342-0.1437\cos(3x) -0.0047\cos(6x) -0.0370\sin(3x) \\
\mathcal{J}_Y^{(1)}&=0\\
\mathcal{H}_Y^{(2)} &=0.0093-0.0147\cos(3x) -0.0020\cos(6x) -0.0045\cos(9x) \\
&\quad +0.0715\sin(3x) -0.0067\sin(6x) \\
\mathcal{J}_Y^{(2,1)} &=0.6305\cos(3x) +0.0618\cos(6x) +0.0634\cos(9x) \\
&\quad +0.0250\cos(12x) -2.3754\sin(3x) -0.1567\sin(6x) +0.0736\sin(9x) \\
\mathcal{J}_Y^{(2,2)} &=0
\end{align*}

\subsubsection{3:1}
\begin{figure}[ht!]
    \centering
    \includegraphics[width=\textwidth]{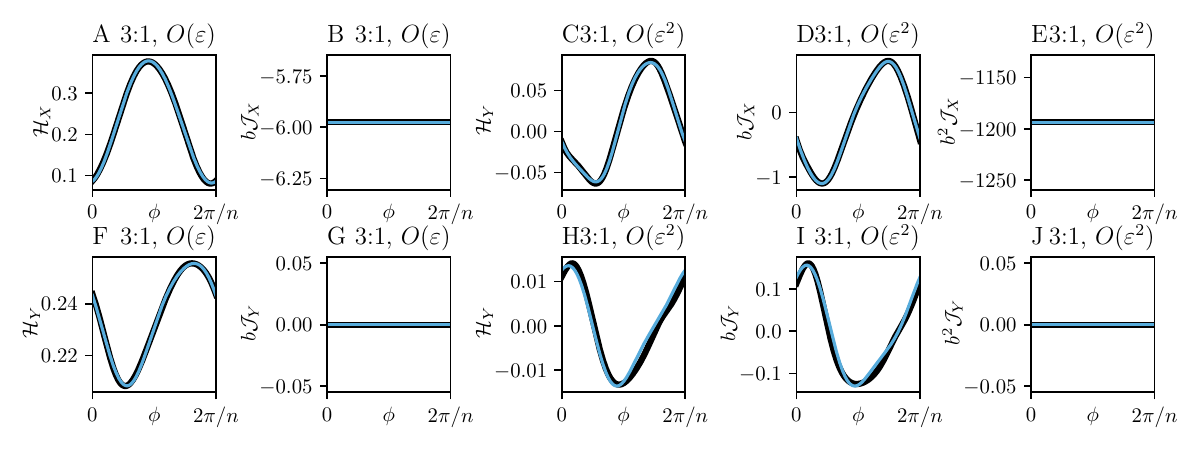}
    \caption{Original $\mathcal{H}$- and $\mathcal{J}$-functions (black) vs the corresponding Fourier approximation (blue) for $3{:}1$ phase-locking for the pair of coupled thalamic models.}
    \label{fig:a:31}
\end{figure}

Plots of the following $\mathcal{H}$ and $\mathcal{J}$ functions are shown in Figure \ref{fig:a:31}.

\begin{align*}
\mathcal{H}_X^{(1)}&= 0.2342-0.1437\cos(3x) -0.0047\cos(6x) +0.0370\sin(3x) \\
\mathcal{J}_X^{(1)}&= -5.9757\\
\mathcal{H}_X^{(2)}&= 0.0093-0.0147\cos(3x) -0.0020\cos(6x) -0.0045\cos(9x) \\
&\quad -0.0715\sin(3x) +0.0067\sin(6x) \\
\mathcal{J}_X^{(2,1)}&= -0.1349-0.2452\cos(3x) -0.0203\cos(9x) -0.8977\sin(3x) \\
&\quad -0.0673\sin(6x) +0.0272\sin(9x) \\
\mathcal{J}_X^{(2,2)}&= -1193.5286\\
\mathcal{H}_Y^{(1)} &=0.2342+0.0062\cos(3x) +0.0028\cos(6x) -0.0226\sin(3x) \\
\mathcal{J}_Y^{(2,1)} &=0\\
\mathcal{H}_Y^{(2)} &=0.0126\cos(3x) +0.0027\sin(6x) \\
\mathcal{J}_Y^{(2,1)} &=0.1276\cos(3x) +0.0375\sin(3x) +0.0312\sin(6x) \\
\mathcal{J}_Y^{(2,2)} &=0
\end{align*}

\subsubsection{2:3}
\begin{figure}[ht!]
    \centering
    \includegraphics[width=\textwidth]{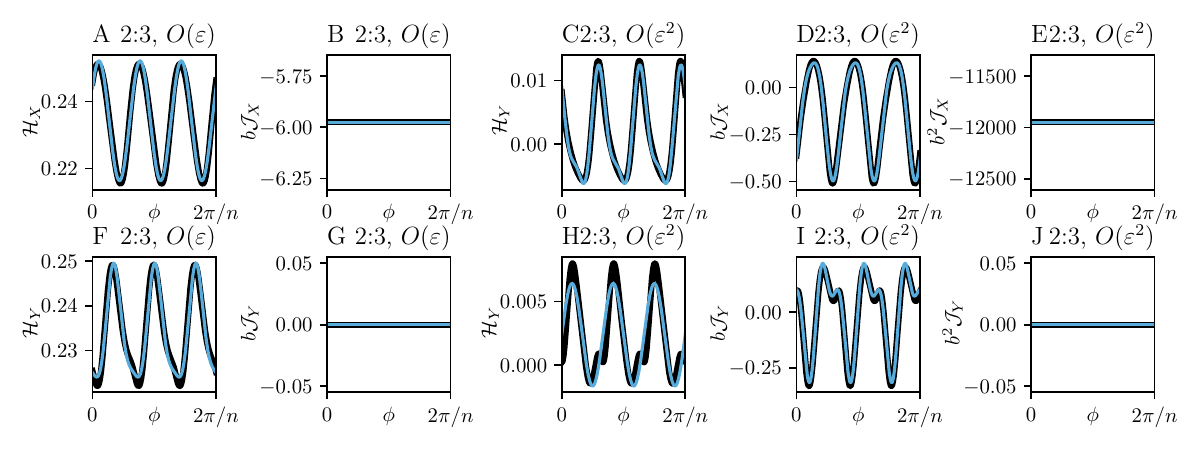}
    \caption{Original $\mathcal{H}$- and $\mathcal{J}$-functions (black) vs the corresponding Fourier approximation (blue) for $2{:}3$ phase-locking for the pair of coupled thalamic models.}
    \label{fig:a:23}
\end{figure}

Plots of the following $\mathcal{H}$ and $\mathcal{J}$ functions are shown in Figure \ref{fig:a:23}.

\begin{align*}
\mathcal{H}_X^{(1)}&= 0.2342+0.0098\cos(6x) +0.0149\sin(6x)\\ 
\mathcal{J}_X^{(1)}&= -5.9757\\
\mathcal{H}_X^{(2)} &=0.0014+0.0075\cos(6x) -0.0039\sin(6x) -0.0028\sin(12x) \\
\mathcal{J}_X^{(2,1)}&= -0.1346-0.2507\cos(6x) +0.1866\sin(6x) +0.0516\sin(12x) \\
\mathcal{J}_X^{(2,2)}&= -11949.8445\\
\mathcal{H}_Y^{(1)}&= 0.2342-0.0122\cos(6x) +0.0030\cos(12x) -0.0022\sin(6x) \\
\mathcal{J}_Y^{(1)}&=0\\
\mathcal{H}_Y^{(2)} &=0.0024+0.0040\sin(6x) \\
\mathcal{J}_Y^{(2,1)}&= 0.0378\cos(6x) +0.0623\cos(12x) -0.2033\sin(6x) \\
&\quad +0.0992\sin(12x) \\
\mathcal{J}_Y^{(2,2)} &=0
\end{align*}

\subsubsection{3:2}
\begin{figure}[ht!]
    \centering
    \includegraphics[width=\textwidth]{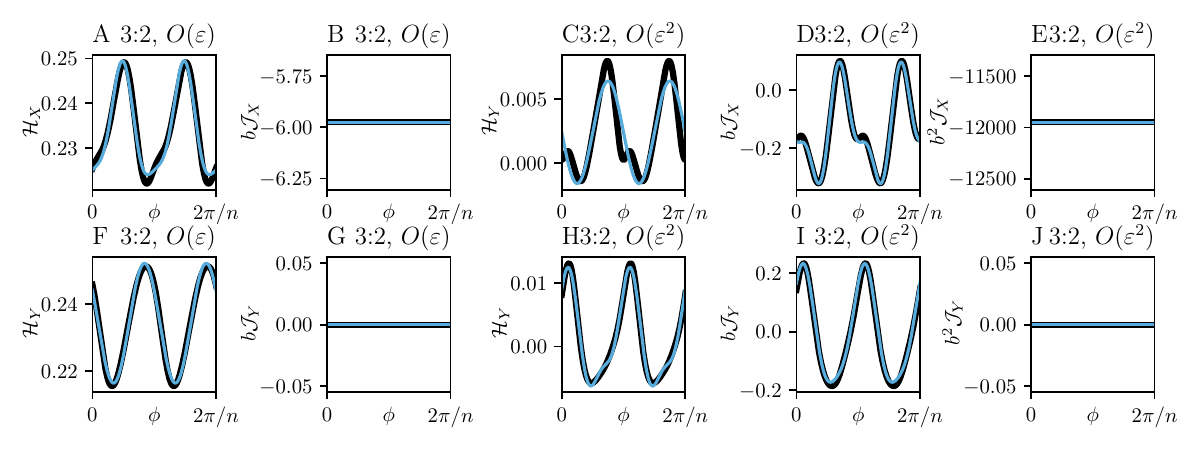}
    \caption{Original $\mathcal{H}$- and $\mathcal{J}$-functions (black) vs the corresponding Fourier approximation (blue) for $3{:}2$ phase-locking for the pair of coupled thalamic models.}
    \label{fig:a:32}
\end{figure}

Plots of the following $\mathcal{H}$ and $\mathcal{J}$ functions are shown in Figure \ref{fig:a:32}.

\begin{align*}
    \mathcal{H}_X^{(1)}&= 0.2342-0.0122\cos(6x) +0.0030\cos(12x) +0.0022\sin(6x) \\
\mathcal{J}_X^{(1)}&= -5.9757\\
\mathcal{H}_X^{(2)} &=0.0024-0.0040\sin(6x) \\
\mathcal{J}_X^{(2,1)} &=-0.1349-0.0424\cos(12x) -0.1655\sin(6x) +0.0695\sin(12x) \\
\mathcal{J}_X^{(2,2)} &=-11949.8408\\
\mathcal{H}_Y^{(1)} &=0.2342+0.0098\cos(6x) -0.0149\sin(6x) \\
\mathcal{J}_Y^{(1)}&=0 \\
\mathcal{H}_Y^{(2)} &=0.0014+0.0075\cos(6x) +0.0039\sin(6x) +0.0028\sin(12x) \\
\mathcal{J}_Y^{(2,1)} &=0.1659\cos(6x) +0.1126\sin(6x) +0.0339\sin(12x) \\
\mathcal{J}_Y^{(2,2)} &=0
\end{align*}

\subsubsection{3:4}
\begin{figure}[ht!]
    \centering
    \includegraphics[width=\textwidth]{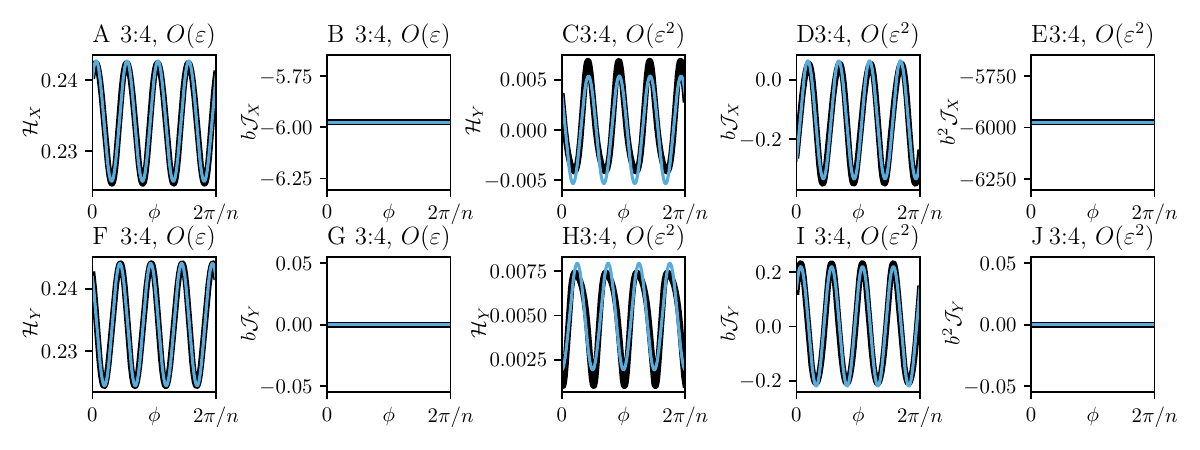}
    \caption{Original $\mathcal{H}$- and $\mathcal{J}$-functions (black) vs the corresponding Fourier approximation (blue) for $3{:}4$ phase-locking for the pair of coupled thalamic models.}
    \label{fig:a:34}
\end{figure}

Plots of the following $\mathcal{H}$ and $\mathcal{J}$ functions are shown in Figure \ref{fig:a:34}.

\begin{align*}
\mathcal{H}_X^{(1)}&= 0.2342+0.0062\cos(12x) +0.0059\sin(12x) \\
\mathcal{J}_X^{(1)}&= -5.9757\\
\mathcal{H}_X^{(2)} &=0.0035\cos(12x) -0.0040\sin(12x) \\
\mathcal{J}_X^{(2,1)}&= -0.1349-0.1352\cos(12x) +0.1448\sin(12x) \\
\mathcal{J}_X^{(2,2)} &=-5974.1118\\
\mathcal{H}_Y^{(1)}&= 0.2342+0.0076\cos(12x) -0.0063\sin(12x) \\
\mathcal{J}_Y^{(1)}&=0\\
\mathcal{H}_Y^{(2)}&= 0.0049-0.0030\cos(12x) \\
\mathcal{J}_Y^{(2,1)}&=0.1398\cos(12x) +0.1691\sin(12x) \\
\mathcal{J}_Y^{(2,2)} &=0
    \end{align*}

\subsubsection{4:3}
\begin{figure}[ht!]
    \centering
    \includegraphics[width=\textwidth]{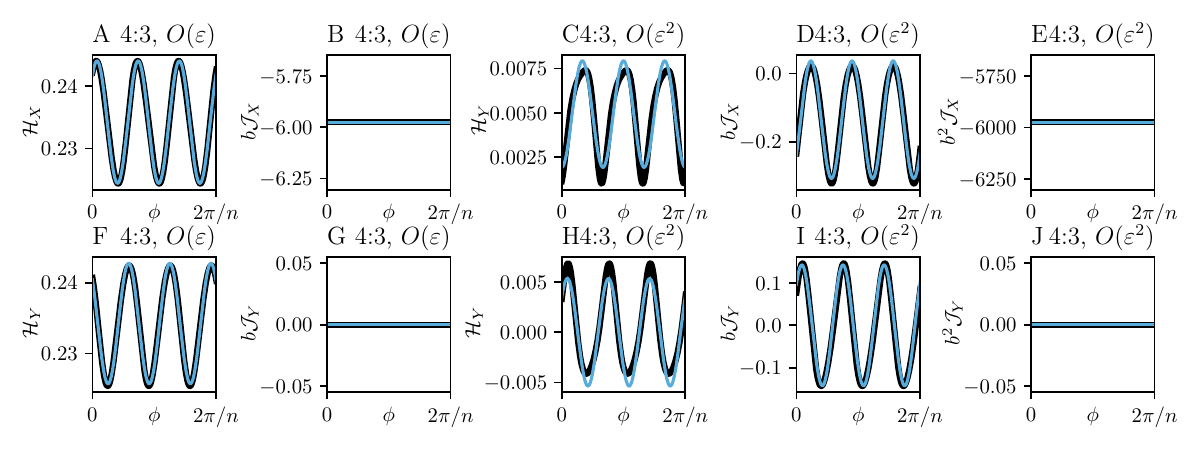}
    \caption{Original $\mathcal{H}$- and $\mathcal{J}$-functions (black) vs the corresponding Fourier approximation (blue) for $4{:}3$ phase-locking for the pair of coupled thalamic models.}
    \label{fig:a:43}
\end{figure}

Plots of the following $\mathcal{H}$ and $\mathcal{J}$ functions are shown in Figure \ref{fig:a:43}.

\begin{align*}
\mathcal{H}_X^{(1)} &=0.2342+0.0076\cos(12x) +0.0063\sin(12x)\\ 
\mathcal{J}_X^{(1)}&= -5.9757\\
\mathcal{H}_X^{(2)}&= 0.0049-0.0030\cos(12x) \\
\mathcal{J}_X^{(2,1)}&= -0.1352-0.1020\cos(12x) +0.1385\sin(12x) \\
\mathcal{J}_X^{(2,2)} &=-5974.1083\\
\mathcal{H}_Y^{(1)} &=0.2342+0.0062\cos(12x) -0.0059\sin(12x) \\
\mathcal{J}_Y^{(1)}&=0\\
\mathcal{H}_Y^{(2)} &=0.0035\cos(12x) +0.0040\sin(12x) \\
\mathcal{J}_Y^{(2,1)} &=0.0973\cos(12x) +0.1039\sin(12x) \\
\mathcal{J}_Y^{(2,2)} &=0
\end{align*}

\yp{\subsection{Van der Pol Coupled to Thalamic}\label{a:vdp_thalamic}}

\subsubsection{1:1}
\begin{figure}[ht!]
    \centering
    \includegraphics[width=\textwidth]{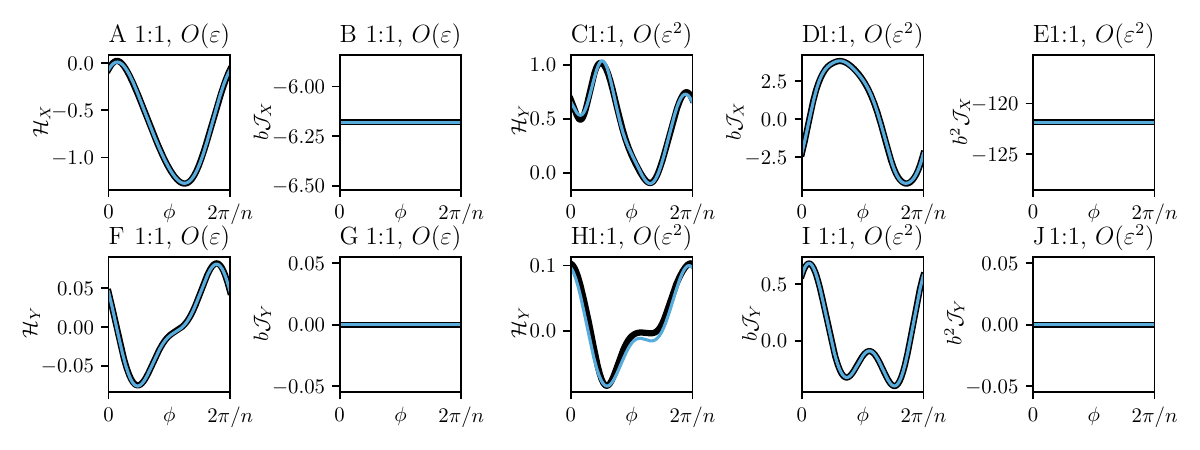}
    \caption{\yp{Original $\mathcal{H}$- and $\mathcal{J}$-functions (black) vs the corresponding Fourier approximation (blue) for $1{:}1$ phase-locking for the Van der Pol oscillator coupled to a thalamic neuron model.}}
    \label{fig:a:vdp11}
\end{figure}

\yp{Plots of the following $\mathcal{H}$ and $\mathcal{J}$ functions are shown in Figure \ref{fig:a:vdp11}.}

\yp{\begin{align*}
\mathcal{H}_X^{(1)} &=-0.6517+0.5244\cos(x) +0.0585\cos(2x) +0.3539\sin(x)\\
\mathcal{J}_X^{(1)}&= -6.1799\\
\mathcal{H}_X^{(2)}&= 0.4623+0.2425\cos(x) -0.0753\cos(2x) +0.0269\cos(4x) \\
&\quad +0.3560\sin(x) -0.1776\sin(2x) -0.1127\sin(3x)\\
\mathcal{J}_X^{(2,1)}&= 0.3413-2.2822\cos(x) -0.0511\cos(2x) -0.1004\cos(3x)\\
&\quad -0.1225\cos(4x) -0.0332\cos(5x) +3.3898\sin(x) +0.7445\sin(2x) \\
&\quad +0.0215\sin(4x)\\
\mathcal{J}_X^{(2,2)} &=-121.8605\\
\mathcal{H}_Y^{(1)}&= 0.0167\cos(2x) -0.0591\sin(x) -0.0188\sin(2x)\\
\mathcal{J}_Y^{(1)}&=0\\
\mathcal{H}_Y^{(2)} &=0.0578\cos(x) +0.0386\cos(2x) -0.0375\sin(x)\\
\mathcal{J}_Y^{(2,1)} &=0.3496\cos(x) +0.2191\cos(2x) +0.0078\cos(3x)\\
&\quad +0.0027\cos(4x) +0.1754\sin(x) +0.1999\sin(2x) -0.0061\sin(3x)\\
\mathcal{J}_Y^{(2,2)} &=0
\end{align*}}

\subsubsection{1:2}
\begin{figure}[ht!]
    \centering
    \includegraphics[width=\textwidth]{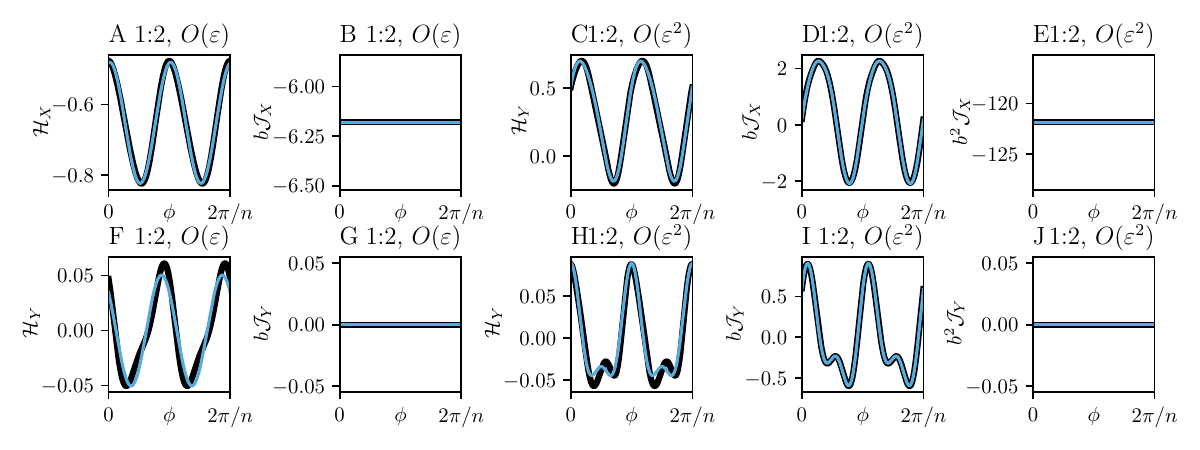}
    \caption{\yp{Original $\mathcal{H}$- and $\mathcal{J}$-functions (black) vs the corresponding Fourier approximation (blue) for $1{:}2$ phase-locking for the Van der Pol oscillator coupled to a thalamic neuron model.}}
    \label{fig:a:vdp12}
\end{figure}

\yp{Plots of the following $\mathcal{H}$ and $\mathcal{J}$ functions are shown in Figure \ref{fig:a:vdp12}}.

\yp{\begin{align*}
\mathcal{H}_X^{(1)} &= -0.6517+0.1696\cos(2x) +0.0214\sin(2x) \\
\mathcal{J}_X^{(1)}&=-6.1799\\
\mathcal{H}_X^{(2)}&= 0.2885+0.1758\cos(2x) +0.0481\cos(4x) +0.3977\sin(2x) \\
\mathcal{J}_X^{(2,1)}&=  0.3413-0.4142\cos(2x) +0.2517\cos(4x) \\
&\quad +0.0240\cos(8x) +2.1403\sin(2x) +0.0971\sin(4x) \\
\mathcal{J}_X^{(2,2)} &=-121.8605\\
\mathcal{H}_Y^{(1)}&= 0.0347\cos(2x) -0.0365\sin(2x) \\
\mathcal{J}_Y^{(1)} &=0\\
\mathcal{H}_Y^{(2)} &=0.0614\cos(2x) +0.0275\cos(4x) \\
\mathcal{J}_Y^{(2,1)} &=0.4239\cos(2x) +0.1694\cos(4x) +0.4149\sin(2x)+0.2794\sin(4x) \\
\mathcal{J}_Y^{(2,2)} &= 0
\end{align*}}

\subsubsection{2:1}
\begin{figure}[ht!]
    \centering
    \includegraphics[width=\textwidth]{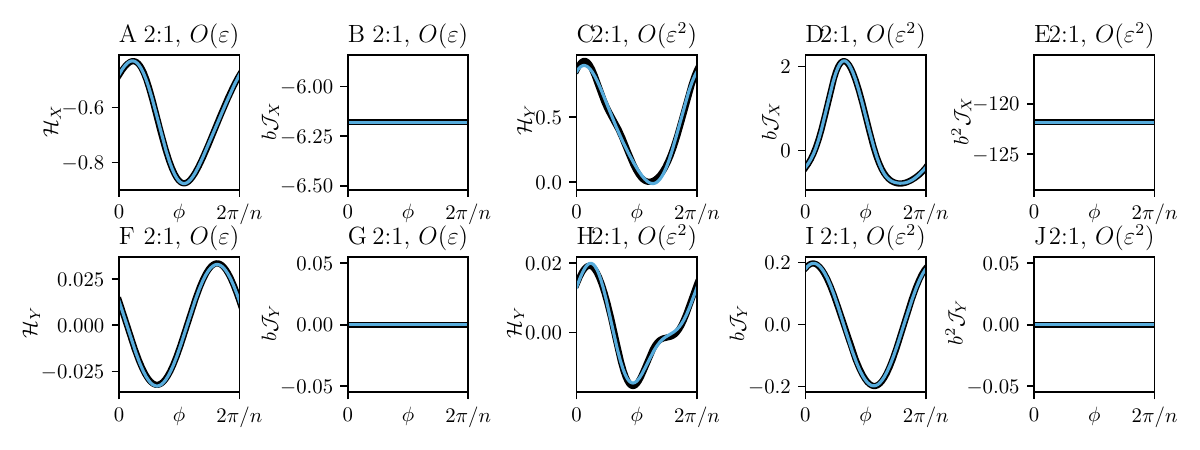}
    \caption{\yp{Original $\mathcal{H}$- and $\mathcal{J}$-functions (black) vs the corresponding Fourier approximation (blue) for $2{:}1$ phase-locking for the Van der Pol oscillator coupled to a thalamic neuron model.}}
    \label{fig:a:vdp21}
\end{figure}

\yp{Plots of the following $\mathcal{H}$ and $\mathcal{J}$ functions are shown in Figure \ref{fig:a:vdp21}.}

\yp{\begin{align*}
\mathcal{H}_X^{(1)} &=-0.6517+0.1930\cos(2x) -0.0231\cos(4x) +0.0983\sin(2x) +0.0138\sin(4x) \\
\mathcal{J}_X^{(1)} &=-6.1799\\
\mathcal{H}_X^{(2)}&=0.4208+0.3725\cos(2x) +0.0572\cos(4x) +0.2409\sin(2x) \\
\mathcal{J}_X^{(2,1)}&=0.3459-0.6404\cos(2x) -0.1585\cos(4x) +0.0287\cos(6x) \\
&\quad +0.0249\cos(8x) +1.2614\sin(2x) -0.3095\sin(4x) -0.0274\sin(6x) \\
\mathcal{J}_X^{(2,2)}&=-121.8370\\
\mathcal{H}_Y^{(1)}&= 0.0132\cos(2x) -0.0300\sin(2x) \\
\mathcal{J}_Y^{(1)}&= 0\\
\mathcal{H}_Y^{(2)} &=0.0017+0.0135\cos(2x) -0.0022\cos(4x) +0.0050\sin(2x) +0.0051\sin(4x)\\
\mathcal{J}_Y^{(2,1)} &=0.1798\cos(2x) +0.0824\sin(2x) \\
\mathcal{J}_Y^{(2,2)} &=0
\end{align*}}

\subsubsection{2:3}
\begin{figure}[ht!]
    \centering
    \includegraphics[width=\textwidth]{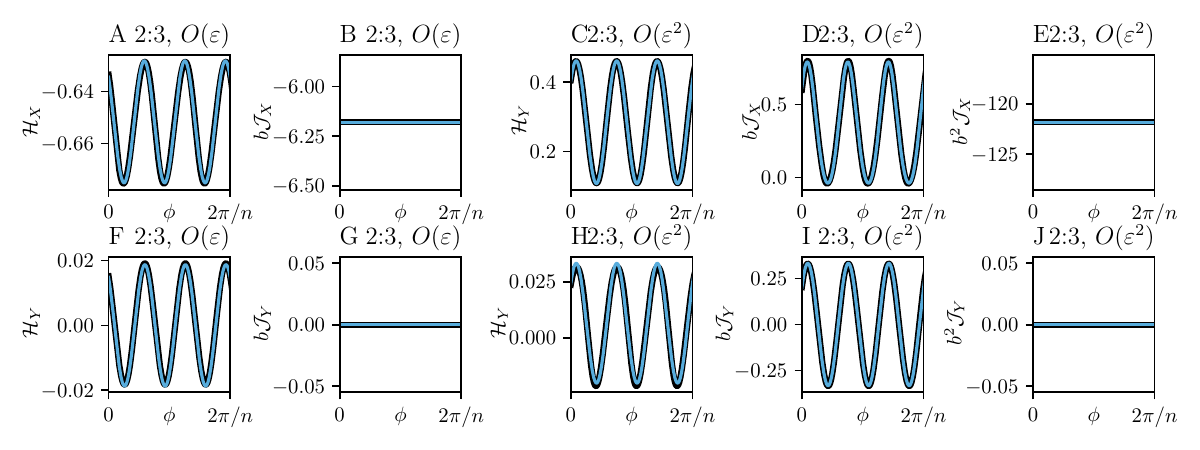}
    \caption{\yp{Original $\mathcal{H}$- and $\mathcal{J}$-functions (black) vs the corresponding Fourier approximation (blue) for $2{:}3$ phase-locking for the Van der Pol oscillator coupled to a thalamic neuron model.}}
    \label{fig:a:vdp23}
\end{figure}

\yp{Plots of the following $\mathcal{H}$ and $\mathcal{J}$ functions are shown in Figure \ref{fig:a:vdp23}.}

\yp{\begin{align*}
\mathcal{H}_X^{(1)}&=-0.6517+0.0173\cos(6x) -0.0161\sin(6x) \\
\mathcal{J}_X^{(1)}&=-6.1799\\
\mathcal{H}_X^{(2)} &=0.2842+0.1183\cos(6x) +0.1299\sin(6x) \\
\mathcal{J}_X^{(2,1)}&=0.3459+0.2733\cos(6x) +0.3142\sin(6x) +0.0312\sin(12x) \\
\mathcal{J}_X^{(2,2)}&=-121.8370\\
\mathcal{H}_Y^{(1)}&=0.0148\cos(6x) -0.0114\sin(6x) \\
\mathcal{J}_Y^{(1)}&=0\\
\mathcal{H}_Y^{(2)} &=0.0065+0.0177\cos(6x) +0.0201\sin(6x)\\
\mathcal{J}_Y^{(2,1)}&=0.2015\cos(6x) +0.2647\sin(6x) \\
\mathcal{J}_Y^{(2,2)} &=0
\end{align*}}

\subsubsection{3:2}
\begin{figure}[ht!]
    \centering
    \includegraphics[width=\textwidth]{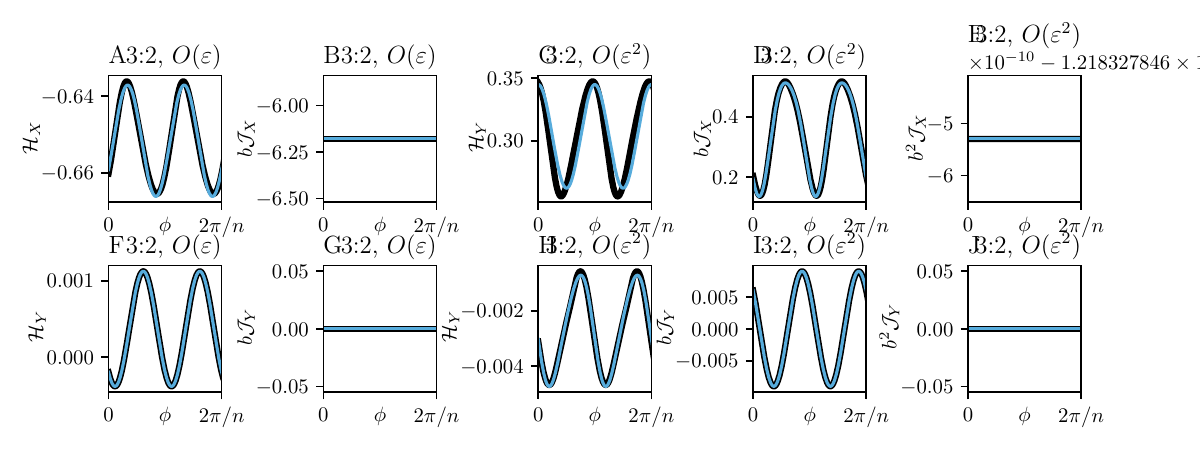}
    \caption{\yp{Original $\mathcal{H}$- and $\mathcal{J}$-functions (black) vs the corresponding Fourier approximation (blue) for $3{:}2$ phase-locking for the Van der Pol oscillator coupled to a thalamic neuron model.}}
    \label{fig:a:vdp32}
\end{figure}

\yp{Plots of the following $\mathcal{H}$ and $\mathcal{J}$ functions are shown in Figure \ref{fig:a:vdp32}.}

\yp{\begin{align*}
    \mathcal{H}_X^{(1)}&=-0.6517-0.0075\cos(6x) +0.0124\sin(6x) \\
\mathcal{J}_X^{(1)}&=-6.1799\\
\mathcal{H}_X^{(2)} &=0.3035+0.0413\cos(6x) \\
\mathcal{J}_X^{(2,1)} &=0.3468-0.1530\cos(6x) -0.1072\sin(6x) -0.0247\sin(12x) \\
\mathcal{J}_X^{(2,2)} &=-121.8328\\
\mathcal{H}_Y^{(1)} &=0.0004-0.0006\cos(6x) -0.0005\sin(6x) \\
\mathcal{J}_Y^{(1)}&=0\\
\mathcal{H}_Y^{(2)} &=-0.0027-0.0004\cos(6x) -0.0019\sin(6x) -0.0002\sin(12x) \\
\mathcal{J}_Y^{(2,1)} &=0.0060\cos(6x) -0.0067\sin(6x) \\
\mathcal{J}_Y^{(2,2)} &=0
\end{align*}}

\end{document}